\RequirePackage[2018-12-01]{latexrelease}
\documentclass[final,review=false]{rQUF2e}
\usepackage{epstopdf}
\usepackage{xcolor}
\usepackage{soul}
\usepackage{bbm}
\usepackage{dsfont}
\usepackage[font={small}]{subcaption}
\usepackage[font={small}]{caption}
\usepackage{float}
\usepackage{pdflscape}
\usepackage{booktabs}
\theoremstyle{plain}

\usepackage[
backend=biber,
style=authoryear,
sorting=nyt,
]{biblatex}
\usepackage{pdflscape}
\usepackage{tabularx}
\theoremstyle{definition}

\theoremstyle{assumption}
\newtheorem{assumption}{Assumption}
\usepackage{soul}
\theoremstyle{remark}

\newcommand{\highlight}{\textcolor{black}}
\newcommand{\highlighttwo}{\textcolor{black}}

\begin{document}


\title{No Tick-Size Too Small: A General Method for Modelling Small-Tick Limit Order Books }

\author{Konark Jain$^{\ast}$$^\dag$$^\ddag$\thanks{$^\ast$Corresponding author.
Email: konark.jain.23@ucl.ac.uk \newline $\dag$ Opinions expressed in this paper are those of the authors, and do not necessarily reflect the view of JP Morgan.}, Jean-François Muzy$^{\P}$, Jonathan Kochems$^\dag$, Emmanuel Bacry$^{\S}$\\
\affil{$\ddag$Department of Computer Science, University College London, London, UK\\
$\P$ SPE CNRS-UMR 6134, Universit\'e de Corse BP 52, 20250 Corte, France\\
$\S$CEREMADE, CNRS-UMR 7534, Universit\'e Paris-Dauphine PSL\\ Place du Mar\'echal de Lattre de Tassigny, 75016 Paris, France\\
$\dag$Quantitative Research, JP Morgan Chase, London, UK}  }
\maketitle

\begin{abstract}
Tick-sizes not only influence the granularity of the price formation process but also affect market agents' behavior. We investigate the disparity in the microstructural properties of the Limit Order Book (LOB) across a basket of assets with different relative tick-sizes. A key contribution of this study is the identification of several stylized facts, which are used to differentiate between large, medium, and small-tick assets, along with clear metrics for their measurement. We provide cross-asset visualizations to illustrate how these attributes vary with relative tick-size. Further, we propose a Hawkes Process model that {\color{black}not only fits well for large-tick assets, but also accounts for }sparsity, multi-tick level price moves, and the shape of the LOB in small-tick assets. Through simulation studies, we demonstrate the {\color{black} versatility} of the model and identify key variables that determine whether a simulated LOB resembles a large-tick or small-tick asset. Our tests show that stylized facts like sparsity, shape, and relative returns distribution can be smoothly transitioned from a large-tick to a small-tick asset using our model. We test this model's assumptions, showcase its challenges and propose questions for further directions in this area of research.
\end{abstract}


\newpage

\section{Introduction}

The rapid electronification of the financial market has made the data structure Limit Order Books the central domain of essentially all trading, particularly for equities. Limit Order Books match sellers with the buyers according to their price priority first, and queue priority second. These markets are classified as order driven instead of quote driven since there exists a \emph{lit}, or in other words publicly visible, list of unmatched orders that any market participant can utilize for their trading needs. There are largely three kinds of orders: Market Orders (MOs), Limit Orders (LOs) and Cancel Orders (COs). Limit orders are representative of a market agent willing to buy / sell a certain quantity of the security at a certain price. Cancel Orders are cancellations of existing LOs by the market agent. Market orders are similar to limit orders however their price is aggressive enough to match quotes on the opposite side of the LOB (for eg. a buy MO will match with the LOs at ask side of the LOB). A lot of variations, provided by the exchange to their customers for a variety of reasons, exist in these three order types however these orders, in their essential components, have an order price and order size. The order price defines the price at which the agent placing the order is willing to buy or sell the asset. 

The granularity of the order price in modern lit LOBs is set by a quantity known as the \emph{tick-size}. By definition it is the minimum difference between two distinct price levels. Tick-sizes vary both by asset classes and markets. Focusing on equities, the tick-sizes are set to be constant in the US equities exchanges uniformly for all stocks\footnote{Certain penny stocks have a tick-size of one hundredth of a cent however their market capitalisation is not significant.} to one cent (or 0.01\$).  However in Europe, the tick-sizes change with the price of each security on a ladder like logic. We will focus on US securities in this work. There are several discussions in the literature on which tick-size regime is optimal for the welfare of market participants (\cite{bonart2017optimal}, \cite{graziani_optimal_2023}, \cite{CHUNG2020879}, \cite{chao2018}, \cite{baldacci2023}). Tick-sizes not only influence the granularity of the price formation process but also affect market agents' behavior (\cite{ohara2018}, \cite{CHAKRABARTY2022100658}, \cite{Dyhrberg_Foley_Svec_2023}). \cite{LaSpada2011} establish certain facts of the long-term behaviour of the price process as they vary the tick-size. We note that the relation between tick-size and short term microstructural dynamics of a asset's LOB has been well documented in the literature (\cite{huang2019glostenmilgromlimitorderbook}, \cite{dayri2015}). \cite{robert2010} showcase that the tick-size granularity price grid has certain zones around it which define where the `realized' price will land on the grid given a continuous `efficient' price of the security. 

We define the ratio of the tick-size (in \$) to the price (in \$) as the {\em    relative tick-size}. We report this ratio in units of bps which is equal to $10^{-4}$. Numerous stylized facts of asset returns, along with their potential dependence on the relative tick-size, have been thoroughly reviewed in \cite{bouchaud2018trades}.
It turns out that large-tick assets (i.e. assets with a large relative tick-size) and small-tick assets (those with a small relative tick-size) exhibit very different behavior as respect to their microstructural properties.
Some securities that possess some traits of a large-tick asset and some other traits of a small-tick asset are categorized as `medium-tick' securities. To differentiate between large- and small-tick asset features, \cite{bouchaud2018trades} analyze various microstructural characteristics, including the bid-ask spread, limit order arrival rates, order size distributions, LOB shape, and top-of-book volumes relative to daily traded volumes. Similarly, \cite{briola2024deep}, also test the distribution of bid-ask spread, volumes at the top of the LOB and the distance between the top of the LOB and the 10th best top of the LOB against the three categories of relative tick-size. They find that each of these three stylized facts differ remarkably across these three categories. Further in \cite{briola2024hlob}, the authors show that the mutual information of the LOB top 10 best queues on both sides of the LOB have significantly different structures across these three categories of relative tick-size. All these findings highlight the need for distinct modeling approaches as respect to the relative tick-size.


\label{hpLitReview} In terms of modelling the LOB dynamics, \cite{Gould2013} conducted a comprehensive review of Limit Order Books (LOBs), examining their characteristics and presenting various models for LOB simulation. Similarly, \cite{cont2011} provided a survey that highlighted the effectiveness of several zero-intelligence models in LOB modeling, offering empirical observations to validate the models' outputs. For a more detailed exploration of the microstructural statistics of the LOB and the associated modeling techniques, the work by \cite{abergel_anane_chakraborti_jedidi_munitoke_2016} serves as a key reference. \cite{jain2024limit} is a more recent review with a breakdown on the basis of the core methodology used, and the stylized facts tested. Particularly, they note that the Hawkes Process has emerged as a promising model for addressing the limitations of Poisson Processes in modeling Limit Order Book (LOB) queueing systems. \cite{Bacry2015} provide a comprehensive review, outlining the fundamental concepts of the Hawkes Process, including its mathematical foundations, key properties, and various applications, with a particular focus on Limit Order Book models. Recent studies, such as those by \cite{Hawkes2018}, have further explored the financial applications of Hawkes Processes, underscoring their utility in modeling a range of market phenomena. The extension to multidimensional Hawkes Processes has gained traction in LOB modeling, leading to the development of various formulations. For instance, \cite{toke2010market} introduces a two-agent model employing one-dimensional Hawkes Processes for market orders and limit orders. \cite{bacry2016estimation} take a different approach by categorizing LOB events and utilizing an eight-dimensional Hawkes Process to model the bid and ask sides separately. \cite{kirchner2017estimation} propose a non-parametric estimation method for Hawkes Processes, optimizing hyperparameters using the AIC statistic. \cite{jain2023hawkes} extend the non-parametric estimation method for slowly decaying kernels. They develop a Compound Hawkes Process with spread closing events being state-dependent to model the LOB while being cognizant of the distribution of individual orders' sizes. They showcase the fact that their methodology is general enough to be applied to all kinds of assets however due to their model assumptions being too strong for small-tick assets, their model fails in matching the stylized facts of small-tick equities.  

To date, the modeling of the Limit Order Book has focused predominantly on large-tick assets, leaving a significant gap in the literature concerning small-tick assets. We categorize an LOB model as a \textit{full} LOB model if the modelling method is general enough to be applied to deeper dimensions rather than just the top of the LOB. We review the mathematical methodology of some of the state of the art in the following. \cite{Abergel2015}, using the exponential kernels of a multivariate Hawkes Process, model a moving frame of the LOB with $K$ price limits on each side of the LOB measured from the opposite best quote. Therefore the frame moves with events at both sides. They develop the generator operator and thereafter long time limits of the price and volume processes from this model. Using a different approach, \cite{cont2021stochastic} propose an SPDE (Stochastic Partial Differential Equations) model for the centered volume density $u(x, t)$, representing volume at $x$ ticks from the mid-price. Limit orders and market orders are modeled as functions of $x$, with cancellations handled in three ways: constant multipliers over Limit Order intensities, diffusion for cancel-and-replace near existing levels, and convection towards the mid-price for liquidity seekers. A multiplicative Wiener process is added to capture high-frequency trading activity. Similarly, \cite{horst2019scaling} develop step functions for instantaneous volume density on each side of the LOB, as functions of time and an absolute price coordinate $y$. Their framework is quite general and they formulate the price and volume dynamics within this framework and study their long time limits.  While these SPDE models are useful for calculations of the steady state behaviour of a general LOB, we still stress the gap in literature on simulating a discrete price grid general LOB which is faithful to the stylized facts of the LOB. 

{\color{black}This study presents visualizations of key stylized facts in market microstructure and seeks to quantify the specific relationship between decreasing relative tick size and the observed trends.} To the best of our knowledge, this is the first work which defines quantifiable statistics of the LOB dynamics and depict their variation with relative tick-size of several assets. We showcase the fact that with the constant tick-size of the US securities and with a wide range of associated prices of the security, the LOB behaves significantly differently with varying relative tick-size. The aim of this discussion is to then highlight the challenges of modelling LOBs in a situation when the relative tick-size is very small. We tackle the problem of simulating a general LOB, regardless of its relative tick-size, by modelling the LOB dynamics using a point process. Our contribution can be summarised as:

\begin{enumerate}
    \item In Section \ref{sty}, we provide evidence for a set of stylized facts and introduce clear metrics for their measurement. We showcase the fact that certain stylized facts vary in a monotonous manner from large-tick to small-tick assets. In order to motivate the interest of using Hawkes Process models across all relative tick-sizes, we notably consider a stylized fact known as \emph{leverage}.
    \item In Section \ref{model}, we propose an extension to the Compound Hawkes Model in \cite{jain2023hawkes} which is also capable of modeling small-tick assets with a sparse LOB and wide spreads. We introduce the notion {\color{black}``meta-queues''} of the LOB in order to maintain model parsimony.
    \item In Section \ref{results}, we showcase the universality of this model (from small to large-tick assets) by depicting various attributes of the LOB by varying the parameters of this model. We further identify certain `critical' parameters which can be controlled to have a large- or small-tick simulated LOB.
    \item In Section \ref{calib}, we discuss the calibration techniques and then  {\color{black}in the appendix (Appendix \ref{assump}),} we challenge the assumptions of our model using empirical evidence. Finally, we propose future research directions to tackle them within our model's framework.
\end{enumerate}

\section{Stylized Facts related to Tick-Size} \label{sty}

\highlight{To assess the variations in microstructural properties of the Limit Order Book (LOB) across different  tick-sizes, we analyze several key aspects of LOB states over a one-year period, and report their empirical characteristics. These characteristics are commonly referred to in the literature as stylized facts, which represent consistent empirical regularities observed in financial markets (\cite{cont01}, \cite{bouchaud2018trades}). The selection of stylized facts is intended to ensure that the most salient features of the LOB microstructure are captured within this statistical framework, thereby facilitating a comprehensive understanding of its dynamic behavior.} 

In this section, we present an analysis of several stylized facts for 15 assets over a whole year's LOB data (from 2nd January 2019 to 30th December 2019) i.e. 250 trading days. \highlight{Some basic statistics related to each asset of this collection are shown in Table \ref{tab:stocks}. For each asset, high-resolution, tick-by-tick LOB data obtained from the LOBSTER provider (\cite{LOBSTER}) is employed. } Let us point out that the assets in our collection are all US equities with large capitalization and therefore have all a constant tick size $\delta=1$ cent \addtocounter{footnote}{+1}\footnote{Some US equities have a tick size of one-hundredth of a cent however their market capitalization is very small compared to the equities considered in this study.}. 

\begin{table}[h]
\centering
\makebox[.8\linewidth][c]{%
\begin{tabular}{|c|c|c|c|c|}
\hline
    \textbf{Asset} & \textbf{Avg Spread ($ \overline{s}$)} & \textbf{Avg Mid-Price ($ \langle P_{mid} \rangle$) } & \textbf{Relative Tick-Size ($ \epsilon $)} & \textbf{Category}  \\
    & \textbf{(in ticks)} & \textbf{(in \$)}& \textbf{(in bps)} & \\
    \hline
    SIRI & 1.03 & 6.07 & 16.47 & Large-tick \\
    BAC & 1.06 & 29.40 & 3.40 &Large-tick \\
    INTC & 1.09 & 45.50 & 2.19 &Large-tick \\
    CSCO & 1.09 & 51.31 & 1.94&Large-tick \\
    ORCL & 1.10 & 54.05 & 1.85&Large-tick \\    
    \hline
    ABBV & 2.30 & 76.86 & 1.30 &Medium-tick\\
    PM & 2.61 & 81.89 & 1.22&Medium-tick\\
    MSFT & 1.25 & 129.44 & 0.77&Medium-tick \\
    IBM & 3.37 & 137.94 & 0.72&Medium-tick\\
    AAPL & 1.97 & 208.62 & 0.48&Medium-tick \\
    \hline
    TSLA & 12.66 & 270.94& 0.36&Small-tick \\
    CHTR & 23.76 & 394.76 & 0.25&Small-tick \\
    GOOG & 57.98 &  1186.57 & 0.08&Small-tick \\
    AMZN & 46.81 & 1777.35 & 0.05&Small-tick \\
    BKNG & 162.39 & 1859.92& 0.05&Small-tick \\
    \hline
\end{tabular}
}
\caption{\textbf{Data Description:} We make use of 15 assets which are the constituents of S\&P500 as of 2nd January 2019 with a wide range of average mid-prices, average spread and relative tick-sizes. We use tick-by-tick data for one calendar year (2nd January 2019 to 30th December 2019) from \cite{LOBSTER}. The assets are categorized into Large- ($\epsilon \in [1.5,[$), Medium- ($\epsilon \in [0.4,1.5[$) and Small-tick ($\epsilon \in [0,0.4[$) according to the relative tick-size $\epsilon$.}
\label{tab:stocks}
\end{table}

In order our study to cover a large variety of dynamics, we chose a collection of assets with a wide range of average mid-price   varying from $\langle P_{mid} \rangle$ =6\$  (SIRI) to $\langle P_{mid} \rangle$ =1800\$ (BKNG). 
The mid-price average was computed over the whole period averaging on all the mid-price change events. In that sense it corresponds to an event-average quantity {\color{black}as per convention in the literature}\footnote{Let us note that, throughout the paper, given a time dependent quantity $a(t)$,  $\langle  a \rangle$ will denote the corresponding event-average (as for the average mid-price $\langle  P_{mid} \rangle$) and $\overline a$ will denote the corresponding time-weighted average (as for the average spread $\overline s$), see Eq. \eqref{eqn:s_bar}.}.

In order to study tick size related stylized facts, we have to choose a universal proxy that we will be used for comparing cross-asset tick-sizes.

\subsection{The relative tick size : a universal proxy to compare cross assets tick-sizes}
Traditionally, in order to compare cross asset tick size, one uses the average bid-ask spread value expressed in ticks (\cite{bouchaud2018trades, briola2024deep}). 
The bid-ask spread $s(t)$ of an asset at a given time $t$ is defined by the difference between the best ask price and the best bid price at time $t$. Thus it is defined  (when expressed in ticks) by : 
\begin{align}
    s(t) := \frac{p_{\text{ask}}(t) - p_{\text{bid}}(t)}{\delta}
\end{align}
The average spread can then be defined as a time-weighted average denoted $\overline{s}$ as follows : 
\begin{align}
   \overline{s} := \frac{\sum^N_{i=1} s(t_i) \Delta t_i}{T}, \label{eqn:s_bar}
\end{align}
where the $t_i$'s correspond to the timestamp of the jumps of the $s(t)$ process and $\Delta t_i = t_{i+1}-t_{i}$ and $T$ the total observed duration. {\color{black}This time-weighted averaging is performed with the objective of making the measurement of average spread to be robust to outliers that persist for a very short duration in the LOB. A simple event based average would likely be biased due to the presence of outliers.}

The average time-spread $\overline s$ of each asset of our assets collection are displayed in Table \ref{tab:stocks}. One sees that it covers a very wide range of value from 1.03 ticks (SIRI) to 162.39 ticks (BKNG).
{
However, although this quantity is commonly used as a proxy for comparing cross-asset tick size, it has two main drawbacks. }
First of all, it does not take directly into account the actual size (in the currency) of the tick, which, on some markets, could vary a lot even for very liquid and highly capitalized assets (e.g., London Stock Exchange, Xetra Frankfurt, etc.). It is natural to expect that this actual size has some impacts on actual statistical properties of the LOB dynamics. Secondly, the average spread size is mechanically floored by one. So, it prevents from comparing the tick size of two different large-tick assets.

Thus, in our study, as a proxy for comparing the tick size of the various assets, we prefer to use the relative tick-size { (measured in basis points, bps)} defined by 
\begin{equation}
\label{eq:epsilon}
 \epsilon := \frac{\delta}{\langle P_{mid} \rangle}\times 10^4 
\end{equation}
As shown in Table \ref{tab:stocks}, the relative tick size of the assets of our collections covers a very wide range of values from 16.47 bps (for SIRI) to 0.05 bps (for BKNG). 

Following this proxy, for the sake of simplicity, we decide to regroup the assets of our collection into 3 groups : 
\begin{itemize}
\item Large-tick assets ($\epsilon \ge 1.5$) : SIRI, BAC, INTC, CSCO, ORCL
\item Medium-tick assets ($\epsilon \in [0.4,1.5[$) : ABBV, PM, MSFT, IBM, AAPL
\item Small-tick assets ($\epsilon < 0.4$) : TSLA, CHTR, GOOG, AMZN, BKNG
\end{itemize}

We are now ready to present the various tick size related stylized facts.  

\subsection{Bid-Ask Spread : Mean and Distribution vs Relative Tick-Size} \label{spread}


\begin{figure}[h]
    \centering
    \includegraphics[width=.6\textwidth]{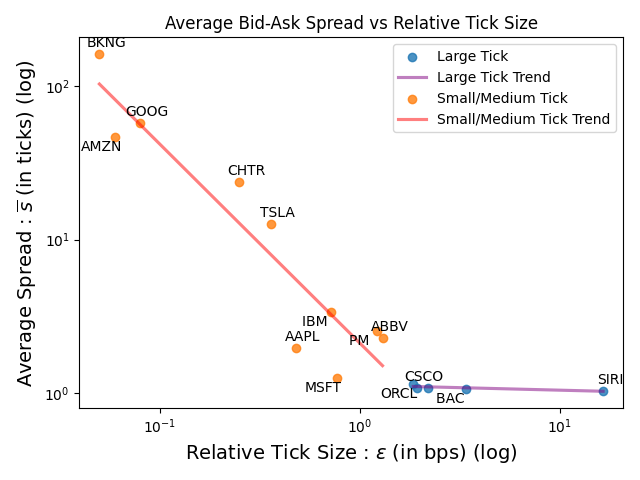}
    \caption{\highlighttwo{\textbf{Average Spread (in ticks) vs Relative Tick Size(in bps)}   { in a log-log scale : 
    Remarkably, the average spread appears to be highly correlated with the relative tick size. The figure reveals clearly two regimes : a regime for large-tick assets (purple plain-line) for which the average spread is basically constant and equal to 1 tick (the minimum constrained value for the spread), and a regime for small and medium-tick assets (orange plain-line) for which (as already pointed out in \cite{bouchaud2018trades}) the spread seems to be a power-law (Eq. \eqref{eq:pl}) of the relative tick-size  with an exponent close to -1  (the slope of the orange plain line is -1.23).}
    }}
    \label{fig:SpreadVsPrice}
\end{figure}

In Fig. \ref{fig:SpreadVsPrice}, we plot, 
{ for each asset of our collection}, the average spread $\overline{s}$ against the relative tick size $\epsilon$ on a log-log scale 
 to explore the relationship between relative tick-size and spread. 
Unsurprisingly, we observe a gradual increase in the mean bid-ask spread as the relative tick-size decreases across assets in the dataset. For instance, the mean spread for INTC, a large-tick asset, is $ 1.02 $ ticks, whereas for AMZN, a small-tick asset, it reaches $ 47.18 $ ticks. 
{ The figure clearly reveals two regimes. A first regime for large-tick assets, for which, unsurprisingly, due to the constraint that the spread cannot be smaller than 1, the average spread is basically constant and equal to 1. A second regime for medium and small-tick assets, for which  $\overline{s}$, as a function of $\epsilon$, seems to follow a power law with an exponent close to $-1$. 
Thus, we state the following approximate relations for some constant $k_s$:
\begin{align}
    \overline{s} \simeq & 
    \begin{cases}
         1, &\text{ for large-tick assets} \\
        k_s\epsilon^{-1}, &\text{ for small- and medium-tick assets} \label{eq:pl}
    \end{cases}
\end{align} 
{
Let us point out a similar relation (to the latter relation) was found in \cite{bouchaud2018trades}.}

  \begin{figure}[h]
    \centering
    \includegraphics[width=.6\textwidth]{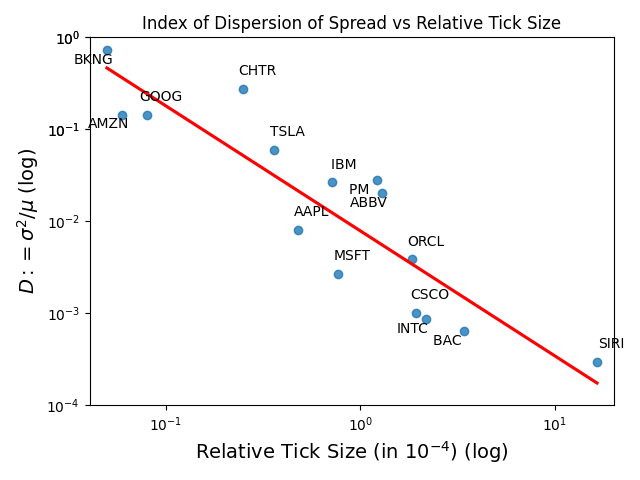}
    \caption{\highlighttwo{\textbf{Spread Distribution's Index of Dispersion vs Relative Tick-size}} { on a log-log scale} : We observe that the index of dispersion $D$ decreases while the relative tick-size increases. There seems to be a single regime for both small and large-tick assets implying a smooth transition of spread distribution's shape. The slope of the linear regression fit is  $-1.36$ corresponding to the power-law regime of Eq. \eqref{eq:dispersion}.
    }
    \label{fig:IoDSpread}
\end{figure}

Appendix \ref{app:spread} (Fig. \ref{fig:spreads}) focuses on the distribution of the instantaneous spread $s(t)$ for 3 different assets  (resp. large, medium and small-tick). The shape appears  to be very different depending on the tick-size. In order to characterize these shapes, we make use of the Index of Dispersion metric (noted $D$ hereafter) defined as the ratio of the variance of the distribution (denoted by $\sigma^2$) with the mean of the distribution (denoted by $\mu$) { i.e. $D := \frac{\sigma^2}{\mu}$.}
This choice is motivated by the empirical { observation} that with decreasing relative tick-size, one observes the mean shifting from one tick to multiple ticks however the tail of the distribution dies quicker as well. Therefore the mean increases while the variance does not seem to increase on an equivalent scale. We empirically measure the Index of Dispersion ($D$) over one calendar year. Fig. \ref{fig:IoDSpread} displays $D$ as a function of the relative tick-size { in a log-log scale}. One can clearly observe a definitive linear trend as we decrease the relative tick-size with a slope of  $-1.36$,  so we have the {\color{black} following approximate} relation 
\begin{equation}
\label{eq:dispersion}
D \propto \epsilon^{-\frac{3}{2}}.
\end{equation}
Thus the shape of the spread distribution becomes less and less dispersed when increasing the relative tick-size of an asset. Let us point out that, unlike the previous figure, there does not seem to be a regime change when we move from large to small-tick assets.
    
\subsection{Trade-driven mid-price changes: Probability Distribution vs Relative Tick-Size} \label{trades}
 \begin{figure}[h]
     \centering
    \includegraphics[width=.6\textwidth]{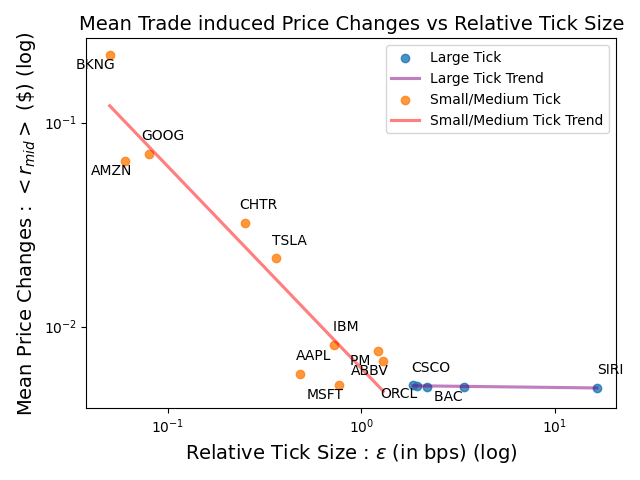}
    \caption{\highlighttwo{\textbf{Mean Mid-Price Changes (in \$) vs Relative tick-size (in bps) } in a log-log scale: The figure reveals clearly two regimes. For large-tick assets mid-price basically always change by half a tick ({\color{black}i.e. 0.005 \$}), this is not the case for small and medium-tick assets. 
    For the latter, a linear behavior (slope of $-0.955$) is found. Thus the trade induced mean price changes are inversely proportional to the relative tick-size (Eq. \eqref{eq:rmid}). As explained in the text, it supports the use of Geometric Brownian Motion models of the price process.}}
    \label{fig:meanRet}
    \end{figure}  
    \begin{figure}[h]
    \centering
    \includegraphics[width=.6\textwidth]{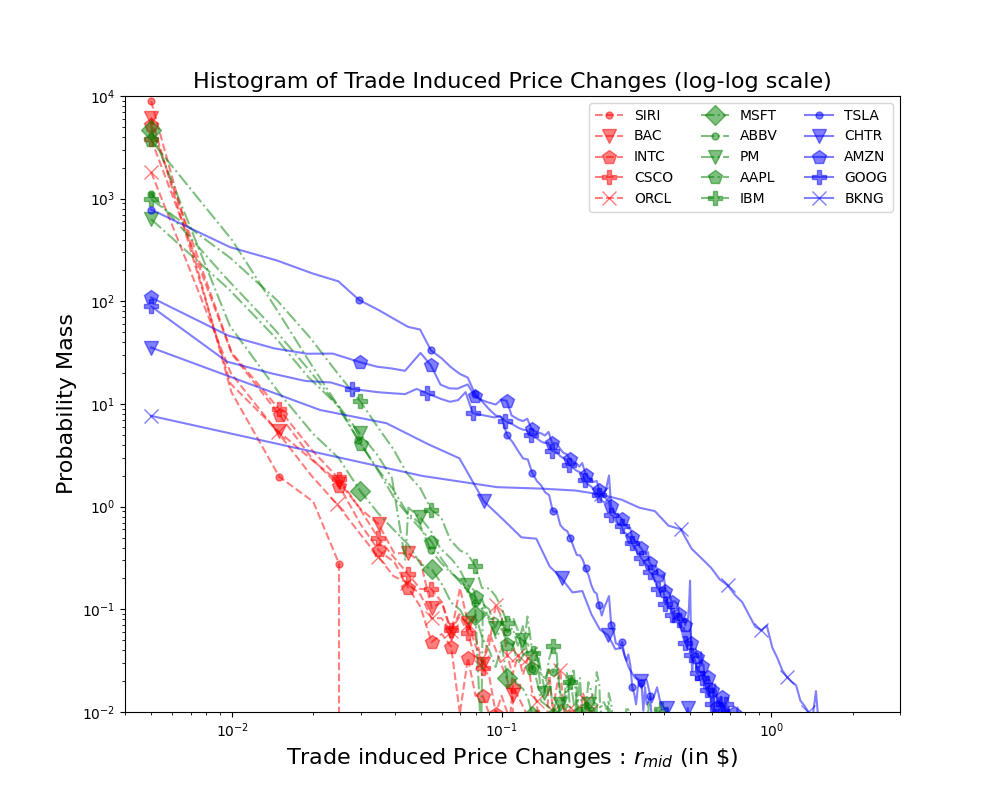}
    \caption{ \highlighttwo{\textbf{Histogram (log-log) of Trade induced Mid-Price Moves (in \$):} Notably the tail of all large (red) and medium (green) tick assets quickly dies quickly after half a tick however the distribution has a higher variance and a higher (second) mode for small-tick assets (blue). This implies that it is not just the mean of the mid-price changes that vary with relative tick-size but the shape of the distribution changes significantly as well. }}
    \label{fig:priceChangeDensity}
    \end{figure}
Mid-price changes can occur either due to a 
 bid/ask order at a new best price
or due to a market  or cancellation order depleting the best bid or best ask queue. Since the price of an asset is restricted to be a multiple of the tick-size, the mid-price change has a minimum constrained value   of half a tick.  In this section, we are focusing on mid-price changes triggered by trades. {\color{black} We ignore the mid-price changes due to cancellations and limit orders placed inside the spread for two reasons. First, we want to decouple the effect of the current bid-ask spread to the mid-price changes. Therefore limit orders placed inside the spread are excluded. Secondly, in order to filter spoofing{\footnote{Spoofing is defined as a market manipulative practice where a trader intentionally places large orders to buy or sell an asset on one side of the order book, with the sole purpose of creating a false impression of market demand or supply, and then quickly cancels those orders before they are executed.}} driven mid-price changes, we exclude the cancellations as well.}    
More precisely, we study
the mean trade induced mid-price moves defined by 
\begin{equation}
\langle r_{mid} \rangle := \mathbb{E}_{(trades)} \Delta p_{mid}(t_{trade}),
\end{equation}
where $p_{mid}(t)$ is the instantaneous mid-price at time $t$ and $\Delta p_{mid}(t_{trade})$ the variation of the mid-price directly induced by the trade occurring at time $t_{trade}$. 

Fig. \ref{fig:meanRet} displays $\langle r_{mid} \rangle$ versus the relative tick-size on a log-log scale.
Again, the figure clearly reveals two regimes. 
For large-tick assets, $\langle r_{mid} \rangle$ is basically constant and equal to its minimum constrained value (1/2 tick = 1/2 cent), whereas for medium and small-tick assets, the two quantities seem to be linked, in a good approximation, by a power law (the orange plain line has a slope very close to the value $-1$). Thus we can  therefore state the following approximate relation for some constant $k_r$:
\begin{align}
\label{eq:rmid}
    \langle r_{mid} \rangle \simeq & 
    \begin{cases}
         \frac{1}{2} , &\text{ for large-tick assets} \\
        k_r\epsilon^{-1} , &\text{ for small- and medium-tick assets}
    \end{cases}
\end{align}

Since the relative tick-size is inversely proportional to the price of the security (see Eq. \eqref{eq:epsilon}), we therefore conclude that the mean of trade driven mid-price changes is almost linearly related to the current price of the security. This experiment supports the use of geometric scaling for price process models (the most popular one being the Geometric Brownian Motion). As mentioned previously, this is not true for large-tick assets due to the flooring of mid-price changes at half-ticks.

Actually, not only the mean of mid-price change induced by market orders depends stongly on the relative tick-size but the distribution too. 
 In Fig. \ref{fig:priceChangeDensity}, we overlay the density of mid-price moves resulting from a market order in the LOB. {\color{black} We note that the tail of the distributions become fatter as we decrease $\epsilon$. Further we can see clearly that while the mode of all distributions is at 1/2 cents, the second mode increases rapidly with decreasing $\epsilon$. These observations imply that the distribution of mid-price changes is heavily shaped by the asset's relative tick-size.} For large-tick assets like SIRI, the probability of a price move exceeding one tick (e.g., one cent on the x-axis) is almost zero. Conversely, for small-tick assets, price moves of hundreds of ticks (i.e. one dollar) are relatively common. 

These results are not very surprising. 
Indeed, on the one hand for large-tick assets,
market participants that send market orders are generally reluctant to deplete more than one level of the LOB at once. The induced ``loss" of 1 (or more) tick for part of the order is significant, since the relative tick-size is large. 
Thus most market orders induce a change of 1/2 tick in the mid-price. On the other hand, when the relative tick-size is small (or medium), a ``loss" of 1 tick for a market participant is not that much significant, thus market participants are much less reluctant to deplete several queues with a single market order. 
As shown in Appendix \ref{app:price} (see Fig. \ref{fig:MO}), this behavior can be confirmed by studying the average market order sizes compared to the best queue size as a function of the relative tick-size of the asset. 

Let us finally point out that the overall phenomenon
is reinforced by the fact that the LOB of small-tick assets is much sparser that the one of a large-tick asset (and sparsity of the LOB, induces larger move in the mid-price). See Section \ref{sparsity} for sparsity related stylized facts.

\subsection{Shape of the LOB: Average Volume Distribution and its Maxima vs Relative Tick-Size}  \label{shape}

\begin{figure}[h]
    \centering
    \includegraphics[width=.7\textwidth]{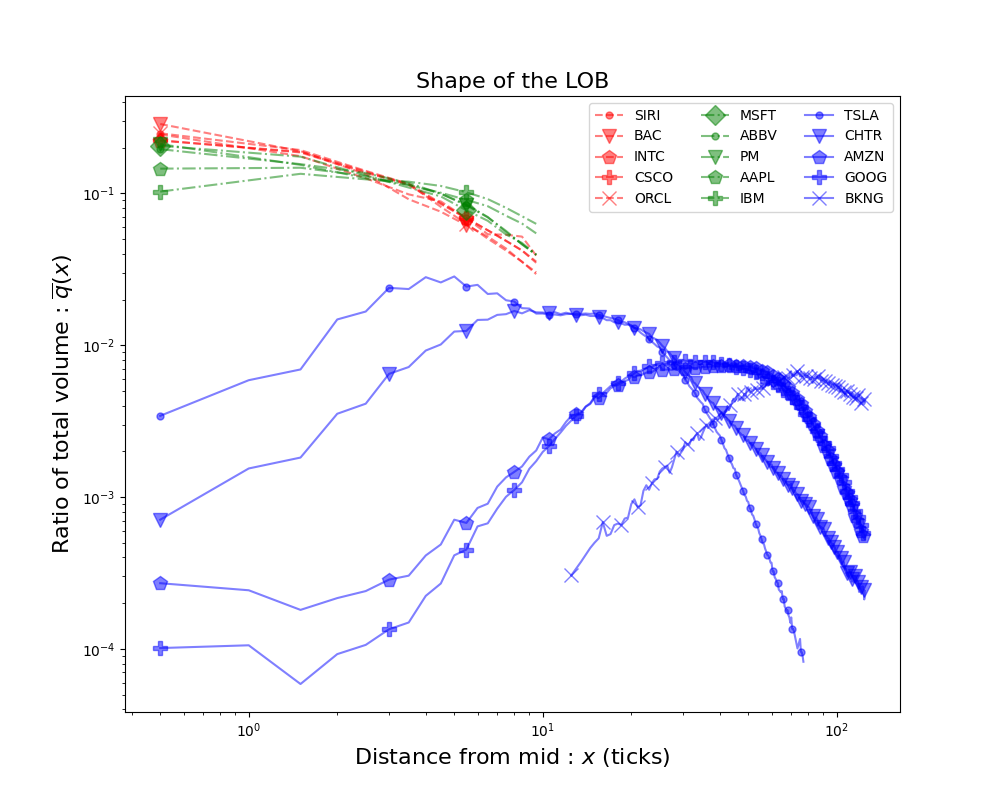}
    \caption{\highlighttwo{\textbf{Shape of the LOB - empirical density (log-log):} {\color{black} We plot the distribution of liquidity across price levels in this figure (Eq. (\ref{eq:shape})).} The tails of large (red) and medium (green) tick assets is insignificant beyond 10 ticks { and hence have been omitted.} 
    { For all large- and most of the medium-tick assets, most of the liquidity is at the top of the LOB. As we decrease the relative tick-size, the distribution shifts progressively from near-top concentrated distribution  to flatter distributions and finally to much deeper concentrated distributions. Actually, for small-tick assets (blue), we can clearly see a shift of the maxima to deeper levels as well as consistent shapes of the LOB. }}}
    \label{fig:shapeLOB}
\end{figure}



\begin{table}[h]
\centering
\begin{tabular}{|c|r|r|r|}
\hline
    \textbf{Asset} & \textbf{1st Quartile} & \textbf{2nd Quartile} & \textbf{3rd Quartile}   \\
    \hline
    SIRI & 1.1 $\pm$ 0.54 & 2.7 $\pm$ 0.88 & 5.2 $\pm$ 1.10 \\
    BAC & 0.7 $\pm$ 0.42 & 2.2 $\pm$ 0.52 & 4.6 $\pm$ 0.62 \\
    INTC & 1.3 $\pm$ 0.40 & 2.6 $\pm$ 0.35 & 5.1 $\pm$ 0.53 \\
    CSCO & 1.3 $\pm$ 0.39 & 2.5 $\pm$ 0.29 & 5.0 $\pm$ 0.53 \\
    ORCL & 1.1 $\pm$ 0.45 & 2.5 $\pm$ 0.33 & 4.6 $\pm$ 0.43 \\ 
    \hline
    MSFT & 1.5 $\pm$ 0.15 & 2.9 $\pm$ 0.43 & 5.5 $\pm$ 0.33 \\     
    AAPL & 1.8 $\pm$ 0.31 & 3.8 $\pm$ 0.43 & 6.5 $\pm$ 0.51 \\
    ABBV & 1.6 $\pm$ 0.38 & 3.5 $\pm$ 0.55 & 6.1 $\pm$ 0.70  \\
    PM & 1.6 $\pm$ 0.40 & 3.6 $\pm$ 0.61 & 6.4 $\pm$ 0.84 \\
    IBM & 2.5 $\pm$ 0.66 & 4.9 $\pm$ 1.04 & 7.9 $\pm$ 1.36\\
    \hline
    TSLA & 7.7 $\pm$ 2.63 & 15.0 $\pm$ 4.37 & 23.4 $\pm$ 5.80  \\
    CHTR & 11.8 $\pm$ 2.51 & 20.4 $\pm$ 4.44 & 33.6 $\pm$ 9.27 \\
    AMZN & 32.1 $\pm$ 6.32 & 49.2 $\pm$ 8.52 & 67.0 $\pm$ 10.45 \\
    GOOG & 31.2 $\pm$ 6.20  &  47.6 $\pm$ 8.41 & 66.9 $\pm$ 10.23 \\
    BKNG & 59.3 $\pm$ 8.01 & 78.6 $\pm$ 7.52 & 98.8 $\pm$ 5.95 \\
    \hline
\end{tabular}
\caption{\textbf{Shape Quartiles (in ticks):} For the 15 stocks we study, we report the three quartiles of the average shape of the LOB ($\overline{q}(x)$) in this table. As we can see the 1st and 2nd quartile for large-tick assets remains constant at one tick and three tick distance respectively from the mid-price. 
This underlines the fact that the available volume for these assets is concentrated at the top of the LOB. This fact is not true for medium and small-tick assets as can be seen by the almost monotonous shifting of the quartiles deeper in the LOB. Therefore the liquidity in medium- and small-tick assets is not concentrated at the top but rather is distributed across various levels deeper in the LOB.}
\label{tab:pctiles}
\end{table}

\begin{figure}[h]
\centering \includegraphics[width=.6\textwidth]{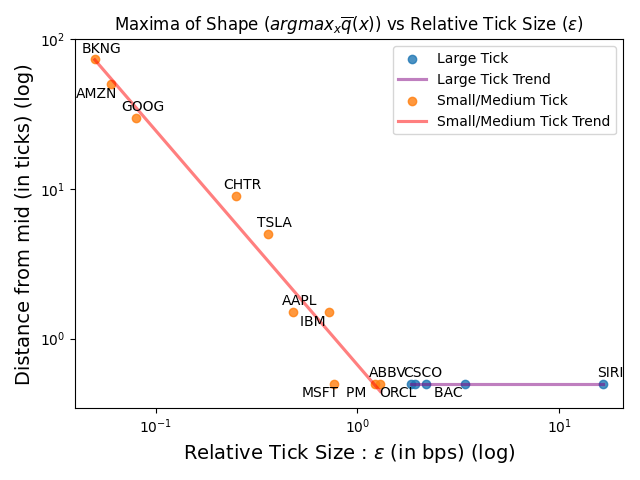}
\caption{\highlighttwo{\textbf{Depth with Maximum Liquidity of the LOB (log-log):} Again we observe two regimes - the large-tick assets have the maximum liquidity at the top of the LOB (which is at a distance of 1/2 tick of the mid-price) for the small-tick assets, the maximum liquidity is correlated with the current relative tick-size {\color{black}(Eq. (\ref{eq:maxshape}))} . This correlation is power law with an exponent over relative tick-size of approximately $-\frac{3}{2}$.}}
    \label{fig:maximaShape}
\end{figure}

Let the volume resting at a price level \( p \) at time \( t \) be denoted as \( Q(p,t) \), which refers to the outstanding quotes at that price. We define the `instantaneous' shape of the LOB at time $t$ as the normalized distribution of the volume \( Q(p,t) \) as a function of the distance of the price level \( p \) from the mid-price \( p_{\text{mid}}(t) \) and denote it by $q(x,t)$ where $x := \frac{p - p_{\text{mid}}(t)}{\delta}$ is the distance from mid-price in ticks. Further we define the average shape of the LOB as the time-weighted average of $q(x,t)$ denote it by $\overline{q}(x)$. Similar to Eq. \eqref{eqn:s_bar}, we have for index $i = 1,..., N$ and time $0 \leq t_i \leq T$, LOB volume distribution $q(x, t_i)$ lasting for $\Delta t_i$ seconds, so the calculation of $\overline{q}(x)$ is:

\begin{align}
    q(x,t) &:= \frac{Q(p_{\text{mid}}(t) + \delta x, t)}{ \sum_y Q(p_{\text{mid}}(t) + \delta y, t)} \\
    \overline{q}(x) &:= \frac{\sum^N_{i=1} q(x, t_i) \Delta t_i}{T} \label{eq:shape}
\end{align}
 The average shapes of the LOB for all 15 assets are displayed in Fig. \ref{fig:shapeLOB}. 
 It clearly shows that the distribution of liquidity shifts from near-top concentrated distribution for large-tick assets to  flatter behavior for medium-tick asset and finally to deeper concentrated distribution for small-tick assets. 
 Table \ref{tab:pctiles} displays the corresponding  quartiles of these distributions. {\color{black} One of the primary observations we have from this table is that around 50\% of liquidity of the entire LOB is in the first $\sim2.5$ price levels (measured from the mid-price) for large-tick assets. This along with the fact that the spread is almost always one tick in these assets implies that around half the liquidity of the LOB lies within the top two price levels of a large-tick LOB. This is clearly not the case for medium- and small-tick assets where the liquidity is more spread out across several deeper price levels.}
 {Further, the table clearly shows that basically the quartiles are increasing with the relative tick-size, indicating that a ``shift" of the mass of the distribution on the right side as the relative tick-size decreases.
 This result is confirmed by Fig. \ref{fig:shapePerStock} 
  in Appendix \ref{app:shape} which illustrates this ``shift" by displaying some per asset daily and averaged over the year shapes of the LOB for a large, medium and small-tick asset.

 The same ``shift" can be illustrated directly plotting the average shape of the LOB profile (i.e. $\text{arg max}_{x} \overline{q}(x)$) as a function of the relative tick-size as done in Fig. \ref{fig:maximaShape}. 
 As usual, the figure reveals two regimes.
}
On the one hand, large-tick assets have the maxima of the shape uniformly at half a tick away from the mid-price (fit shown in purple line). On the other hand we see an almost { power-law} 
trend wherein the maximum liquidity shifts deeper in the LOB as we decrease the relative tick-size (fit shown in red line). \highlighttwo{The slope of the red line is $-1.513$, therefore the relation is approximately power law with an exponent of $-\frac{3}{2}$ over the relative tick-size}. 
{For some constant $k_q$, these results can be summarized as :} 
\begin{align}
    \text{arg max}_{x} \overline{q}(x) \simeq & 
    \begin{cases}
         \frac{1}{2} , &\text{ for large-tick assets} \\
        k_q\epsilon^{-\frac{3}{2}} , &\text{ for small- and medium-tick assets}
    \end{cases}
    \label{eq:maxshape}
\end{align}

\begin{figure}[h]
    \centering
    \includegraphics[width=.75\textwidth]{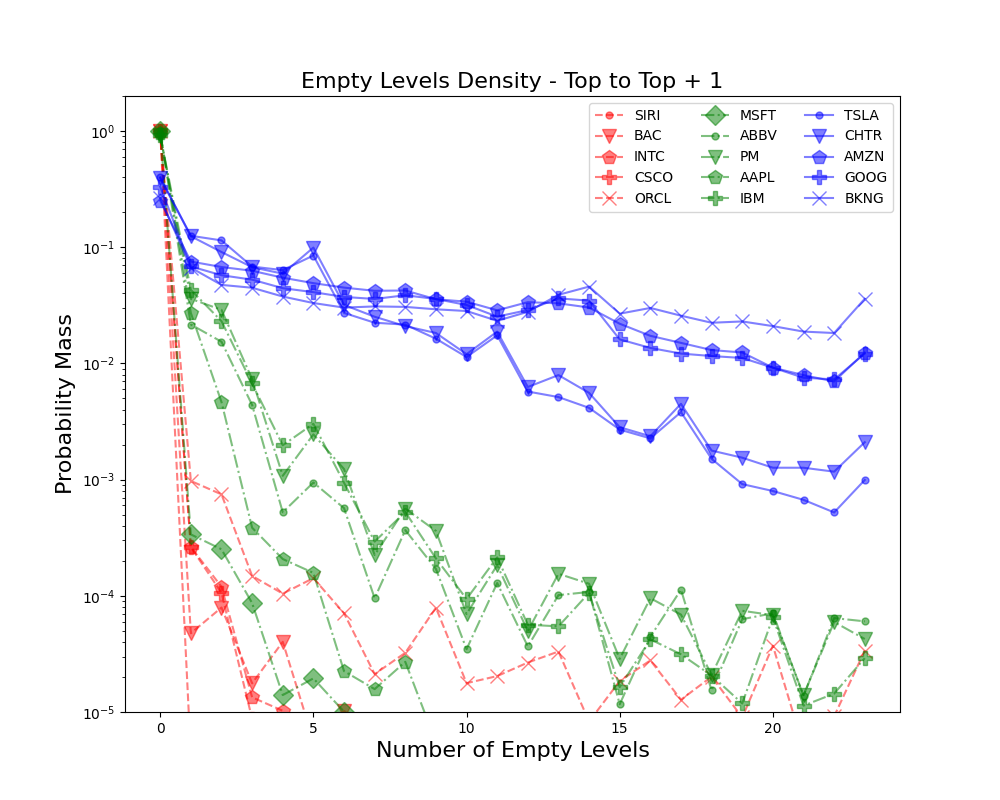}
\caption{\highlighttwo{\textbf{Empty Levels between Top and Top + 1 (log density) : } This depicts that the distribution of the number of empty levels becomes flatter as we decrease the relative tick-size. Indeed we can clearly observe that for large-tick assets (red), the distribution is very narrow and most of the mass is at 0. However for small-tick (blue), there is a significant probability of more than 10 empty levels in the LOB. }}
\label{fig:emptyLevels}
\end{figure}

Let us point out that \cite{bouchaud2018trades} mention the same fact that  most liquidity in small-tick assets actually lies much deeper in the LOB however to the best of our knowledge, our work is the first to form a statistical relation between this proxy of `market depth' and relative tick-size or mid-price. Since $x$ is measured from the mid-price, and since the spread is scaling with the exponent of $-1$, the maximum volume's distance from the respective top should scale in a square-root manner with the inverse of relative tick-size from our observations. 
In Appendix \ref{app:shape}, we also plot the empirical density of the instantaneous maxima to analyze the depth at which maximum liquidity lies in the LOB in Fig. \ref{fig:pdfMaxima}. {\color{black} We observe that not only the maxima of the average shape of the LOB is shifting deeper in the LOB but also the distributions of maxima of the instantaneous shape of the LOB are remarkably different. As we decrease $\epsilon$, we clearly see the instantaneous maxima's modes shifting deeper as well.} { We can therefore conclude that the top of the LOB is not necessarily the best place to post a passive limit order for these categories of assets since the maximum liquidity most likely lies much deeper in the LOB. }

\subsection{Sparsity of the LOB: Empty Price Levels and their effect on the instantaneous LOB's Shape} \label{sparsity}

\begin{figure}[h]
\centering
\includegraphics[width=.7\textwidth]{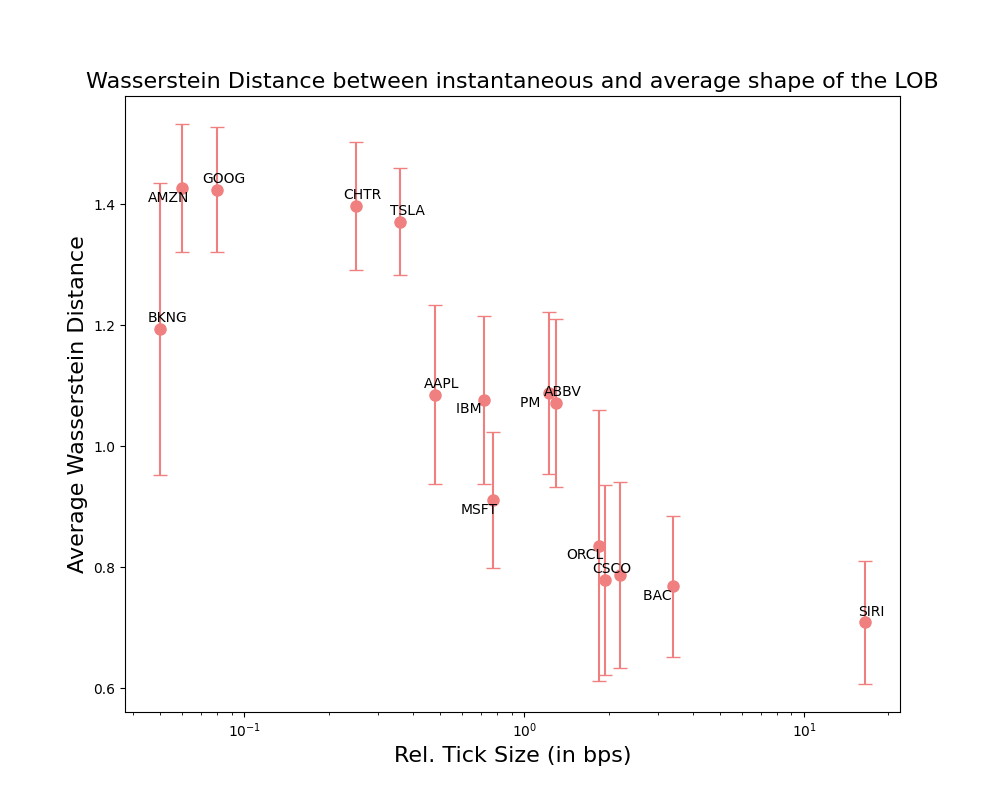}
    \caption{\highlighttwo{\textbf{Typical vs Average Shape of the LOB:} This measure of Wasserstein Distance {\color{black}(Eq. (\ref{eq:wass}))} 
  indicates how different, on a time-weighted average scale, is the instantaneous shape of the LOB from the average shape. This difference indicates whether the shape of the LOB remains persistent in the LOB or does the LOB's liquidity vary a lot owing to the sparse levels in the LOB. As we can see clearly here, there is an increasing trend as we decrease the relative tick-size however this relation seems to be non-linear. The error bars denote the variance of the Wasserstein Distance observed in our dataset. }}
    \label{fig:wass}
\end{figure}  

A well known feature in small-tick assets' LOB is that not every price level in the LOB will have liquidity. In this section we focus on the sparsity of the LOB as a function of the relative tick-size. 

The most straightforward way to  measure the sparsity of some LOB is to count the 
number of empty levels between successive quotes. 
 Following this idea, let $n_e(i,t)$ denote the number of empty levels between the two successive quotes ranked $i$ and $i + 1$, at time $t$ (for each rank $i$, we add up the numbers on bid and ask sides). 
Fig. \ref{fig:emptyLevels} displays the empirical distribution (log-scaled) of 
$n_e(1,t)$ (number of empty levels between the best quote and the second best quote) sampled at each time $t$.
Notice that as the relative tick-size decreases, the distribution becomes flatter and heavier-tailed. This clearly implies that while for large-tick and medium-tick assets, the LOB has liquidity in almost all levels near the top of the LOB, for small-tick assets, we can have dozens of price levels empty between the best and the second best quote. In  Appendix \ref{app:sparsity}, we confirm that this pattern persists for deeper levels in the LOB (Fig. \ref{fig:emptyLevelsPerStock}).

However, previous measurements can be biased by{\color{black}, for example, a single unit volume order }standing alone in their queue. 
To avoid these biases, as a sparsity proxy, we choose to measure the time average (and the corresponding variance) of the Wasserstein distance (or Earth-Mover distance) $l(\overline{q},q(.,t_i))$ between the average 
shape of the LOB $\overline{q}(x)$ and the shape of the LOB $q(x,t_i)$ at a timestamp $t_i$ : 
\begin{align}
    l(\overline{q}, q(t_i)) &:= \sum_x | \overline{q}(x) - q(x,t_i) | \Delta x \label{eq:wass}
\end{align}
Not surprisingly, Fig. \ref{fig:wass} shows that roughly,  the larger  Wasserstein distance the smaller the relative tick-size. However, let us point out the relation between the two is not linear and there are a few outliers (e.g., BKNG, MSFT).

\subsection{LOB Events and their Endogenous Excitation: Structural Similarities across Relative Tick-Sizes }\label{leverage}

\highlight{As previously discussed, a Limit Order Book can be modeled as a queuing system, where various types of events contribute to the formation and evolution of these queues. Empirical observations suggest that the realizations of these events persistently exhibit both cross-correlation with other events and autocorrelation with themselves. Notable works in this direction include \cite{lillo2004long}, \cite{bouchaud2009how}, \cite{yamamoto2010order}, and, \cite{toth2015why}. This phenomenon is commonly referred to as market endogeneity (\cite{bouchaud2018trades}), reflecting the idea that market activity is influenced by its own past behavior. Our focus lies in analyzing the conditional probability distribution of these events, given the history of prior events, in order to capture the feedback mechanisms inherent in the dynamics of the LOB.} 
Inspired by \cite{LuAbergel2018}, in this section, we study how the occurrence of certain events in the LOB affects the likelihood of other events happening shortly afterward. 

Let us denote an event $e$ happening at $x$ ticks distance from the respective top of the LOB by $e_x$. 
We use positive (resp. negative) values  for $x$ to refer to ask (resp. bid) side events.
In the following we will consider four types of events $e$\footnote{Of course we are aware there exist several more order types offered by the exchanges (for eg. fill-or-kill, stop loss etc). However these four types of events are the most basic ones. And the more exotic ones can generally be thought, mathematically, as a combination of these four basic order types.} : 
\begin{itemize}
\item Passive Limit Orders ($e=LO$), i.e., limit orders which do not change the mid-price
\item Active or `In-Spread' Limit Orders ($e=IS$) i.e. limit orders which are posted at an improved price than the respective top price, 
\item Cancel Orders ($e=CO$), 
\item and Market Orders ($e=MO$)
\end{itemize}
Let us note that market orders can only occur at best levels, thus $MO_x$ is only valid for $x=0$.

 Given two consecutive events $e^{(1)}_x$ and $e^{(2)}_y$, we compute the ratio of the conditional probability (i.e., the chance of \( e^{(2)}_y \) given \( e^{(1)}_x\)) to the unconditional probability of \( e^{(2)}_y\) occurring i.e. $\frac{P(e^{(2)}_y | e^{(1)}_x)}{P(e^{(2)}_y)}$. This ratio is known as Conditional Probability Leverage or simply `leverage'. 
We illustrate the results in Fig. \ref{fig:leverageTop_1} for two assets : a large-tick asset (BAC) on the left  and small-tick asset (GOOG) on the right. {\color{black}For some more plots for a medium-tick asset (IBM) and another small-tick asset (AMZN), we refer the reader to Fig. \ref{fig:leverageTop} in the Appendix A.5.}
To facilitate a qualitative analysis, we plot the leverage for all these events on a 2D grid with the x-axis representing an event \( e^{(1)}_x\)  and the y-axis the successive event \( e^{(2)}_y\). As indicated by the colormap, a red (resp. blue) shade corresponds to a leverage higher than one which corresponds to an \emph{excitation} (resp. \emph{inhibition}) of $e^{(2)}_y$ by $e^{(1)}_y$. 
 For the large-tick asset we use a linear scale for $x$ and $y$ with $x,y\in[0,10[$. For the small-tick asset we use a linear scale for $x$ and $y$ up to $x,y< 10$ and a log scale for $x,y \ge 10$ and smaller than the 95\% quantile of the shape of the LOB (this is done to enforce high resolution near the top of the LOB).

\begin{figure}[h]
\centering
\begin{subfigure}[c]{.49\textwidth}
\includegraphics[width=\textwidth]{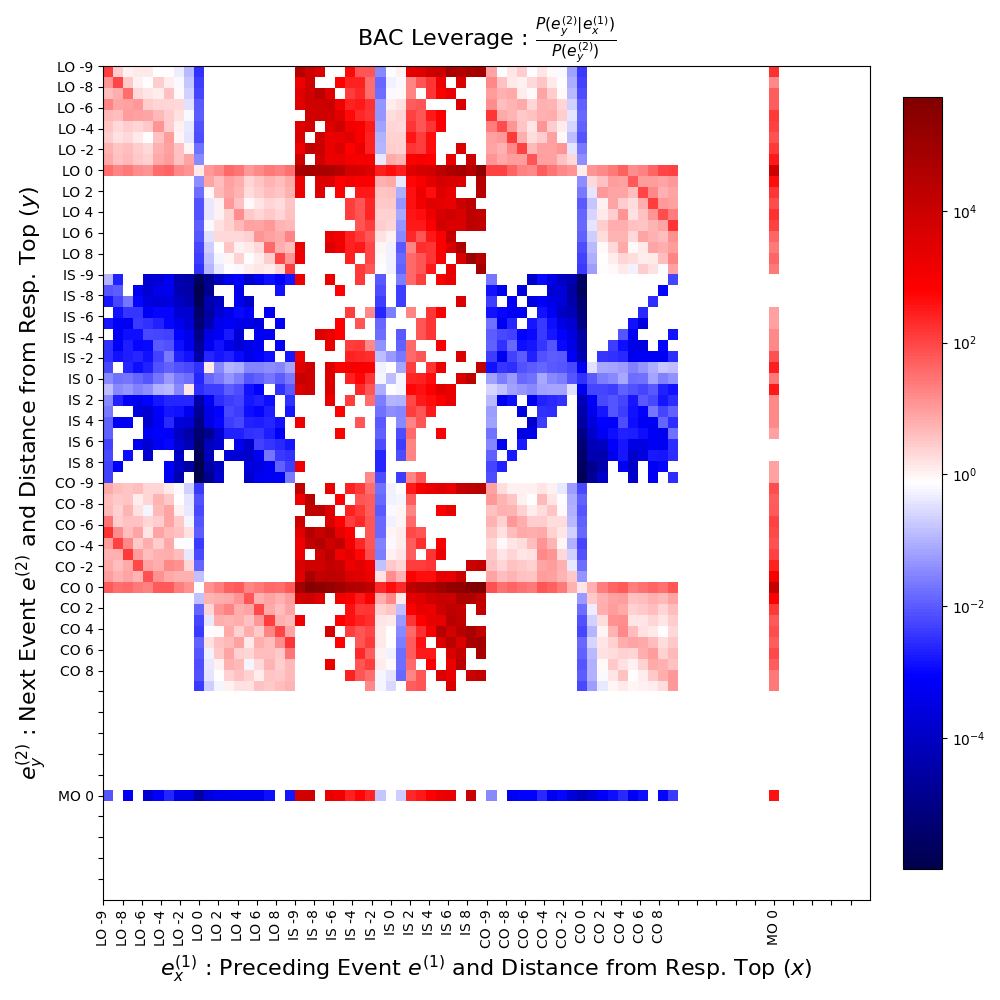}
\caption{BAC (Large-tick) }
\label{fig:baclev}
\end{subfigure}
\begin{subfigure}[c]{0.49\textwidth}
    \includegraphics[width=\textwidth]{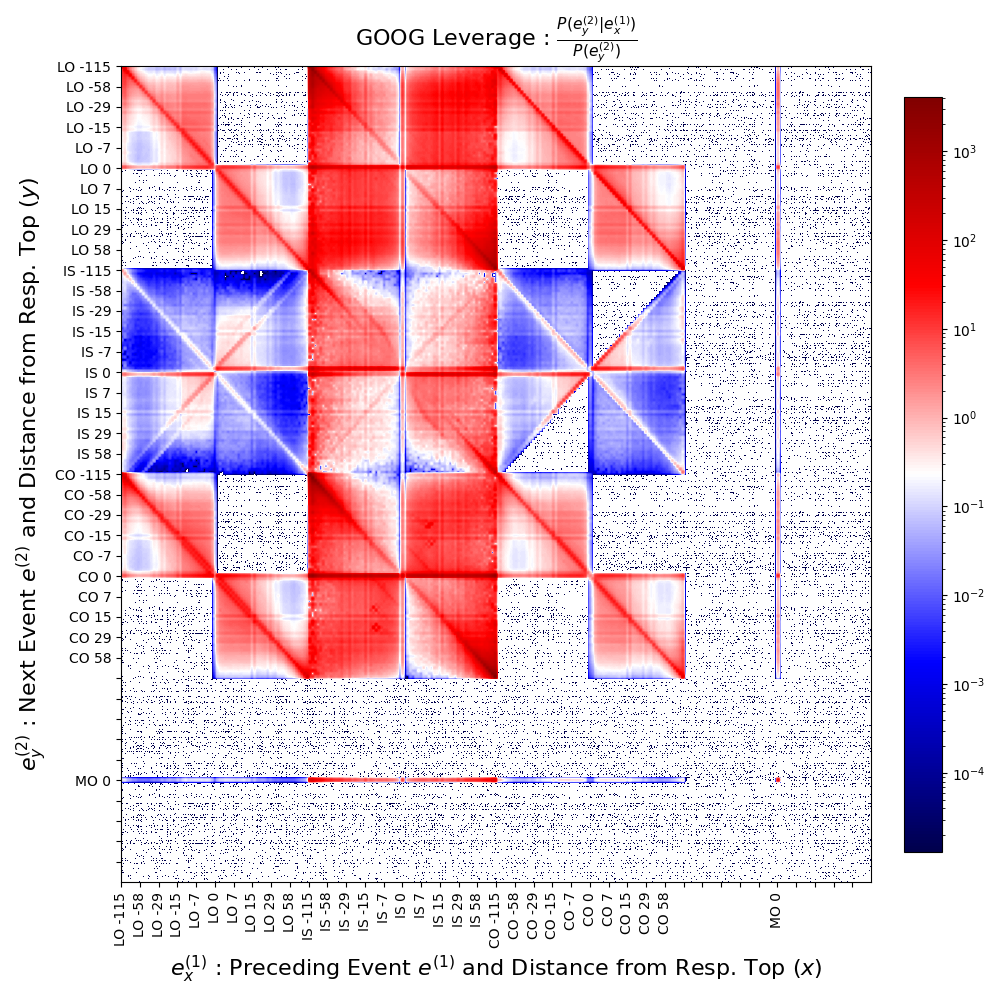}
    \caption{GOOG (Small-tick) }
    \label{fig:googlev}
\end{subfigure}
 \caption{
\textbf{Leverage (from top of the LOB):  Red shade implies Excitation, White implies Independence and Blue implies Inhibition -} {\color{black}The x-axis represents the preceding event, and the y-axis represents the future event. This figure highlights both common and distinct features across different assets. Notably, similar patterns are observed in LO-LO, CO-CO, and MO-MO excitations, while differences are evident in CO-LO and LO-CO blocks between large-tick (Fig. 9a) and small-tick (Fig. 9b) assets. For a detailed analysis, see the main text.} }
\label{fig:leverageTop_1}
\end{figure}


Fig. \ref{fig:leverageTop_1}  shows that most of the main patterns of the leverage  do not seem to depend strongly on the relative tick sizes (albeit with different scaling).
Thus, the way events influence each other is consistent, up to a linear-log scaling, regardless of how large or small the tick-size is. These findings provide insights about the dynamics of event interactions within the LOB. Let us point out some of the main patterns that do not depend on relative tick-size aand some of the main differences that do depend on relative tick-size :
\begin{enumerate}
    \item Whatever the relative tick-size is, we observe that LOs excite COs and vice-versa. This is not surprising as Cancels are causally related to previously posted LOs, and, Cancels usually imply posting a new LO at a possibly different depth from the mid-price.
    \item Whatever the relative tick-size is, we observe that the LOs at the top of the LOB behave remarkably differently from the ones at deeper levels. For instance, in Fig. \ref{fig:baclev}, $e^{(1)}_x = LO_0$ mostly inhibits other orders (i.e. the vertical line is blue colored) however $x = LO_2$ excites $e^{(2)}_y = LO_{0/1/2}$. A similar pattern is observed in Fig. \ref{fig:googlev}. 
    \item For small-tick assets, COs and LOs behave very similar to each other except for the off-diagonal terms which indicates that deeper LOs lead to cancellations nearer to the top. Additionally deeper COs lead to LOs nearer to the top as well. However the pattern is really difference for large-tick assets (see Fig. \ref{fig:googlev}). Fig. \ref{fig:ibmlev} in the Appendix \ref{app:endogeneity} shows that the transition is smooth going from small to medium to large-tick assets.
    \item ISs are both self excited and cross excited by opposite end LOs and COs in a triangular fashion. This is not visible in Fig. \ref{fig:baclev} due to the fact that the probability of a higher than one tick distance from mid-price IS event is very low, but in Fig. \ref{fig:googlev}, one can clearly see the self excitation and the cross excitation by LOs and COs in the red anti-diagonal. This is an interesting un-intuitive finding as deeper LOs and COs seem to be affecting opposite side in-spread LOs. 
    \item ISs excite other events very similarly between small and large-tick assets. They seem to excite themselves both on the same and the opposite side of the LOB, and in general looks to increase all kinds of events' occurrence probability. Notable exceptions are when IS happens very close to the current top of the LOB.
    \item As already mentionned, MOs naturally only occur at the top of the LOB, however we can see very similar patterns across all types of assets for market orders. They excite all other kinds of orders and are themselves inhibited by previous LOs and COs.
\end{enumerate}



\section{A Meta-Queue Hawkes Model for Limit Order Books} \label{model}

In this section, we utilize the information from the study of stylized facts reviewed in previous section to propose a simple extension to the Compound Hawkes Process model proposed in \cite{jain2023hawkes} that accounts for both small- and large-tick assets.

\highlight{As it is well known (see e.g. \cite{Bacry2015}, \cite{toth2015why}, \cite{bacry2016estimation}, \cite{LuAbergel2018}, \cite{huang2015}) and as also seen in Section \ref{leverage}, the events 
in an LOB are endogenously excited by themselves. This stylized fact, along with the general applicability and versatility of the Hawkes Process as discussed in Section \ref{hpLitReview}, is the main motivation of using a self exciting point process as our LOB model.} More precisely, we propose a model of LOB dynamics where events are driven by a multidimensional Hawkes process as well as by some state variables (such as the top of the LOB volume) whose dynamics are itself driven by a point process. 

However, as discussed in \cite{jain2023hawkes}, a Hawkes model by itself is not really suited to capture dynamics of a small-tick LOB. Indeed, in order to keep track of the activity in each queue of the LOB, a Hawkes model has to consider as many components as the number of such active queues. To avoid  a model with a dimension of several hundreds 
(as advocated in  \cite{delattre2014high}, the calibration becomes exponentially harder as one increases the dimension of the process),
 one needs to assume that very few queues are really active. Thus,  in the literature, Hawkes models for LOB modelling  are generally assuming such things like : best bid/ask levels of the LOB containing most of the information of the LOB dynamics, a dense LOB, in-spread orders occurring uniformly one tick away from the previous best quote, etc.
As seen in Section 2, all these assumptions basically fit a large-tick asset and are clearly not suitable for a small- or medium-tick asset. 


In this section, we propose to create a model of the LOB which not only captures the top of the LOB dynamics but also is able to model the deeper price levels' events without modelling precisely each single queue dynamics (i.e., without increasing the dimension dramatically). This will essentially be made possible by grouping LOB queues into {\em meta-queues}, regrouping in that way similar dynamics. 
Based on the analysis in Section \ref{leverage}, we identify five groups of limit order queues, each exhibiting similar statistical dynamics for specific order types :
the in-spread queues, the best ask (resp. bid) limit queues and the deeper ask (resp. bid) queues.   
Therefore we propose to model each of these groups as a meta-queue, i.e., a group of queues of the LOB that are merged together. The exact ranges of queues that are covered by these meta-queues will be associated with dynamic quantities controlled by state variables which evolve with the events in the LOB.  In the next section, we fully describe this model. 



\subsection{LOB components and Model State Variables:}  
\begin{figure}[h]
    \centering
    \includegraphics[width=0.75\linewidth]{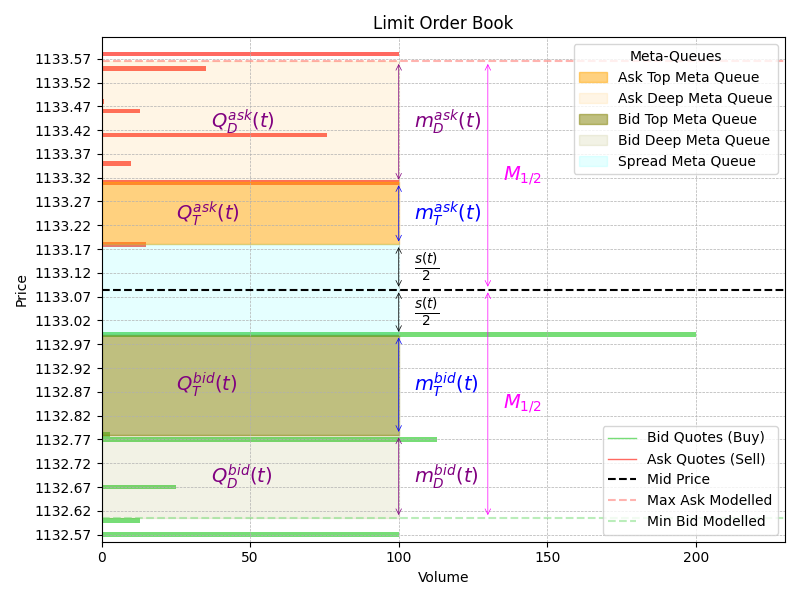}
    \caption{\textbf{Example LOB meta-queues and state variables :} Each horizontal line corresponds to the size of a single queue of a snapshot of GOOG's LOB. Let us point the high sparsity of the LOB (i.e., it is full of empty queues). The  colored rectangles (not to scale and just for illustrative purposes) indicate the meta-queues that we consider for our model. The corresponding state-variables of each meta-queue are listed on the right side of the figure (in purple).}
    \label{fig:statevar}
\end{figure}

Following the previous discussion, our model structures the  LOB in 5 different meta-queues (we encourage the reader to look at Fig. \ref{fig:statevar} while reading the next paragraph) each of them being characterized by one or several state variables, namely : 
\begin{itemize} 
\item The {\em spread meta-queue} (light blue rectangle in Fig. \ref{fig:statevar}) corresponding to all the queues in the spread. It is characterized by a single state variable : 
\begin{itemize} 
\item the spread $s(t)$ (i.e., the number of price levels in the spread). 
\end{itemize}
\vskip .15cm
\item The {\em top ask (resp. bid) meta-queue} (orange (resp. green) rectangle) that corresponds to a group of queues starting from the best ask (resp. bid) quote and ending just before the second best quote. It is associated with 2 state variables : 
\begin{itemize} 
\item $Q_T^{ask}(t)$ (resp. $Q_T^{bid}(t)$) corresponding to the total volume at the best limit 
\item $m_T^{ask}(t)$ (resp. $m_T^{bid}(t)$) corresponding to the number of price levels in the group (not including the second best quote).  
\end{itemize}
In Fig. \ref{fig:statevar}, for example, the best ask quote is at 1133.17 \$ and the 
second best ask is at 1133.31 \$, thus $m^{(ask)}_{T}(t) = 14$ ticks .
\vskip .15cm
\item The {\em depth ask (resp. bid) meta-queue}  (light orange (resp. light green) rectangle) corresponding to a group of queues characterized by 
\begin{itemize} 
\item two state variables : 
\begin{itemize} 
\item $Q_D^{ask}(t)$ (resp. $Q_D^{bid}(t)$) corresponding to the total volume in the group
\item $m_D^{ask}(t)$ (resp. $m_D^{bid}(t)$) corresponding to the number of queues in the group (including the second best quote)
\end{itemize}
\item one hyper-parameter 
\begin{itemize} 
\item $M_{1/2}$ : that defines the maximum depth of the meta-queue, assuming orders beyond this point have negligible impact on top-of-book dynamics due to low fill probabilities. This leads to the 4 constraints 
\begin{align}
    M_{1/2} \geq \frac{1}{2} s(t) + m^{(\zeta)}_{T}(t) + m^{(\zeta)}_{D}(t), ~~{\mbox{and}}~~ 
m^{(\zeta)}_{D}(t) \geq 1,~~~\forall \zeta \in \{bid, ask\}. \label{eq:constr}
\end{align}
In the following we will fix this parameter to be the median of the LOB's average shape (i.e., 50\% quartile in Table \ref{tab:pctiles}) minus 1,
\end{itemize} 
\end{itemize}
In Fig. \ref{fig:statevar}, since the the 
second best ask is at 1133.31 \$ and $M_{1/2}$ has been chosen to be 49 ticks, one gets 
$m^{(ask)}_{T}(t) = 26$ ticks .

\end{itemize}

{\color{black} Let us note that,  we do not need to directly model the dynamics of the mid-price. Its changes are a natural mechanical consequence of our events dynamics modelization. }

\subsection{Hawkes Process Model for LOB Events and State Variables Dynamics}
Given the structure we just define, there are a total 12 types of events that can happen in the overall meta-queues, namely :  bid or ask LOs and COs in the {\color{black} 4 meta-queues with pre-existing volume} (so that sums up to 8 different type of orders),  {\color{black} bid or ask in-spread LOs in the spread meta-queue (therefore 2 different types of orders),} and bid or ask MOs in the top meta-queues (which correspond to 2 different types of orders). Thus the set of events is:\textbf{}
\begin{align*}
\mathcal{E} := &\{LO_{\text{ask}_{D}}, CO_{\text{ask}_{D}}, LO_{\text{ask}_{T}}, CO_{\text{ask}_{T}}, MO_{\text{ask}_{T}}, LO_{\text{ask}_{IS}}, \\ &LO_{\text{bid}_{IS}}, LO_{\text{bid}_{T}}, CO_{\text{bid}_{T}}, MO_{\text{bid}_{T}}, LO_{\text{bid}_{D}}, CO_{\text{bid}_{D}}\}
\end{align*}

Before describing the Hawkes model that will be in charge of modelling, 
at any time $t$, the arrival intensity of each type of orders in each of the 5 meta-queues, we need to first specify the impacts on the state variables of any order arrival. 

\subsubsection{Detailed order arrival impacts on the state variables}
For each order we have to precisely specify:  
\begin{itemize}
    \item how to determine the exact queue (i.e., price) the order arrives at within the meta-queue.
    \item how to determined the size of the order
    \item what are the necessary induced updates on the state-variables.
\end{itemize}
This is the purpose of the next list of items (in the following $\zeta$ is used to stand indifferently for $ask$ or $bid$) :

\begin{enumerate}
    \item \textbf{In-Spread Limit Orders - }{\boldmath$LO_{\zeta_{IS}}$}: \\
\begin{itemize}
\item  {\bf Queue specification.} Let $\eta_{IS} \in \mathbb{N}$ be defined as the number of ticks from the best side $\zeta$ where this new in-spread order arrives at. We choose to sample $\eta_{IS}$ from a calibrated stationary distribution 
    $\Pi_{\eta}(\hat{\eta}_{IS})$\footnote{A note on notation: We use $\Pi_a(\hat{a}_{e})$ to refer to a given calibrated stationary, parametric probability distribution function with parameter $\hat{a}_{e}$ for the random variable $a_e$ affecting the event $e$'s dynamics. We index the distribution by $a$ since it is shared by all $a$-type random variable (in this case for all the random variables related to a number of ticks $\eta$, e.g., $\eta_{IS}$, $\eta_{T}$, ...). The choice of each distribution and their calibration will be specified in the next section. These choices  will be directly driven by the our analysis of the stylized facts in the previous section. 
    }
    where $\hat{\eta}_{IS}$ is a calibrated parameter, i.e., 
    \begin{equation} 
    \label{eta_IS}
    \eta_{IS} \sim \Pi_{\eta}(\hat{\eta}_{IS})
    \end{equation} 
    {\color{black}For instance, one good choice for $\Pi_{\eta}(\hat{\eta}_{IS})$ is the Geometric distribution with $\hat{\eta}_{IS} = 0.5$ for a small-tick asset (more calibration results in Section \ref{calib}).} $\eta_{IS}$ is capped at the current spread in ticks and floored by one tick. {\color{black}These bounds are set by sampling from an unconstrained distribution and then rejecting and resampling when these bounds are violated.}
        
    \item {\bf Order size specification.}
    In the same way, the order size  $\kappa_{IS} \in \mathbb{N}$ of this new order is sampled from a calibrated stationary distribution :  
    \begin{equation} 
    \kappa_{IS} \sim \Pi_{\kappa}(\hat{\kappa}_{IS})
    \end{equation} 
    

    \item {\bf State variables update.}   { When a, IS limit order arrives, one needs to update the new top of the LOB and consequently the deeper meta-queues. Thus, the new width of the top meta-queue is equal to $\eta_{IS}$, and the new meta-queue size of the top meta-queue is equal to $\kappa_{IS}$ of this IS order. The spread moves by $\eta_{IS}$. Mathematically,} the following changes happen to the five state variables mentioned above. 
       \begin{align}
        m^{(\zeta)}_{IS}(t^+) &= 0.5 (max(0, m^{(\zeta)}_{IS}(t) - \eta_{IS}) + m^{(\Bar{\zeta})}_{IS}(t))\\
        m^{(\zeta)}_{T}(t^+)&=  \eta_{IS} \label{m_T_IS} \\
        m^{(\zeta)}_{D}(t^+) &= M_{0.5} -m^{(\zeta)}_{D}(t) + m^{(\zeta)}_{IST}(t)
        \label{m_D_IS}\\
        Q^{(\zeta)}_T(t^+) &= \kappa_{IS}\\
        Q^{(\zeta)}_D(t^+) &= Q^{(\zeta)}_T(t) + Q^{(\zeta)}_D(t) - \Delta Q^{(\zeta)}_D(t)\\
        s(t^+) &= s(t) - \eta_{IS} 
    \end{align}
    An example of such an update is given in Fig. \ref{fig:postLOIs}.

    \begin{figure}[h]
    \centering
    \begin{subfigure}[c]{.49\textwidth}
    \includegraphics[width=\textwidth]{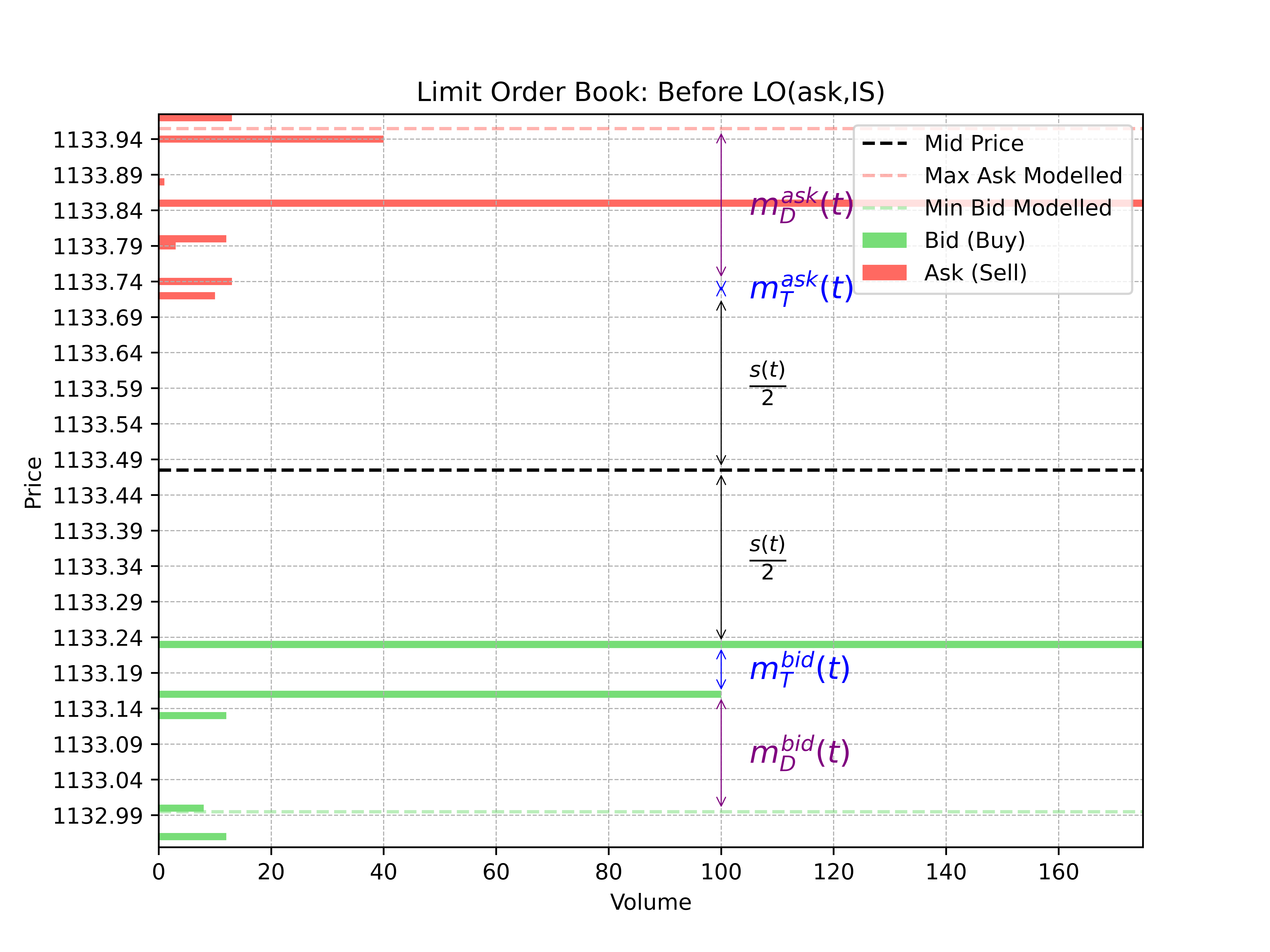}
    \caption{Pre-event }
    \label{fig:preLOIS}
    \end{subfigure}
    \begin{subfigure}[c]{.49\textwidth}
    \includegraphics[width=\textwidth]{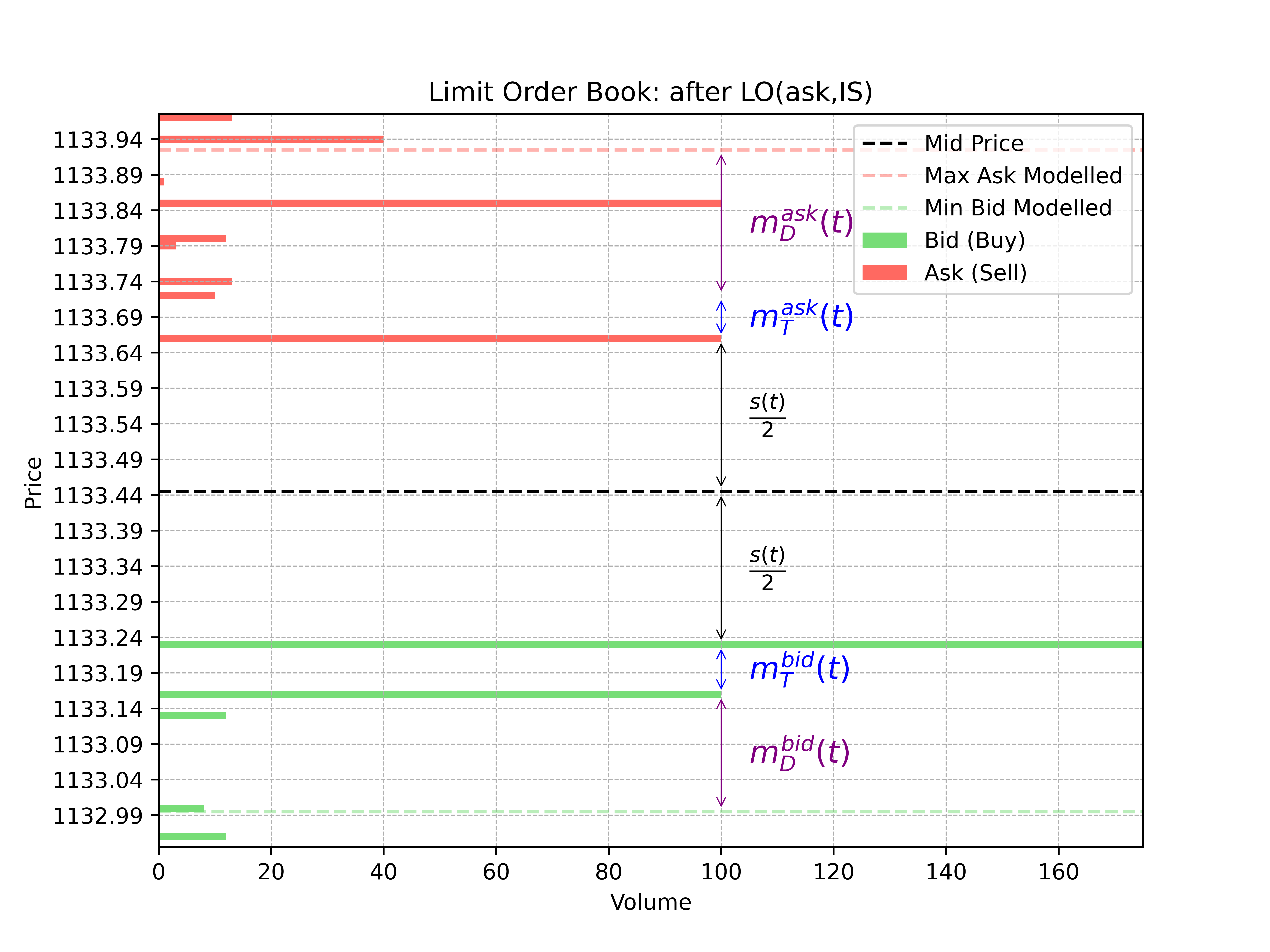}
    \caption{Post-event }
    \label{fig:postLOIs}
    \end{subfigure}
     \caption{\highlighttwo{\textbf{$LO_{\zeta_{IS}}$}: Ask side In-Spread LO occurs at 1133.66 \$. We remove the overlaid meta-queue rectangles for clarity. $m^{(\zeta)}_{T}(t^+)$ increases to 6 ticks from the previous 2 ticks following the logic given in Eqs \ref{m_T_IS} and \ref{m_D_IS}.  }}
    \label{fig:loIS}
    \end{figure}
    
    \item {\bf Constraint check. }
    However, since this event is an event which increases $m^{(\zeta)}_{D}(t)$, there is a chance the first constraint mentioned in Eq. \ref{eq:constr} is no longer satisfied. If this is the case, we need to ``purge" the deep meta-queue to make it less deep. More precisely, we first update the width of the deep meta-queue in order to saturate the constraint, i.e., (we use the $t^{++}$ to refer to an update of the previous update) 
    \begin{equation}
         m^{(\zeta)}_{D}(t^{++}) = M_{1/2} + 1 - \frac{1}{2} s(t^+) - m^{(\zeta)}_{T}(t^+)
    \end{equation}
    then we have to update its corresponding volume $Q^{(\zeta)}_D(t^+)$. Since the model is not tracking each LOB entry individually we have to decide on a somewhat arbitrary rule to make this update. 

    \begin{align}
        Q^{(\zeta)}_D(t^{++}) &= Q^{(\zeta)}_D(t^+) - \Delta Q^{(\zeta)}_D(t^+) \\
        \Delta Q^{(\zeta)}_D(t^+) &= \Xi(Q^{(\zeta)}_D(t^+), m^{(\zeta)}_{D}(t^{++}) - m^{(\zeta)}_{T}(t^+), m^{(\zeta)}_{D}(t^+)) \label{xi}
    \end{align}

    
    {\color{black} In order to purge the deep meta-queue, we naturally need to remove certain volume from the meta-queue. The removed volume } $\Delta Q^{(\zeta)}_D(t^+)$ is defined as a {\color{black}fraction} of the existing $Q^{(\zeta)}_D(t^+)$. This {\color{black}fraction} directly depends on the number of levels in the {\color{black} purged }fraction ($m^{(\zeta)}_{D}(t^{++}) - m^{(\zeta)}_{T}(t^+)$) and the number of levels in the parent ($m^{(\zeta)}_{D}(t^+)$). {\color{black} Therefore, we define a function $\Xi$ which gives us the fraction of purged volume as shown in Eqn. \ref{xi}.}  In the following sections with simulation studies, we make use of a uniform $\Xi$ function {\color{black} i.e. $\Xi(Q, a, b) = \frac{Qa}{b}$}. This function can be calibrated according to the shape of the LOB, as outlined in Section \ref{shape} however for simplicity we choose not to do so.\\
    \end{itemize}
    
    \item \textbf{Arrival of Limit Orders at Top - }\textbf{$LO_{\zeta_{T}}$}: \\
    \begin{itemize}
        \item \textbf{Queue specification.}  Let $\eta_{T} (\in \mathbb{N}) \sim \Pi_{\eta}(\hat{\eta}_{T})$ be defined as the number of ticks from the top at side $\zeta$ where this new order arrives at. $\eta_T$ is capped at $m^{(\zeta)}_{T}(t)$ and floored at zero.
        \item \textbf{Order size specification.} Similar to the previous event, $\kappa_{T} (\in \mathbb{N}) \sim \Pi_{\kappa}(\hat{\kappa}_{T})$ is the order size of this LO.
        \item \textbf{State variables update.} If $\eta_T$ is 0, the LO is simply added to the existing meta-queue at the top and no changes occur in the state variables except $Q^{(\zeta)}_T(t^+) = Q^{(\zeta)}_T(t) + \kappa_T$. Otherwise, { the LO hits some queue within the the best and the second best quotes. Therefore, the new LO becomes the new second best quote and is absorbed into the deeper meta-queue. We change the width of the top meta-queue according to the distance of the new LO from the top of the LOB and }the following changes happen to the random processes mentioned above:
    \begin{align}
    \label{eta_T}
    \eta_{T} &\sim \Pi_{\eta}(\hat{\eta}_{T}) \\
    \text{if } \eta_{T} \text{ is } 0, \qquad & \nonumber\\
    Q^{(\zeta)}_T(t^+) &= Q^{(\zeta)}_T(t) + \kappa_T \\
    \text{else, } \qquad \qquad& \nonumber\\
    m^{(\zeta)}_{T}(t^+) &= \eta_{T}\\
    m^{(\zeta)}_{D}(t^+) &= m^{(\zeta)}_{D}(t) + m^{(\zeta)}_{T}(t) -  \eta_{T}\\
    Q^{(\zeta)}_D(t^+) &= Q^{(\zeta)}_D(t)  + \kappa_T
    \end{align}
    \end{itemize}
    \item \textbf{Arrival of Cancel Orders at Top - }\textbf{$CO_{\zeta_{T}}$}: \\
    \begin{itemize}
        \item \textbf{Queue specification.} Since there's only one queue with volume in the top meta-queue, we deplete that queue with this cancellation.
        \item \textbf{Order size specification.} Since we cannot track individual LOs in the existing queues owing to the loss of fine-grained simulation of the LOB in our model, we sample $\kappa_{CO} (\in \mathbb{N}) \sim \Pi_{\kappa}(\hat{\kappa}_{T})$ as the cancellation order size.
        \item \textbf{State variables update.} The top meta-queue size changes as follows $Q^{(\zeta)}_T(t^+) = Q^{(\zeta)}_T(t) - \kappa_{CO}$. If this CO leads to a queue depletion (QD), \highlight{we need to calculate the new top's width as well as deplete the deeper meta-queue by this new top's total volume. Assuming a sparse structure within the depth meta-queue (informed by section \ref{sparsity}), we sample a random new top width denoted by $\eta_{T+1}$.  Here, $\eta_{T+1} (\in \mathbb{N}) \sim \Pi_{\eta}(\hat{\eta}_{T+1})$ is defined as the number of levels the new top has after the queue depletion event. To calculate the total volume at this new level, we make use of the partition function mentioned earlier to deplete the depth meta-queue.} Mathematically, we perform the following to redefine a new top of the LOB.
    \begin{align}
    { \text{if QD:}} & \\
    \label{eta_Tp1}
    \eta_{T+1} &\sim \Pi_{\eta}(\hat{\eta}_{T+1})\\
    m^{(\zeta)}_{T}(t^+) &= \eta_{T+1} \\
    m^{(\zeta)}_{D}(t^+) &= m^{(\zeta)}_{D}(t) - m^{(\zeta)}_{T}(t^+)\\
    Q^{(\zeta)}_T(t^+) &= \text{Partition}(Q^{(\zeta)}_D(t), m^{(\zeta)}_{T}(t^+), m^{(\zeta)}_{D}(t))\\
    Q^{(\zeta)}_D(t^+) &= Q^{(\zeta)}_D(t)  - Q^{(\zeta)}_T(t^+)\\
    s(t^+) &= s(t) +   m^{(\zeta)}_{T}(t) 
    \end{align}
    \end{itemize}
   
    \item  \textbf{Arrival of Market Orders - } \textbf{$MO_{\zeta}$}: \\
    \begin{itemize}
        \item \textbf{Queue specification.} MOs naturally always hit the top meta-queue.
        \item \textbf{Order size specification. } The size of the market order is given by $\kappa_{MO} (\in \mathbb{N}) \sim \Pi_{\kappa}(\hat{\kappa}_{MO})$
        \item \textbf{State variables update.} MOs, mathematically, behave exactly the same as Cancel Orders at top and we report a trade. If the MO leads to a queue depletion, the same logical waterfall as the one above is followed. If a market order depletes more than one price level, we repeat the process until we completely execute the MO quantity. Thus a MO can deplete multiple price levels using this methodology. \highlight{This clearly means that the stylized facts related to the price moves induced by Section \ref{trades} is modelled implicitly.}\\
    \end{itemize}
    \item  \textbf{Arrival of Deep Limit Orders - }\textbf{$LO_{\zeta_{D}}$}: \\
      \begin{itemize}
        \item \textbf{Queue specification.} Since we do not keep track of the fine-grained orders or queues in the deeper dimension, we do not specify the exact queue the LO hits at.
        \item \textbf{Order size specification.} The new LO size is $\kappa_{D} (\in \mathbb{N}) \sim \Pi_{\kappa}(\hat{\kappa}_{D})$.
        \item \textbf{State variables update. } We simply increment the total volume present in the depth meta-queue like so: $Q^{(\zeta)}_D(t^+) = Q^{(\zeta)}_D(t)  + \kappa_D$.\\
    \end{itemize}
    \item \textbf{Arrival of Deep Cancel Orders - }\textbf{$CO_{\zeta_{D}}$}: \\
    \begin{itemize}
        \item \textbf{Queue specification.} Again, since we do not keep track of the fine-grained queues in the deeper dimension, we do not specify the exact queue the cancellation is at.
        \item \textbf{Order size specification.} A similar logic described in the Top CO section applies here. The CO size is $\kappa_{CO_D} (\in \mathbb{N}) \sim \Pi_{\kappa} (\hat{\kappa}_{D})$.
        \item \textbf{State variables update.} Again, if there is no queue depletion event, we simply remove the CO volume from the existing $Q^{(\zeta)}_D(t)$. In case of a QD event, we need to sample the queue size and width from the unmodelled deeper part of the LOB. An important thing to note here is that sampling queue sizes depends on the meta-queue width variable since a wider meta-queue will be expected to have a higher size than a thinner meta-queue. Therefore we introduce this notation for the distribution of unseen meta-queue sizes : $\Pi_{Q}(\hat{Q}_D, m^{(\zeta)}_{D}(t^+))$. This has two parameters: first being $\hat{Q}_D$ which is calibrated according to the empirical data and the second being the width of the new meta-queue $m^{(\zeta)}_{D}(t^+)$. 
    \item \textbf{Constraint check.}
    In the case the width of the deep meta-queue $m^{(\zeta)}_{D}(t)$ is not violating the first constraint of Eq. \ref{eq:constr}, we must absorb the now empty levels in $m^{(\zeta)}_T(t)$ and create a new depth meta-queue with the following width and volume:
    \begin{align}
    m^{(\zeta)}_{T}(t^+) &= m^{(\zeta)}_{T}(t) + m^{(\zeta)}_{D}(t) \\
    m^{(\zeta)}_{D}(t^+) &=  M_{1/2} - m^{(\zeta)}_{T}(t) -  \frac{1}{2} s(t)\\ 
    Q^{(\zeta)}_D(t^+) &\sim \Pi_Q(\hat{Q}_D, m^{(\zeta)}_{D}(t^+))      
    \end{align}
    In the case the constraint is violated,  we set $m^{(\zeta)}_D(t^{++})$ to 1, sample $Q^{(\zeta)}_D(t^{++})$ from $\Pi_Q(\hat{Q}_D,1)$ and let the deeper meta-queue have just one non empty level since we cannot go deeper than $M_{1/2}$ in total depth of the LOB.
    \end{itemize}
    
\end{enumerate}
It is now time to describe the Hawkes process model that is in charge of modeling the arrival intensities of each type of orders. 

\subsubsection{The Hawkes process model for order arrivals  }


We model the dynamics of these 12 events as a  $d=12$ dimensional counting process $\{N^{(i)}(t)\}_{i \in \mathcal{E}}$ with corresponding  intensity  $\{\lambda^{(i)}(t)\}_{i \in \mathcal{E}}$. We choose this process to be a Hawkes process in the sense that each intensity function $\lambda^{(i)}$ can be basically seen as the sum of an  exogenous intensity $\mu^{(i)}$ and a mutually-exciting term governed linearly by a family of some kernel causal functions  $\{\phi^{(j \rightarrow i)}(t)\}_{j \in \mathcal{E}}$   that, for a given $j$,  accounts for the impact of the past events of type $j$ (i.e., the past jumps of $N_j(t)$) on the current intensity $\lambda_i(t)$. Since the kernel functions could take negative values, we use the classical non-linear version of the Hawkes Process (enforcing the intensity to be greater than 0). Moreover in order to take into account that in-spread orders are more likely to happen when the spread is large, for this specific events, we follow the model introduce in \cite{jain2023hawkes}.

More precisely we consider two cases : 
\begin{itemize}
\item For the intensity of events that are not In-Spread orders : 
\begin{equation}
\label{eq:Hawkes1}
    \lambda^{(i)}(t)  = max \bigg(0,  \mu^{(i)}  + \sum^d_{j=1} \int^t_{0} \phi^{(j \rightarrow i)}(t - u) dN^{(j)}(u) \bigg),~~~\forall i \in \mathcal{E},~ i\neq \mbox{IS}
\end{equation}
\item For the intensity of In-Spread orders : 
\begin{equation}
     \lambda^{(IS)}( t)  =	max \bigg(0,  \bigg(\frac{\delta(s(t)-1)}{\alpha}\bigg)^\beta \times  \bigg( \mu^{(IS)} + \sum^d_{j=1} \int^t_{0} \phi^{(j \rightarrow IS)}(t - u) dN^{(j)}(u) \bigg) 	\bigg) \label{lamb_IS},
\end{equation}
 where $s(t)$ is the spread (in ticks), $\delta$ is the tick-size (in \$), 
  $\alpha$ (in \$) is a scale parameter and $\beta$ a shape parameter.
\end{itemize}


 
{In summary, the model we propose is a model that extends to the median of the shape of the LOB, which can be several dozens of ticks from mid-price while being parsimonious : we have a total of 12 types of events spanning 5 meta-queues and 9 state variables. {\color{black}As mentioned previously, in order to account for the patterns of self- and cross-excitation discussed in Section \ref{leverage} through the conditional probability leverage plots, we model the events as a 12D mutually exciting Hawkes Process.} We now will try to show through simulations studies that such meta-queue Hawkes models can effectively reproduce several stylized facts we studied in Section 2.}

{\color{black}
\section{Simulation Results: Ergodicity, Critical Parameters, and Empirical Validation} \label{results}

The aim of this section is to illustrate, through examples and numerical experiments, that firstly, the class of order book models, {\color{black} i.e. the Meta-Queue Hawkes (MQH) model}, we propose is, within a certain range of parameters, stable and ergodic ({\color{black}Section \ref{ex_erg}}). 
We also aim to identify the critical parameters that determine whether the system falls into the `small-tick' or `large-tick' category ({\color{black}Section \ref{control_rel}}). {\color{black} For brevity, we call this exercise `controlling relative tick-size'. Finally,} we compare the empirical stylized facts' patterns discussed in Section~\ref{sty} with those produced by simulations across varying relative tick sizes, highlighting the model’s ability to replicate key market microstructure phenomena ({\color{black}Section \ref{compare_sty}}).

Given the large number of parameters in the model, from a methodological standpoint, we only consider specific variations around the values proposed by \cite{jain2023hawkes} for the Hawkes parameters in Eqs \eqref{eq:Hawkes1} and \eqref{lamb_IS}.
As for the probability distributions, these were determined based on a simple calibration experiment using empirical data. Regarding the full calibration of the model based on order book data, this question will be discussed in Section \ref{calib}.
}%

\subsection{{\color{black}Meta-Queue Hawkes Model's Small-tick Dynamics and Ergodicity}} \label{ex_erg}

\begin{figure}[h]
\centering
\begin{subfigure}[c]{.49\textwidth}
\includegraphics[width=\textwidth]{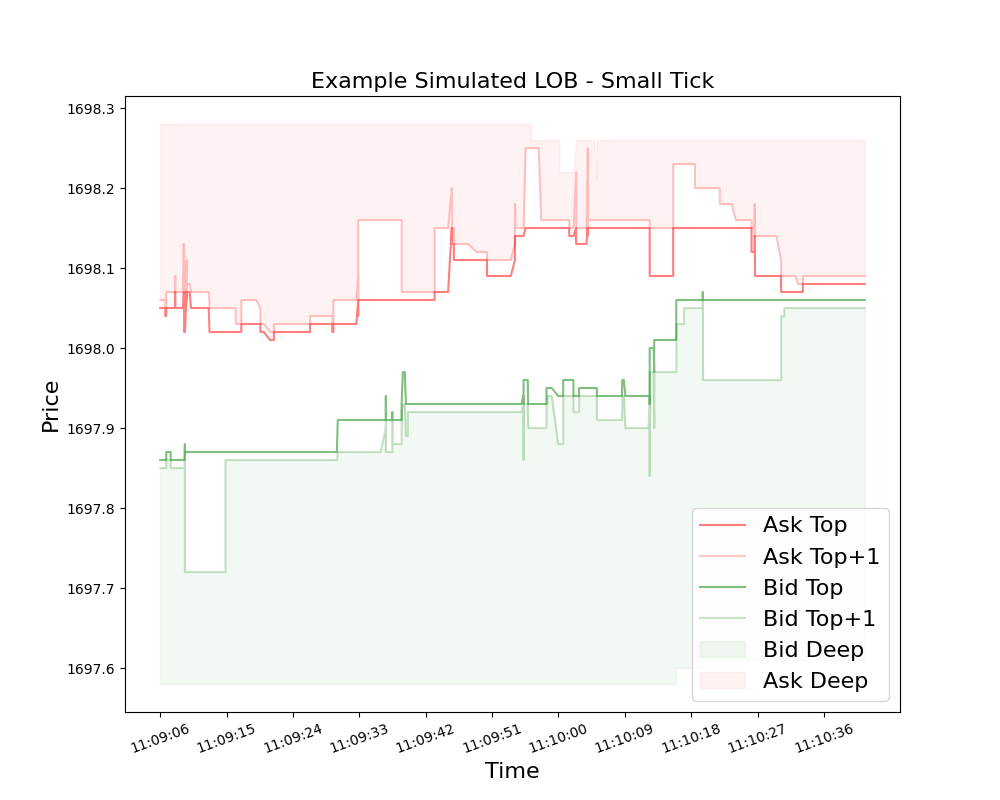}
\caption{Example Simulated LOB Dynamics }
\label{fig:eg}
\end{subfigure}
\begin{subfigure}[c]{.49\textwidth}
\includegraphics[width=\textwidth]{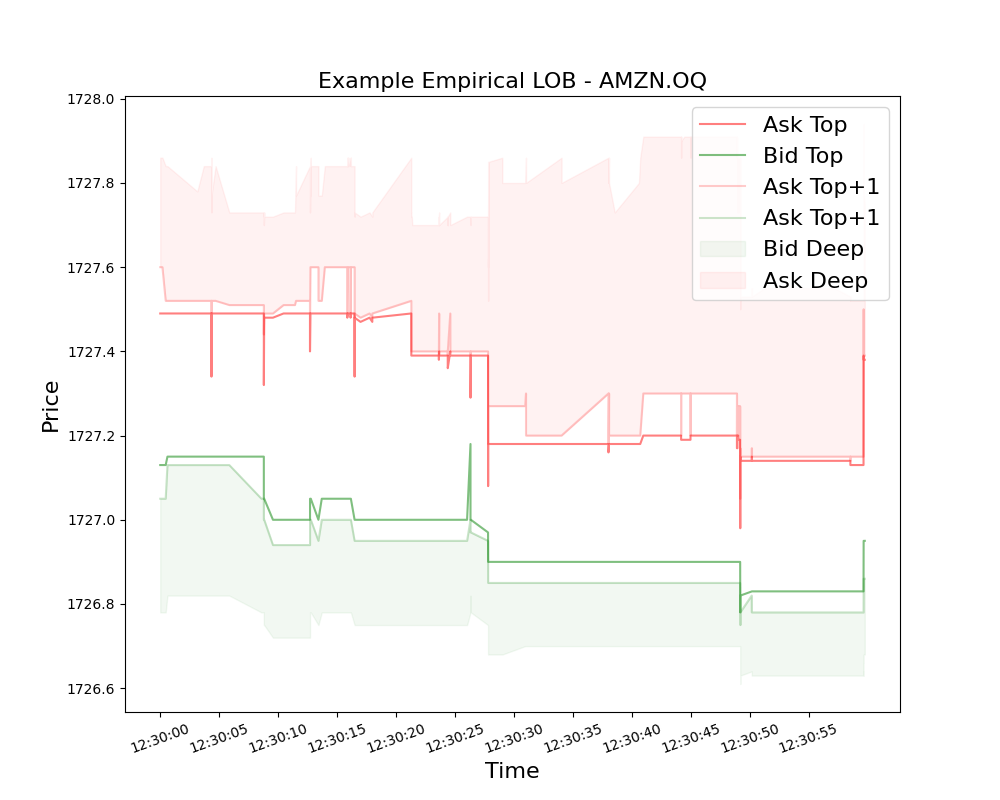}
\caption{Example Empirical LOB Dynamics }
\label{fig:egEmp}
\end{subfigure}
\caption{\highlighttwo{\textbf{{\color{black}Example of simulated LOB dynamics:}} Fig. (a) showcases an example simulated coarse grained LOB. Fig. (b) shows an empirical LOB (AMZN.OQ) with similar `Meta-Queues'. Notice the similarity in the two figures. Several of the stylized facts we discussed in Section 2 are faithfully being replicated: Bid-Ask Spread is persistently much higher than one tick, the best asks and bids change in multiple tick moves (and therefore so does the mid-price), and the LOB is sparse (notice the white gap between Top and Top+1 i.e. the second best quote).}}
\label{fig:SimStudy0}
\end{figure}

Using the model specified above, we simulate a LOB with random Hawkes and LOB Parameters. In Fig. \ref{fig:eg}, we show a sample of the simulated LOB for a small-tick configuration of the parameters which is specified in the following. We select the Hawkes Parameters by taking a power law kernel of the form $\phi^{(.)}(t) = a (1+ ct)^{-b}$ with values for parameters $a,b,c$ for the kernels and the exogenous intensities \highlighttwo{chosen from the calibrated results on AMZN.OQ from \cite{jain2023hawkes}. The exact parameter specification is given in Appendix \ref{params}. Note that this choice is purely driven by having a baseline LOB which makes logical sense (for {\color{black}example,} for a stable spread process, we require the rates of MOs and COs to approximately nullify the rates of LOs)}. { Therefore the following discussion can be replicated for any logically sensible set of Hawkes parameters.} We reflect the parameters of the Ask side to the Bid side to maintain model's symmetry. For In-spread events, we choose $\alpha = 0.95, \beta = 0.6$ to be corresponding to a small-tick asset for this study. Following \cite{jain2023hawkes}, we choose geometric distributions for order sizes $\kappa_{(.)}$ with spikes at round numbers like 1, 10, and 100. Similarly for unseen queue sizes, we use the geometric distribution with spikes at 1, 10, 100, 500 and 1000. Next, for $\eta_{(.)}$, we choose the geometric distribution again (with no spikes) since empirical observations (Fig. \ref{fig:emptyLevels}) show that the distance between top and top + 1 (i.e. the second best quote) levels seem to follow a geometric distribution. {\color{black}We showcase an example of the simulated LOB from the MQH model using a small-tick configuration for the parameters in Fig. \ref{fig:eg} and for comparitive purposes show an empirical small-tick LOB's dynamics (AMZN) in Fig. \ref{fig:egEmp}.}{ The similarity in the two figures in Fig. \ref{fig:SimStudy0} should be noticed. Several of the stylized facts discussed in Section 2 are being faithfully replicated: the Bid-Ask Spread is persistently much higher than one tick, multiple tick changes are observed in the best asks and bids (and consequently in the mid-price), and the LOB appears sparse (as indicated by the white gap between Top and Top+1, i.e., the second-best quote).}

\begin{figure}[h]
\centering
\begin{subfigure}[c]{0.49\textwidth}
    \includegraphics[width=\textwidth]{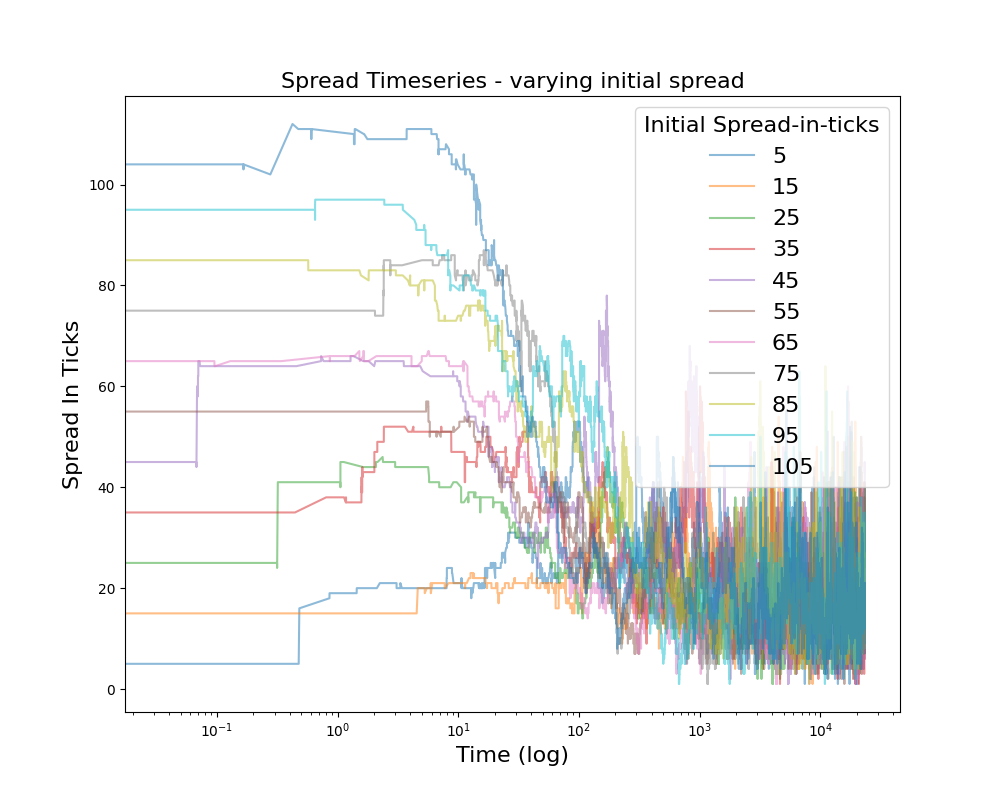}
    \caption{Spread Timeseries: Varying $s(0)$ in the simulations }
    \label{fig:spr0}
\end{subfigure}
\begin{subfigure}[c]{0.49\textwidth}
    \includegraphics[width=\textwidth]{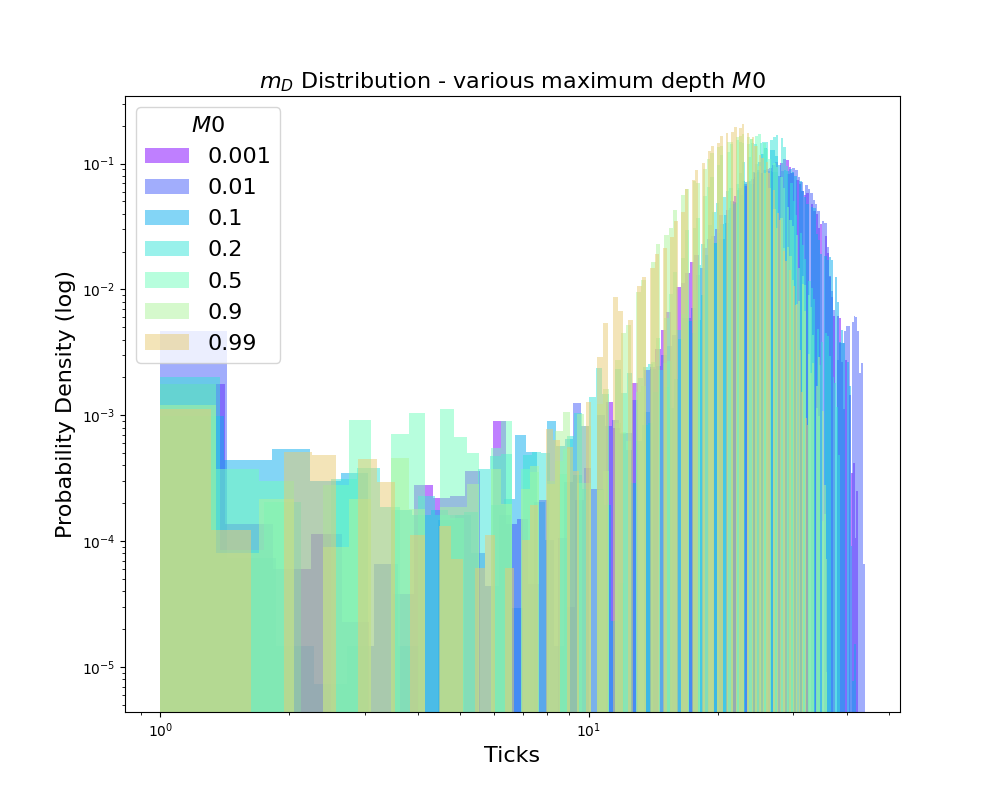}
    \caption{Width of deep levels: Varying initial sparsity in the simulations }
    \label{fig:M0}
\end{subfigure}
 \caption{\highlighttwo{\textbf{{\color{black}Ergodicity of the MQH model:}} In (a) we do an ergodicity study for the spread time series by perturbing the initial spread of the simulation. Similarly in (b) we show the invariance of the distribution of $m_D(t)$ with various initial levels of sparsity (denoted by $M0$). A similar trend is observed for other state variables' distribution with varying initialisations.}}
\label{fig:SimStudy1}
\end{figure}

{\color{black}We now focus on ergodicity of our MHQ model.} Since proving ergodicity analytically for {\color{black}various LOB quantities like for instance the spread process} under a Hawkes Counting Process for 12 types of events with various cross-interacting state variable dynamics is mathematically very difficult and beyond the scope of this paper, we instead rely on a simulation-based analysis to illustrate the model’s ergodic behavior. 
For that purpose,
we vary the initialization of the LOB parameters and check for stationarity the dynamics at long term. In Fig. \ref{fig:spr0}, we vary the initial spread from 5 ticks to 105 ticks for a small-tick parameter configuration and observe that the spread quickly converges to the mean spread of the LOB process which for this experiment is around 15 ticks. In Fig. \ref{fig:M0}, we vary the initial sparsity of the LOB model by varying the parameter $M0 := (M0_T, M0_D)$. We initialise the LOB's $m_T(0), m_D(0)$ using the geometric distribution with parameters $M0_T, M0_D$ respectively. In this experiment we set $M0_T = M0_D$ and we vary this quantity from near zero (which implies a uniform distribution for the initial sparsity) to near one (which implies a very narrow distribution for the initial sparsity and therefore implies no sparsity at all). We plot the the distribution of $m_D(t)$ and see that the distribution remains invariant of the choice of $M0$. { Similarly, for other quantities of interest, we see an ergodic simulation with varying initializations.} These two figures reinforce our hypothesis that the process formed from the dynamics illustrated above is ergodic.

{\color{black}
In the following sections, we focus on identification of critical parameters in our model which decide whether the simulated LOB is a large-, medium- or small-tick asset. Even though we start our analysis with a small-tick configuration, we vary each of these parameters on a range of values to showcase the fact that this model is versatile enough to simulate characteristics of all kinds of assets regardless of their relative tick-size. Note that this analysis is done by keeping all other parameters to be the same as presented in Appendix B and thus, does not represent a true calibration for {any} asset. Even so, in this section we will show that controlling certain {critical parameters can simulate a LOB very similar to a large-tick asset even with rest of the parameters in the small-tick setting. }}


\subsection{Controlling Relative Tick-size and associated Stylized Facts in the MQH Model} \label{control_rel}

\begin{figure}[h]
    \centering
    \includegraphics[width=0.7\textwidth]{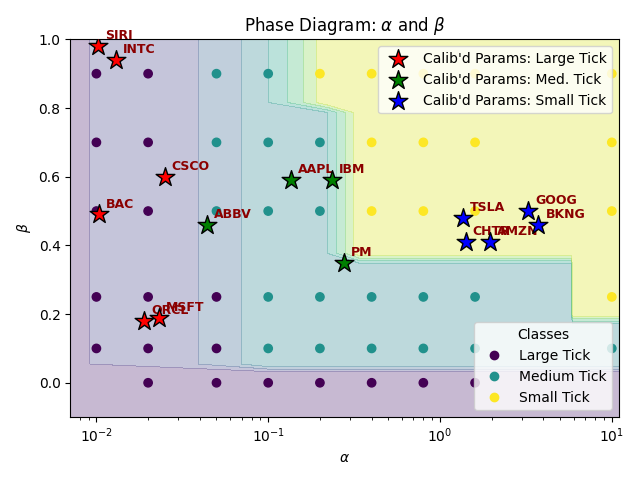}
    \caption{\textbf{Phase Diagram for various tick-size regimes:} This phase diagram shows the classification of simulated tick-size regimes as a function of the In-Spread Hawkes parameters $(\alpha, \beta)$. Each simulated LOB is categorized into a small-, medium-, or large-tick regime based on the average spread ($\bar{s}$). Calibrated parameters for 15 empirical assets are marked as star symbols.}
    \label{fig:phaseDiag}
\end{figure}

\begin{figure}[h]
\centering
\begin{subfigure}[c]{0.49\textwidth}
    \includegraphics[width=\textwidth]{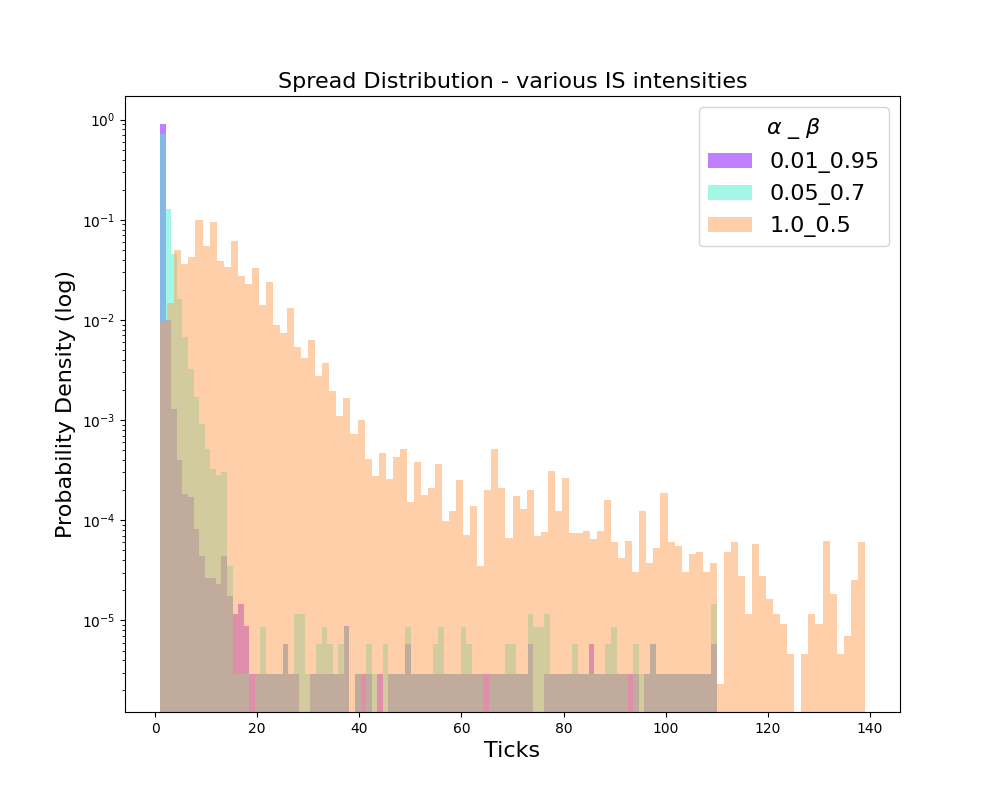}
    \caption{Spread distribution for different $(\alpha, \beta)$ values.}
    \label{fig:ISHP_spr}
\end{subfigure}
\begin{subfigure}[c]{0.49\textwidth}
    \includegraphics[width=\textwidth]{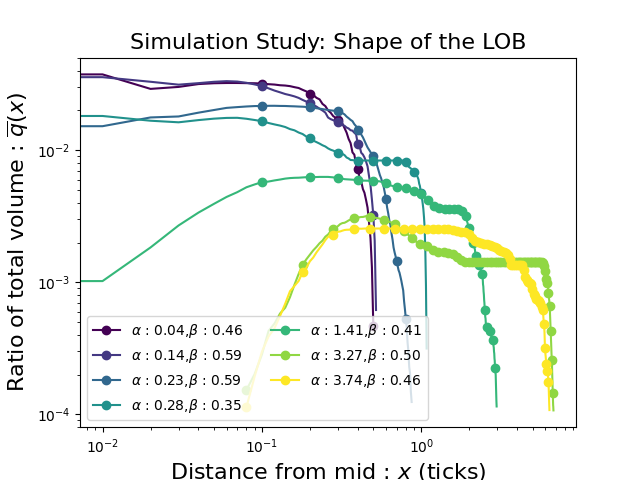}
    \caption{LOB shape for different $(\alpha, \beta)$ values.}
    \label{fig:ISHP_shape}
\end{subfigure}
\caption{\textbf{Effect of varying the IS Hawkes parameters on the spread distribution and LOB shape:} As $(\alpha, \beta)$ increase, the spread distribution transitions from exponential-like to log-normal-like, and liquidity shifts from the top of the book to deeper levels.}
\label{fig:SimStudy2}
\end{figure}

This section investigates the model’s ability to reproduce a variety of relative tick-size regimes and associated stylized facts, as observed empirically in Section \ref{sty}. The focus is on identifying which parameters of the MQH model drive changes in market microstructure features typically associated with small-, medium-, and large-tick assets. {\color{black}As for the question of comparisons with the empirically observed variations of these stylized facts with relative tick-size, this will be discussed in Section \ref{compare_sty}.}

From Eq. \eqref{lamb_IS}, the in-spread intensity contains two additional parameters, $\alpha$ and $\beta$, which influence the frequency and responsiveness of spread-narrowing events. Since these directly affect the spread width, they are natural candidates for controlling the simulated tick-size. Fig. \ref{fig:phaseDiag} presents a phase diagram where LOBs simulated across a grid of $(\alpha, \beta)$ values are classified based on the mean spread into three tick-size regimes. The calibrated values for 15 empirical stocks are superimposed and align well with the simulated classification boundaries, suggesting that these parameters are indeed critical in reproducing the empirical tick-size structure. Fig. \ref{fig:SimStudy2} supports this classification by showing that variations in $(\alpha, \beta)$ lead to systematic changes in the spread distribution and the average LOB shape. The spread distribution shifts from exponential-like to log-normal-like as the $\alpha$ increases and $\beta$ decreases, while the liquidity profile moves from being concentrated at the top of the book to being more dispersed across deeper levels. These trends mirror those observed in empirical data for large- and small-tick assets respectively.

While $\alpha$ and $\beta$ are effective in modulating the spread and the overall shape of the LOB, they are not sufficient to reproduce other important stylized facts {\color{black} on returns and sparsity (eg. those in sections \ref{trades} and \ref{sparsity}).} In particular, small-tick assets exhibit fatter-tailed returns and greater queue sparsity, features which are not captured by $(\alpha, \beta)$ alone. This motivates the introduction of an additional parameter, $\hat{\eta}_{(.)}$, which governs the distribution of empty levels between adjacent queues and thereby affects the microstructural sparsity of the LOB. We note that in the following we use the geometric distribution for $\Pi_{\eta}(.)$ and use a common value of $\hat{\eta}$ for all three parameters $\hat{\eta}_{IS}, \hat{\eta}_{T}, \hat{\eta}_{T+1}$. 

\begin{figure}[h]
\centering
\begin{subfigure}[c]{0.49\textwidth}
    \includegraphics[width=\textwidth]{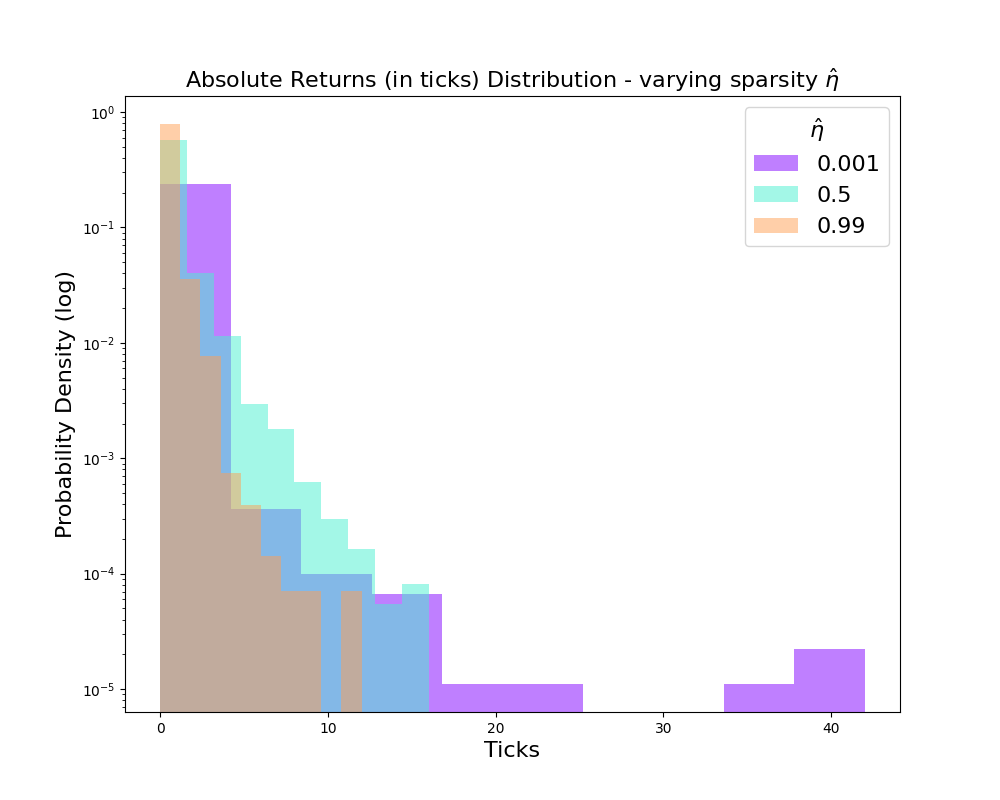}
    \caption{Absolute returns as a function of $\hat{\eta}$.}
    \label{fig:retEta}
\end{subfigure}
\begin{subfigure}[c]{0.49\textwidth}
    \includegraphics[width=\textwidth]{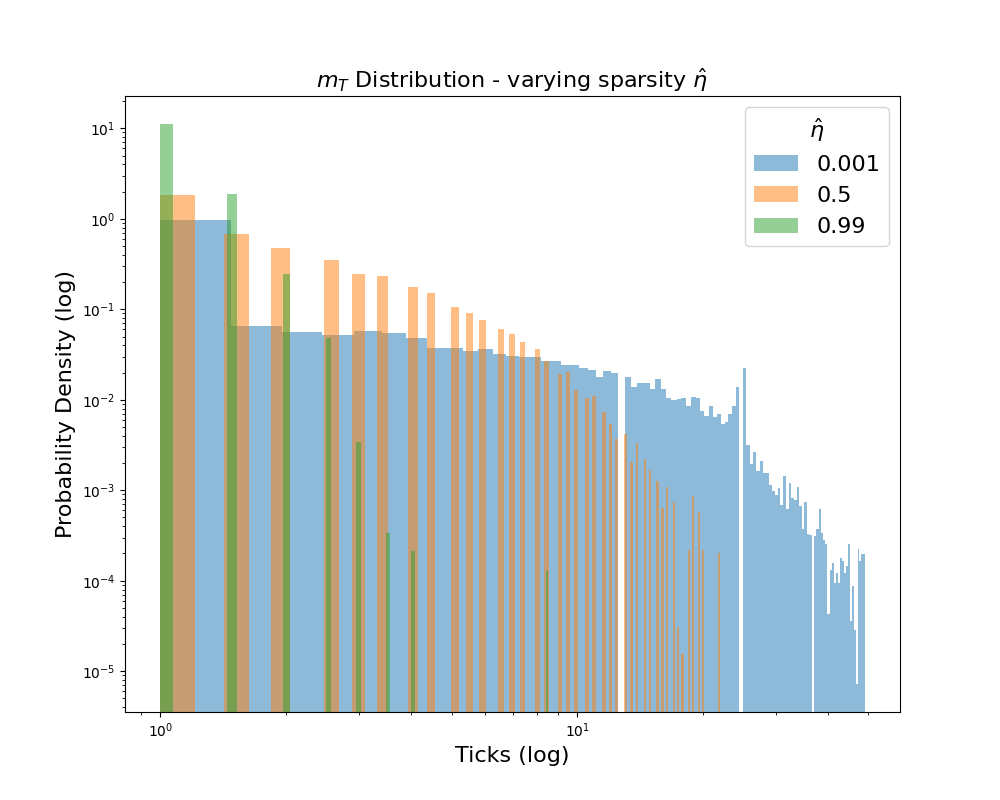}
    \caption{Top-level queue distance $m_T(t)$ for varying $\hat{\eta}$.}
    \label{fig:mTeta}
\end{subfigure}
\begin{subfigure}[c]{0.49\textwidth}
    \includegraphics[width=\textwidth]{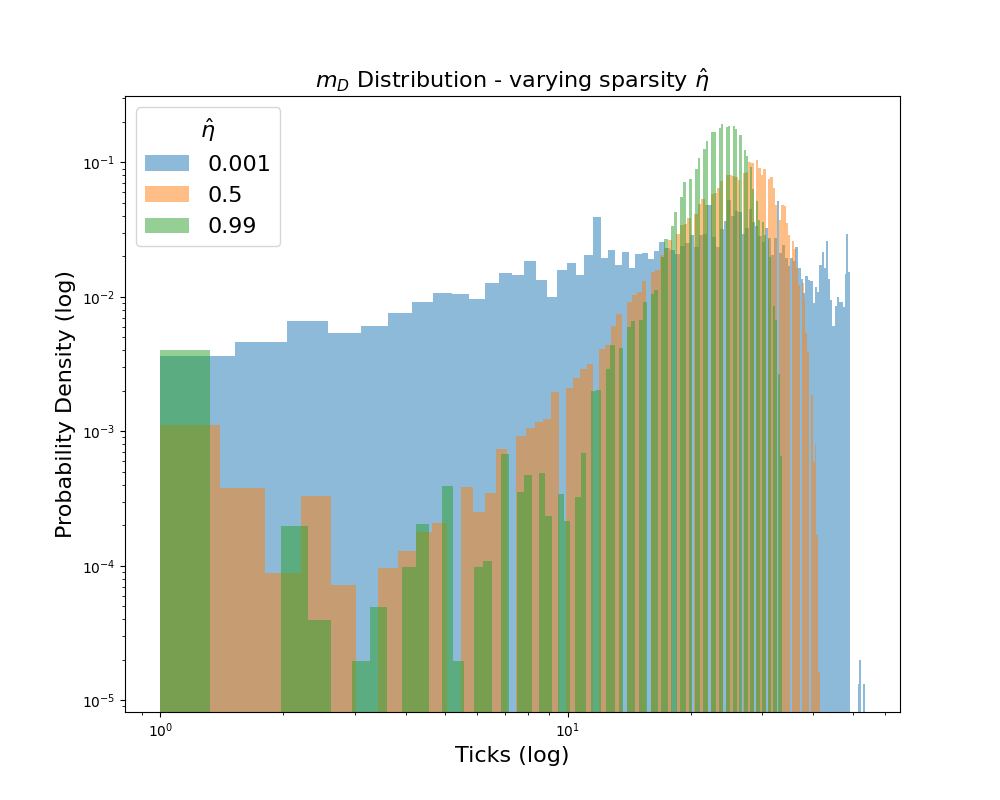}
    \caption{$m_D(t)$ from simulations.}
    \label{fig:mDsim}
\end{subfigure}
\begin{subfigure}[c]{0.49\textwidth}
    \includegraphics[width=\textwidth]{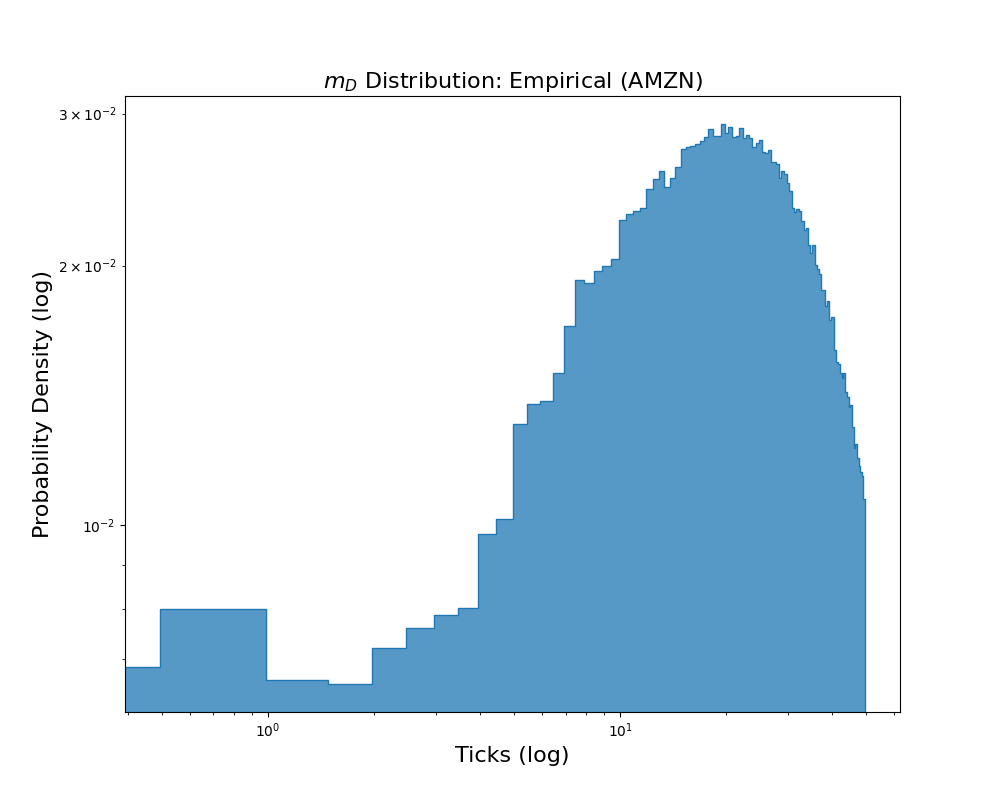}
    \caption{Empirical $m_D(t)$ for AMZN.}
    \label{fig:mDemp}
\end{subfigure}
\caption{\textbf{Effect of varying the dynamic sparsity parameter $\hat{\eta}$ on returns and sparsity:} Simulations are performed at fixed $(\alpha, \beta)$ corresponding to a small-tick regime. Lower $\hat{\eta}$ values induce fatter tails in the returns distribution, increased top-level sparsity, and deeper queue spacing, consistent with empirical observations.}
\label{fig:SimStudy}
\end{figure}

Fig. \ref{fig:SimStudy} illustrates the impact of varying $\hat{\eta}$ while keeping $(\alpha, \beta)$ fixed to values that correspond to a small-tick setting. As $\hat{\eta}$ decreases, indicating greater sparsity, the returns distribution becomes more heavy-tailed and the spacing between the top levels of the book increases. These changes are visible in the distribution of the top-level distance variable $m_T(t)$ and the deeper queue spacing variable $m_D(t)$. The shape of the simulated $m_D(t)$ closely resembles that of the empirical distribution shown for AMZN, suggesting that $\hat{\eta}$ effectively captures a key empirical feature of small-tick LOBs.

These findings demonstrate that accurately simulating LOBs across tick-size regimes requires the joint calibration of all three parameters: $\alpha$, $\beta$, and $\hat{\eta}$. While the former two control the spread and shape of the book, the latter is essential for reproducing observed variations in returns and sparsity. Together, they enable the MQH model to replicate a broad range of empirical microstructural phenomena associated with different tick-size conditions.

{\color{black}
\subsection{Comparing MQH LOB Stylized Facts to Empirical Evidence Across Relative Tick-sizes } \label{compare_sty}

\begin{figure}[h]
\centering
\begin{subfigure}[c]{0.49\textwidth}
    \includegraphics[width=\textwidth]{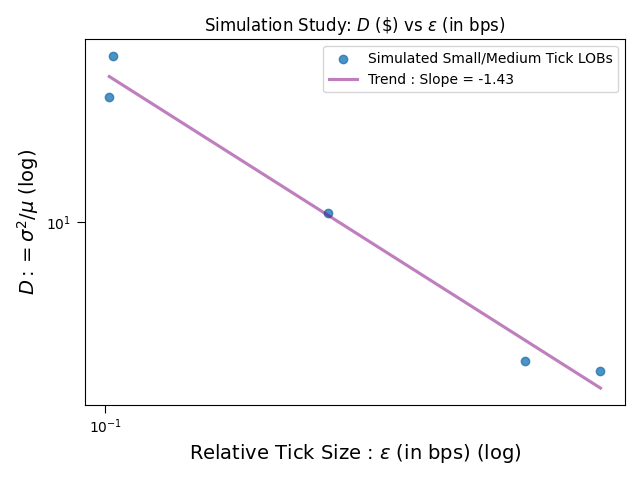}
    \caption{Index of Dispersion of Spread Distribution vs Rel. Tick-Size }
    \label{fig:IoD_SimEta}
\end{subfigure}
\begin{subfigure}[c]{0.49\textwidth}
    \includegraphics[width=\textwidth]{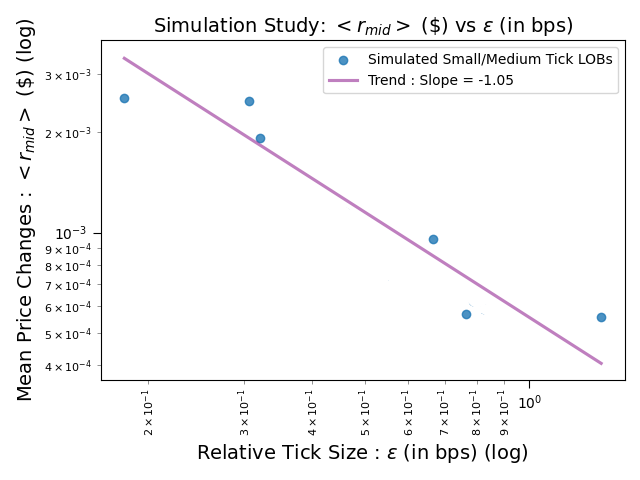}
    \caption{Avg Mid-Price Changes vs Rel. Tick-Size}
    \label{fig:ret_SimEta}
\end{subfigure}
\begin{subfigure}[c]{0.49\textwidth}
    \includegraphics[width=\textwidth]{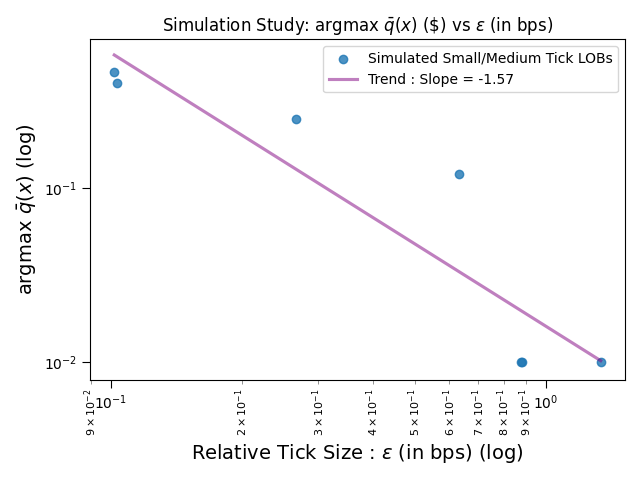}
    \caption{Maxima of Shape of the LOB vs Rel. Tick-Size }
    \label{fig:max_SimEta}
\end{subfigure}
\begin{subfigure}[c]{0.49\textwidth}
\includegraphics[width=\textwidth]{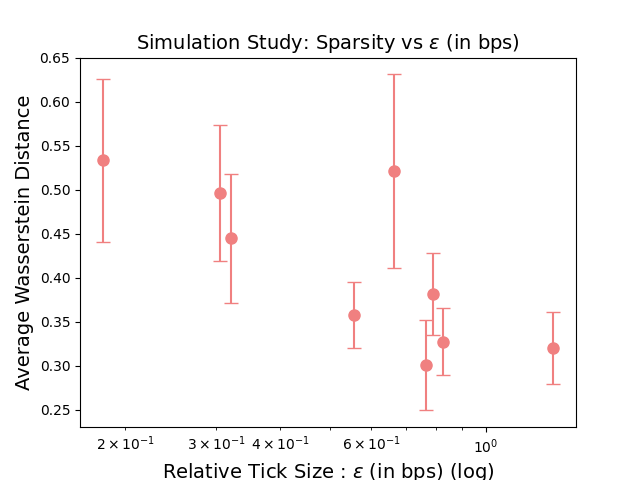}
    \caption{Sparsity Measure (Wasserstein Metric) vs Rel. Tick-Size }
    \label{fig:sparse_SimEta}
\end{subfigure}
 \caption{{\color{black}\textbf{Stylized Facts (Simulated)}: The simulated $\epsilon$ is compared to various stylized facts measures from Section \ref{sty} for small and medium-tick LOBs. On log-log scale, we clearly see the smooth regressions of $D$ (index of dispersion for spread), $\langle r_{mid} \rangle$ (mean mid-price changes), and $\text{argmax }\bar{q}(x)$ (maxima of shape of the LOB) as we saw in Fig.s \ref{fig:IoDSpread}, \ref{fig:meanRet}, \ref{fig:maximaShape}, respectively. Further, the slopes of the fitted trends are approximately the same as the slopes we calibrated from empirical data. Finally, we also see the sparsity measure, introduced in Section \ref{sparsity}, also increases with decreasing $\epsilon$ as noticed in Fig. \ref{fig:wass}.}}
\label{fig:compSty}
\end{figure}

This section assesses whether the MQH model is able to reproduce the key stylized facts of the limit order book documented empirically in Section~\ref{sty}.

Before studying how stylized facts vary with the relative tick size $\epsilon$ in the simulated MQH model, we must first define a way to estimate $\epsilon$ in simulations. An important feature of the MQH model is that the absolute scale of prices plays no role in the MQH model’s dynamics. Consequently, it is not possible to compute $\epsilon$ directly from Eq.~\ref{eq:epsilon}, which requires a price scale for its definition. Instead, we estimate $\epsilon$ in simulations using the empirical power-law relation between the average spread $\bar{s}$ and $\epsilon$ established in Eq.~\ref{eq:pl} and illustrated in Fig.~\ref{fig:SpreadVsPrice}. This relation provides a way to infer relative tick size from observed spread statistics. However, this method is only valid for small- and medium-tick LOBs, as it breaks down in the large-tick regime. Therefore, in what follows, we restrict our comparison to simulated LOBs that fall into the small- and medium-tick classes based on this proxy.

We begin with the spread distribution. As shown in Fig.~\ref{fig:IoD_SimEta}, the index of dispersion $D$ of the spread decreases systematically with increasing $\epsilon$, following a clear power-law trend. This is consistent with the empirical observation in Fig.~\ref{fig:IoDSpread}, where a similar monotonic dependence was reported. The simulated scaling exponent $\alpha_D^{\text{sim}} = -1.43$ is in close agreement with the empirical value $\alpha_D^{\text{emp}} = -1.36$.

We now turn to mid-price changes. Fig.~\ref{fig:ret_SimEta} shows that the mean absolute mid-price return $\langle r_{\text{mid}} \rangle$ scales with $\epsilon$ in a manner that closely follows the empirical trend shown in Fig.~\ref{fig:meanRet}. The estimated slope from simulations is $\alpha_r^{\text{sim}} = -1.05$, which compares well to the empirical slope of $-0.95$.

Next, we study the shape of the order book. Fig.~\ref{fig:max_SimEta} displays the variation in the location of the peak of the average LOB shape $\arg\max \bar{q}(x)$ as a function of $\epsilon$. As in the empirical case (Fig.~\ref{fig:maximaShape}), the peak shifts deeper into the book as $\epsilon$ decreases. The simulated slope $\alpha_q^{\text{sim}} = -1.57$ again aligns well with its empirical counterpart $\alpha_q^{\text{emp}} = -1.50$.

\begin{table}[ht]
\centering
\begin{tabular}{lccc}
\toprule
\textbf{Metric} & \textbf{Symbolic Form} & \textbf{Simulated Slope} & \textbf{Empirical Slope } \\ &&&\textbf{(from Section \ref{sty})} \\
\midrule
Index of Dispersion & \( D \sim \epsilon^{\alpha_D} \) & \( \alpha_D^\text{sim} = -1.43 \) & \( \alpha_D^\text{emp} = -1.36 \) \\
Mean Mid-Price Change & \( \langle r_{\text{mid}} \rangle \sim \epsilon^{\alpha_r} \) & \( \alpha_r^\text{sim} = -1.05 \) & \( \alpha_r^\text{emp} = -0.95 \) \\
Maxima of Avg. LOB Shape & \( \arg\max \bar{q}(x) \sim \epsilon^{\alpha_q} \) & \( \alpha_q^\text{sim} = -1.57 \) & \( \alpha_q^\text{emp} = -1.50 \) \\
\bottomrule
\end{tabular}
\caption{\textbf{Comparison of scaling exponents obtained from simulations and empirical analysis (Section~\ref{sty}) for key LOB metrics:} Table \ref{tab:slopes} summarizes the comparison between scaling exponents measured in real order‐book data (Section \ref{sty}) and those obtained from our MQH simulations. In Section \ref{sty} we established three power‐law relationships between relative tick size and (i) the index of dispersion of the spread, (ii) the average mid‐price change, and (iii) the location of the peak in the average LOB shape. Repeating the same regressions on simulated LOBs—using our proxy estimate of relative tick size—yields slopes that closely match the empirical values. This agreement demonstrates that the MQH model faithfully reproduces the key microstructural scaling laws observed in real markets.}
\label{tab:slopes}
\end{table}

Finally, we examine the sparsity of the LOB. As shown in Fig.~\ref{fig:sparse_SimEta}, the Wasserstein-based sparsity metric introduced in Section~\ref{sparsity} increases with decreasing $\epsilon$, confirming the empirical trend illustrated in Fig.~\ref{fig:wass}. This demonstrates that the MQH model is able to replicate not only return- and volume-based features, but also structural characteristics of queue configurations.

In conclusion, across all metrics considered—spread, mid-price return, LOB shape, and sparsity—the MQH model demonstrates strong consistency with the empirical scaling behaviors documented in Section~\ref{sty}. This validates the model’s capacity to realistically replicate LOB microstructure for small- and medium-tick assets when appropriately calibrated.

}%

\section{ Model Calibration:} \label{calib}

\begin{figure}[h]
\centering
\begin{subfigure}[c]{0.32\textwidth}
\includegraphics[width=\textwidth]{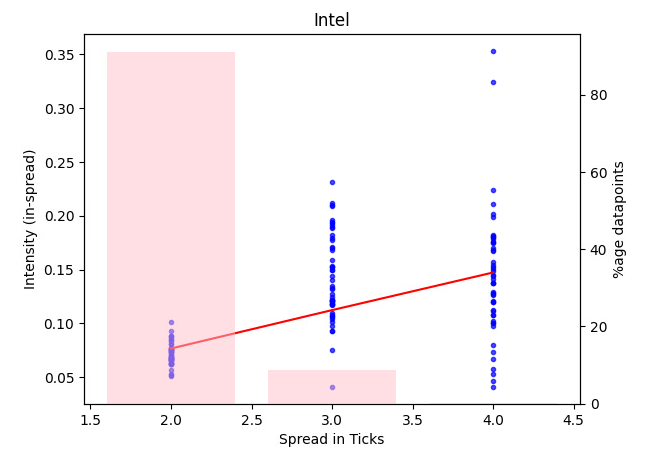}
\caption{INTC }
\label{fig:alphabeta_INTC}
\end{subfigure}
\begin{subfigure}[c]{0.32\textwidth}
\includegraphics[width=\textwidth]{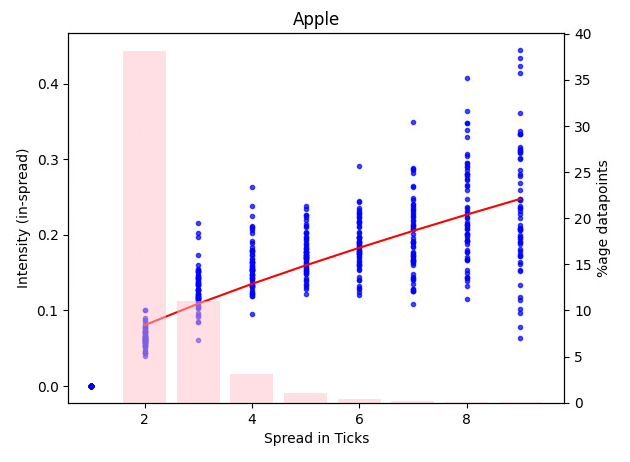}
\caption{AAPL }
\label{fig:alphabeta_AAPL}
\end{subfigure}
\begin{subfigure}[c]{0.32\textwidth}
\includegraphics[width=\textwidth]{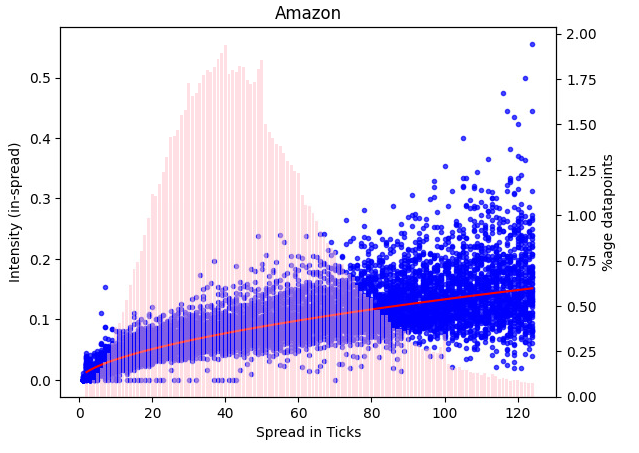}
\caption{AMZN }
\label{fig:alphabeta_AMZN}
\end{subfigure}
 \caption{{\color{black}\textbf{$\alpha, \beta$ Calibration: } The order intensity is approximated by counting in-spread order arrivals within 0.01-second intervals and plotted against the current spread in ticks using scatter plots. The translucent red bars indicate the distribution of these observations. A linear regression, depicted by the red line, on the log-log transformed data (excluding spreads of 1 tick) yields a power-law. The fitted coefficients are summarized in Table \ref{tab:calib_critical}. We can clearly see the power-law behaviour persists for all three categories of stocks depicted here.}}
\label{fig:alphabeta_calib}
\end{figure}

\begin{figure}[h]
\centering

\includegraphics[width=.6\textwidth]{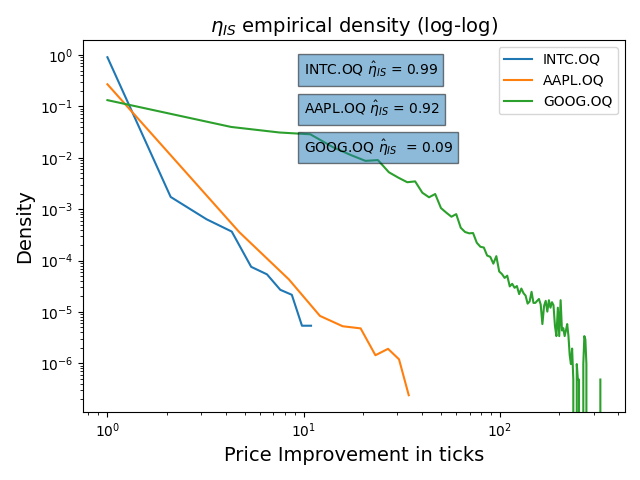}
\label{fig:eta_is_calib}
 \caption{\highlighttwo{\textbf{$\hat{\eta}_{IS}$ Calibration:} $\hat{\eta}_{IS}$ is the parameter which controls the sparsity of the LOB. Here we show three examples, one from each category of assets, of empirical histograms of the sparsity levels. On a log-log scale it can be clearly seen that the distribution is similar to the geometric distribution. Inset we also show the calibrated parameters with this prior for all three assets - we observe the calibrated parameters move from near one to near zero as we decrease the relative tick-size. This indicates an increase in sparsity with a decrease in relative tick-size. }}
\label{fig:eta_calib}
\end{figure}

\begin{table}[h]
\centering
\renewcommand{\arraystretch}{1.2}
\begin{tabular}{|lrrr|lrrr|lrrr|}
\hline
\multicolumn{4}{|c|}{\textbf{Large-tick}} & 
\multicolumn{4}{c|}{\textbf{Medium-tick}} & 
\multicolumn{4}{c|}{\textbf{Small-tick}} \\ \hline
\textbf{Asset} & \(\alpha\) & \(\beta\) & \(\hat{\eta}\) & 
\textbf{Asset} & \(\alpha\) & \(\beta\) & \(\hat{\eta}\) & 
\textbf{Asset} & \(\alpha\) & \(\beta\) & \(\hat{\eta}\) \\ \hline
SIRI & 0.0102 & 0.98 & 0.99 & 
ABBV & 0.0440 & 0.46 & 0.79 & 
TSLA & 1.3610 & 0.48 & 0.19 \\
BAC  & 0.0103 & 0.49 & 0.98 & 
PM   & 0.2750 & 0.35 & 0.77 & 
CHTR & 1.4100 & 0.41 & 0.15 \\
INTC & 0.0130 & 0.94 & 0.98 & 
AAPL & 0.1350 & 0.59 & 0.92 & 
AMZN & 1.9630 & 0.41 & 0.09 \\
CSCO & 0.0250 & 0.60 & 0.99 & 
IBM  & 0.2350 & 0.59 & 0.71 & 
GOOG & 3.2670 & 0.50 & 0.09 \\
ORCL & 0.0190 & 0.18 & 0.97 & 
     &        &      &      & 
BKNG & 3.7410 & 0.46 & 0.03 \\
MSFT & 0.0230 & 0.19 & 0.98 & 
     &        &      &      & 
     &        &      &      \\ \hline
\end{tabular}
\caption{Calibrated critical parameters \(\alpha\), \(\beta\), and \(\hat{\eta}\) for the 15 assets from Table \ref{tab:stocks}.}
\label{tab:calib_critical}
\end{table}

Since the Hawkes Dynamics is assumed to be independent of the LOB state variables, and since the distributions of the order sizes, and order distances from top are unconditional on any state variable and stationary, we can decouple the calibration to a parametric estimation of the distributions and a non-parametric estimation of the Hawkes kernels. We suggest the methodology described in \cite{jain2023hawkes}, combining \cite{kirchner2017estimation} and \cite{bacry2016estimation}, for the latter. { A brief summary of the method is as follows. They approximate the expectation of the intensity's integral with the observed bin counts of each of the 12 events. With certain constraints to control the stationarity of the Hawkes Process and the stability of the fitted kernels, they formulate an AR problem and solve it using a Quadratic Programming solver.} 

{\color{black}Following \cite{jain2023hawkes}, to calibrate the critical parameters $\alpha, \beta$, we analyze empirical high-frequency order book data. Specifically, we focus on the relationship between the order intensity and the spread measured in ticks. We approximate the instantaneous order intensity by counting the number of in-spread limit orders arriving within very short time intervals of 0.01 seconds. Since it is impossible to have in-spread limit orders with one tick spread, we remove this datapoint from the visualization and the regression. In Fig. \ref{fig:alphabeta_calib}, we visualize this relationship using scatter plots, where each point corresponds to the observed order intensity at a given spread value. The distribution of these observations is also represented by translucent red bars, which help illustrate the density and variability of the data points at different spread levels. To uncover the functional form governing this relationship, we perform a regression analysis on the logarithm of both the spread and the estimated intensity, i.e., a log-log regression (red line). 
}%

{\color{black}For the parametric distributions, }we suggest making use of the Maximum Likelihood method with a prior that all these distributions are Geometric distributions. In Fig. \ref{fig:eta_calib}, we showcase the empirical $\eta_{IS}$ distributions (note the shape looks exactly like the Geometric distribution) and their calibrated parameter: $\hat{\eta}_{(.)}$ across three assets: INTC (Large-tick), AAPL (Medium-tick) and GOOG (Small-tick). Optionally, following \cite{jain2023hawkes, LuAbergel2018}, one can add spikes in the probability distribution function to account for the traders' preference of round numbers in the order sizes and queue sizes. {\color{black} We present the calibrated critical parameters in Table \ref{tab:calib_critical} for the 15 assets we studied throughout this paper. As we can clearly see, the trend of decreasing $\epsilon$ with increasing $\alpha$ and decreasing $\beta$ and $\hat{\eta}$ remains consistent in empirical calibration as well. Additionally, as mentioned previously, we overlay the calibrated $\alpha, \beta$ over the phase diagram plot in the previous section (Fig. \ref{fig:phaseDiag}) to show that the calibrated parameters indeed fall consistently into their respective categories.} Finally, for the partition functions, {\color{black}$\Xi$ (Eq. \eqref{xi})} we made use of a simple linear partition function - i.e. we assume that the shape of the LOB in the deeper LOB is flat {\color{black}for purging queues}. This is clearly a strong assumption (Section \ref{shape}), and a better choice could be using the average shape of the LOB.

\section{Discussion and Conclusion}

In this work we outline how, statistically, does the LOB change when the granularity of price levels is kept constant with a constant tick-size while the price of the security increases from a few dollars to a few thousands of dollars. We quantify this change by the change in relative tick-size $\epsilon$ which is defined as the ratio between the tick-size and the price of the security. One of our key contributions is to highlight several stylized facts, which are then used to differentiate between {large-, medium- and small-tick} assets, and define clear metrics to measure them. We also provide a set of cross-asset visualizations of the stylized facts and show how different (or similar) these attributes are with varying relative tick-size. We found  different power law relations in these stylized facts. Firstly, the average spread of a security and the average absolute returns of the security are both approximately proportional to the inverse of the relative tick-size. Secondly, the depth at which we find the maximum liquidity in the LOB follows a power law over the relative tick-size with the exponent roughly equal to $\frac{3}{2}$. In addition to that we see definitive trends in the shape of the spreads distribution as well as the sparsity in the LOB as well decrease the relative tick-size. We make use of a stylized fact known as \emph{leverage}, which seems to be universally similar in structure across all assets we studied, to model the LOB as a Hawkes Process with the top of the LOB as a separate dimension and the rest of the LOB ({roughly }upto the median of the volume shape of the LOB) as a separate dimension. 

We propose a {\color{black}meta-queues} mutivariate Hawkes model extending the one proposed in \cite{jain2023hawkes} to account for sparsity, multi-tick level price moves and the humped shape of the LOB for small-tick assets. We showcase the versatility of the model by performing several simulation studies. Thereafter, we identify the in-spread Hawkes Parameters and the sparsity variables as the critical variables which define if the simulated LOB will be similar to a large-tick or a small-tick asset. We perform several tests to showcase that a number of stylized facts like sparsity, shape, and relative returns' distribution can be varied smoothly from that of a large-tick to small-tick asset using our model. {\color{black}Further we show that the stylized facts' metrics variation with relative tick-sizes is preserved in the model and is inline with empirical observations.}

{\color{black}We note that to maintain analytical tractability and facilitate calibration, several simplifying assumptions were made during model development. Testing the underlying assumptions of any quantitative finance model is essential to ensure its practical relevance and robustness. These assumptions have been explicitly identified and their empirical validity rigorously examined in the Appendix \ref{assump}, providing a critical assessment of the model’s foundations.} We show that almost all of the model's assumptions are violated in the real data and conclude by proposing questions for further research in this area in this following. 

\begin{enumerate}
    \item Is there a mathematically tractable way of incorporating the relationship between queue sizes, width of the top of the LOB and width of the deeper part of the LOB, and the order arrival rates' dynamics?
    \item How to model order sizes such that their correlation structure observed in empirical data is preserved?
    \item How can we calibrate the models above in (i) to a reasonable level of uncertainty in the model parameters? 
\end{enumerate}

Building on these foundations, future research will aim to enhance the adaptability and accuracy of models, leading to more robust and practical applications in limit order book modeling.

{\section*{Disclaimer}}
Opinions and estimates constitute our judgement as of the date of this Material, are for informational purposes only and are subject to change without notice. This Material is not the product of J.P. Morgan’s Research Department and therefore, has not been prepared in accordance with legal requirements to promote the independence of research, including but not limited to, the prohibition on the dealing ahead of the dissemination of investment research. This Material is not intended as research, a recommendation, advice, offer or solicitation for the purchase or sale of any financial product or service, or to be used in any way for evaluating the merits of participating in any transaction. It is not a research report and is not intended as such. Past performance is not indicative of future results. Please consult your own advisors regarding legal, tax, accounting or any other aspects including suitability implications for your particular circumstances. J.P. Morgan disclaims any responsibility or liability whatsoever for the quality, accuracy or completeness of the information herein, and for any reliance on, or use of this material in any way.
Important disclosures at: www.jpmorgan.com/disclosures \\

\section*{Acknowledgements}

The author KJ would like to acknowledge JP Morgan Chase \& Co. for his PhD scholarship. This work benefited from state aid managed by l’Agence Nationale de la Recherche under France 2030 bearing the reference “ANR-23-IACL-0008". We are thankful to Julius Bonart, Tanmay Satpathy, Nick Firoozye and Philip Treleaven for their comments and feedback for this work. Finally,  we  are  grateful  to  the anonymous reviewers for their constructive feedback.

\section*{Disclosure Statement}

No potential conflict of interest was reported by the author(s).

\section*{Funding}

This work was supported by JP Morgan Chase \& Co.

\renewcommand*{\bibfont}{\small}
\printbibliography[
heading=bibintoc,
title={References}
]

\newpage
\appendix


\section{{ Stylized Facts additional illustrations}}
\label{app:spread}
{ In this appendix we include some additional illustrations of the different stylized facts discussed in the paper

\begin{figure}[H]
\centering
\begin{subfigure}[c]{0.4\textwidth}
    \includegraphics[width=\textwidth]{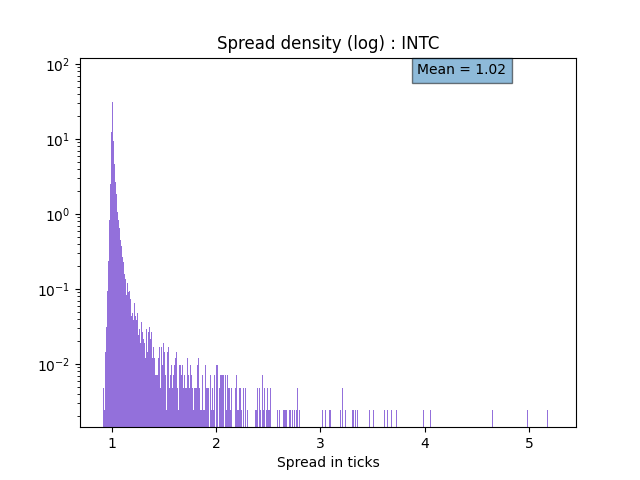}
    \caption{INTC (Large-tick) - PDF Log}
    \label{fig:kernelNorms}
\end{subfigure}
\begin{subfigure}[c]{0.4\textwidth}
    \includegraphics[width=\textwidth]{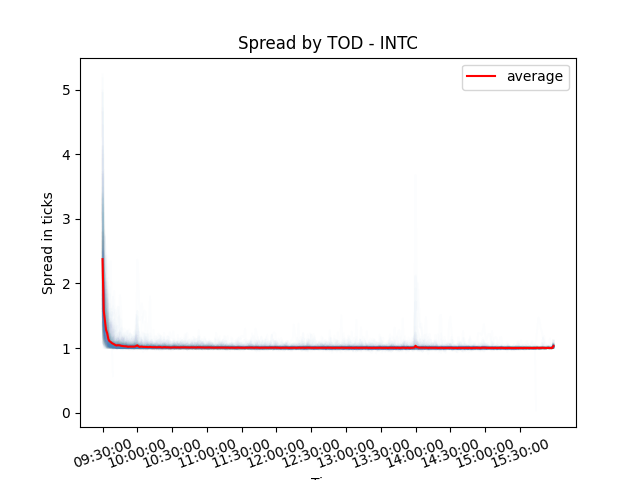}
    \caption{INTC (Large-tick) - Mean by TOD}
    \label{fig:kernelNorms}
\end{subfigure}
\begin{subfigure}[c]{0.4\textwidth}
    \includegraphics[width=\textwidth]{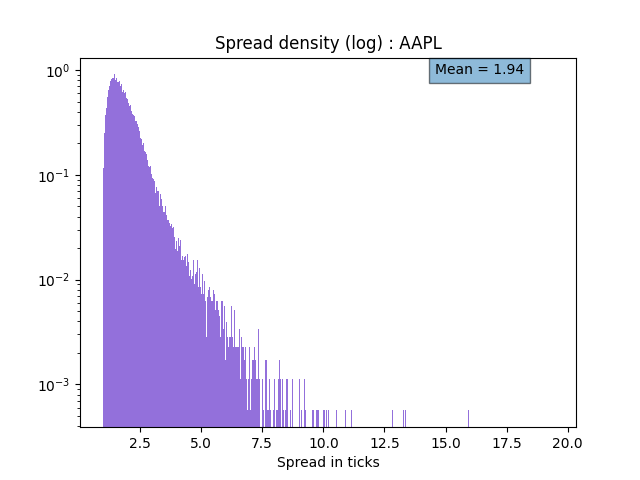}
    \caption{AAPL (Medium-tick) - PDF Log}
    \label{fig:kernelNorms}
\end{subfigure}
\begin{subfigure}[c]{0.4\textwidth}
    \includegraphics[width=\textwidth]{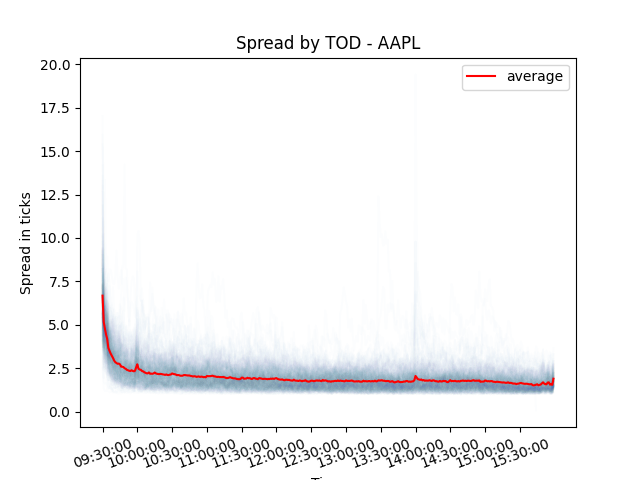}
    \caption{AAPL (Medium-tick) - Mean by TOD}
    \label{fig:kernelNorms}
\end{subfigure}
\begin{subfigure}[c]{.4\textwidth}
\includegraphics[width=\textwidth]{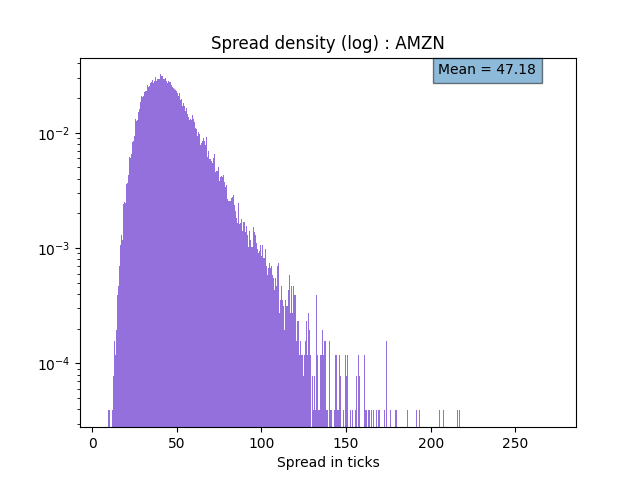}
\caption{AMZN (Small-tick) - PDF Log}
\label{fig:tslaNorms}
\end{subfigure}
\begin{subfigure}[c]{0.4\textwidth}
\includegraphics[width=\textwidth]{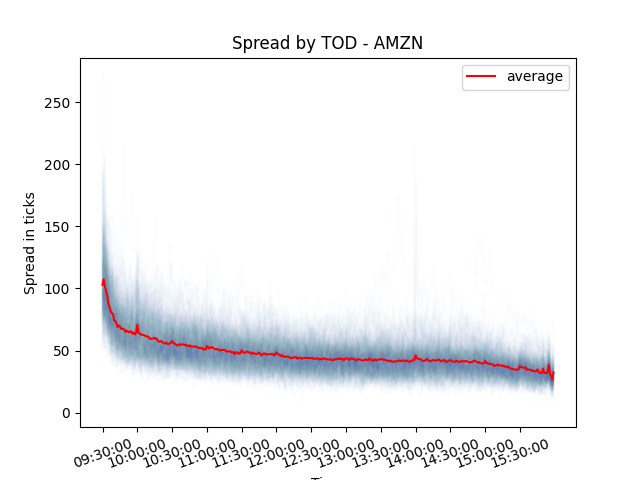}
\caption{AMZN (Small-tick) - Mean by TOD}
\label{fig:kernelNorms}
\end{subfigure}
\caption{
{ Spread log empirical probability distribution function (PDF, figures on the first column) and average spread as a function of the time of the day (TOD, figures on the second column) for 3 different assets (resp. large, medium and small-tick) Bid-Ask Spread}}
\label{fig:spreads}
\end{figure}

\subsection{Additional illustrations on the spread}}

{ 
In Fig. \ref{fig:spreads}, we display statistics on the spread $s(t)$ of three assets INTC, AAPL, and AMZN, resp. a large, medium and small-tick asset. Each row in the figure corresponds to one of the three assets. In each row, the left figure displays the empirical distribution of the spread $s(t)$ whereas the right figure displays the average spread $\overline{s}$ as a function of the time of the day as defined by \eqref{eqn:s_bar}. 

Unsurprisingly, we clearly see that spread is higher at open time and for large-tick assets, except for open time, the spread is basically equal to its constrained minimum (1 tick).  
As far as the empirical distribution is concerned, the shape appears to be very different for small, medium and large-tick assets. In order to measure the difference in shapes we make use of the index of dispersion (i.e., variance divided by mean) in Section \ref{spread}.}


{
\subsection{Additional illustrations on the mid-price change}}
\label{app:price}
\begin{figure}[h]
\centering
\includegraphics[width=.6\textwidth]{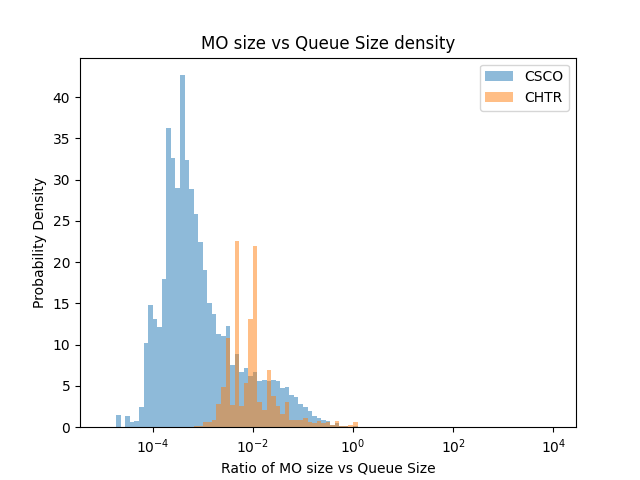}
\caption{Market Order Size vs Instantaneous Volume at Best: Probability density of the 
ratio of the average market order size with the average corresponding best limit for different tick-size assets. This ratio is clear higher for small-tick assets, implying thin quoting at the top of the LOB leading to multiple tick level mid-price moves.}
\label{fig:MO}
\end{figure}
In Section \ref{trades} we saw that the average mid-price change induced by a market order is much larger for small-tick assets than for large-tick assets. An indirect way of illustrating { one of the probable causes of} the same stylized fact consists of studying the ratio of the average market order size with the average corresponding best limit for different tick-size assets. 
This quantity clearly drives the queue-depletions which lead to the so-called trade-induced mid-price changes. Fig. \ref{fig:MO} shows the distribution of this ratio for two assets (one small-tick : CHTR and one large-tick : CSCO). It clearly displays { that for small small-tick assets market orders deplete a larger portion of the top of the LOB quoted volume than for large-tick assets. This implies that the quoted volume at the top for small-tick assets is relatively thinner than that of large-tick names. Large-tick assets typically have several queued up orders at the best, as we will illustrate in Section \ref{shape}, which leads to lower depletions of the top level of the LOB by incoming market orders. }

\subsection{Additional illustrations on the shape of the LOB}
\label{app:shape}\textbf{}
\begin{figure}[H]
\centering
\begin{subfigure}[c]{.32\textwidth}
\includegraphics[width=\textwidth]{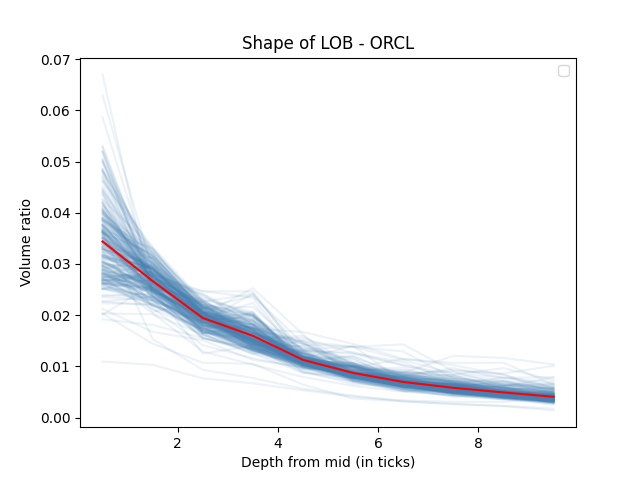}
\caption{ORCL (Large-tick) }
\label{fig:orclShape}
\end{subfigure}
\begin{subfigure}[c]{0.32\textwidth}
    \includegraphics[width=\textwidth]{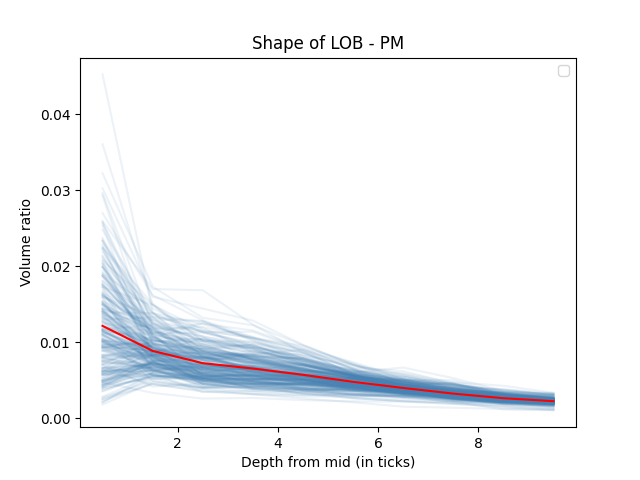}
    \caption{PM (Medium-tick) }
    \label{fig:pmShape}
\end{subfigure}
\begin{subfigure}[c]{0.32\textwidth}
    \includegraphics[width=\textwidth]{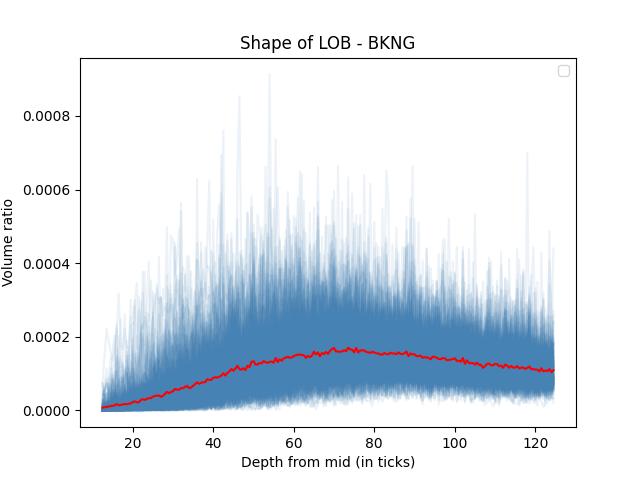}
    \caption{BKNG (Small-tick) }
    \label{fig:bkngShape}
\end{subfigure}

\caption{Per-asset Shape of the LOB: light blue lines are per day shapes and red is the average shape across one calendar year. { As we can clearly see in this set of figures, the shape of the LOB both per-day and over the year are quite similar leading to the notion of stationarity of the shape across calendar days. Secondly, we note that the trend we observed in Section \ref{shape} of the shape of the LOB getting flatter as we decrease the relative tick-size can be again seen here. }}
\label{fig:shapePerStock}
\end{figure}

{ In Fig. \ref{fig:shapePerStock}, we show the per-day and per-year average shape of three assets : ORCL, PM and BKNG resp. examples of large-, medium- and small-tick assets. First observation is that the yearly aggregated shape of the LOB seems very close to the per-day shape and therefore we can assume that the average shape of the LOB as shown in Section \ref{shape}, is stationary. Secondly, }the shape of the LOB decays from its maximum at half a tick from the mid-price for large and medium-tick assets, as illustrated in Fig.s \ref{fig:orclShape} and \ref{fig:pmShape}. Notably, the shape profile decays more rapidly for large-tick assets compared to medium-tick ones. On the other hand, for small-tick assets such as the one shown in Fig. \ref{fig:bkngShape}, the liquidity distribution is wider, with the maximum liquidity lying several ticks below the mid-price.

\begin{figure}[H]
\centering
\includegraphics[width=.75\textwidth]{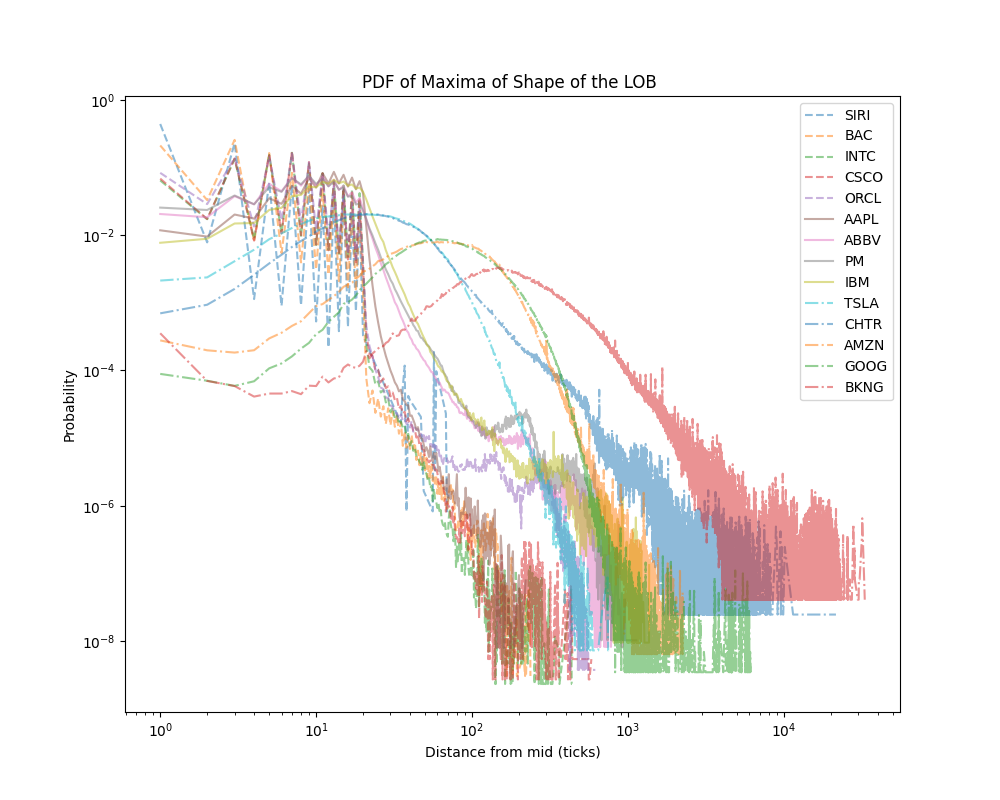}
\caption{Distribution of Shape Maxima (log-log): { this figure shows the empirical density of liquidity maxima, with greater variance as relative tick-size decreases. Despite frequent maxima near the mid-price in small-tick assets, Fig. \ref{fig:spreads} shows a low probability of a one-tick wide spread.}}
\label{fig:pdfMaxima}
\end{figure}

The empirical density of the maxima of the liquidity as a function of distance from the mid-price, shown in Fig. \ref{fig:pdfMaxima}, indicates that the variance in the location of maximum liquidity increases as the relative tick-size decreases. The variability in the location of this maximum provides insights into how deep the liquidity is distributed in the LOB. Even in small-tick assets, there are significant instances where the maximum liquidity lies just half a tick from the mid-price. However, as depicted in Fig. \ref{fig:spreads}, the time-weighted likelihood of a one-tick wide spread is quite low.

\subsection{{Additional illustrations on the sparsity of the LOB}}
\label{app:sparsity}
\begin{figure}[h]
\centering
\begin{subfigure}[c]{.49\textwidth}
\includegraphics[width=\textwidth]{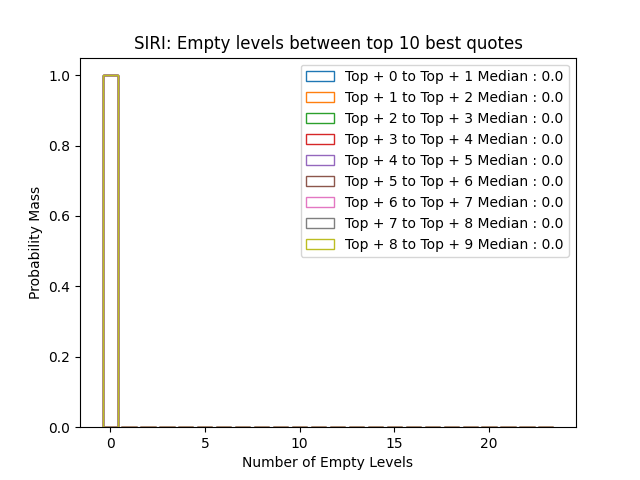}
\caption{SIRI (Large-tick) }
\label{fig:siriSparsity}
\end{subfigure}
\begin{subfigure}[c]{0.49\textwidth}
    \includegraphics[width=\textwidth]{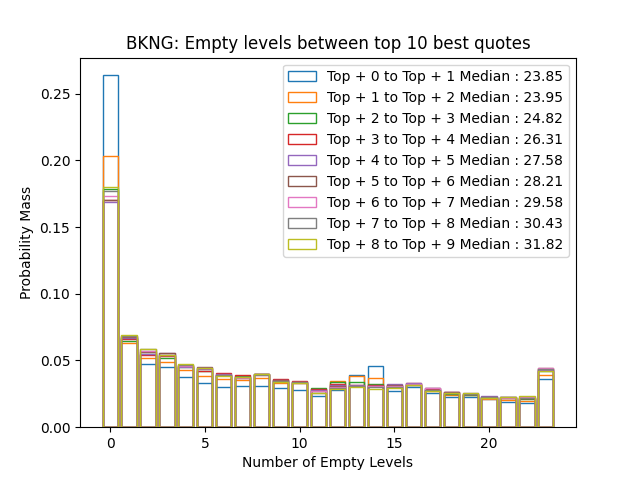}
    \caption{BKNG (Small-tick) }
    \label{fig:bkngSparsity}
\end{subfigure}
\caption{Sparsity of the LOB of a given asset : time-weighted distribution of the 
values $n_e(i,t)$ for the top 10 quotes ($i\in[0,10[]]$) for (a) a large-tick asset (SIRI) and (b) a small-tick asset (BKNG). We see the distributions hardly depend on the level of the quotes and that the smaller the relative tick-size, the sparser the LOB : the distributions are somewhat flat for small-tick assets and very concentrated around 0 for large-tick assets.  \st{(?? We don't see the numbers on the axis and the labels ??)}
}
\label{fig:emptyLevelsPerStock}
\end{figure}
\begin{figure}[h]
\centering
    \includegraphics[width=.6\textwidth]{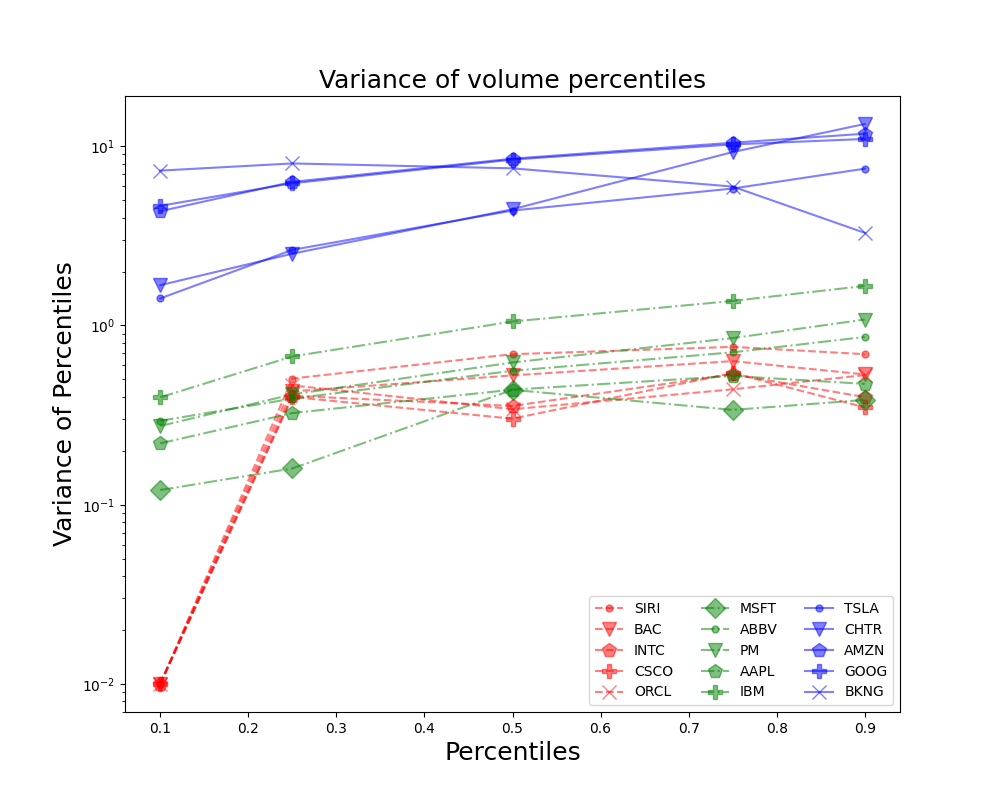}
    \caption{Variance of shape of 4 LOB quantiles (10\%, 25\%, 75\%, and 90\%) (log scale) for the 15 assets. The smaller the tick, the higher the variance, the sparser the LOB. 
     }
    \label{fig:varPctiles}
\end{figure}
In this Section we show that the sparsity pattern that was illustrated in Fig. \ref{fig:emptyLevels} studying $n_e(1,t)$ (i.e., the number of empty queues between the best quote and the next non empty quote) is confirmed for deeper levels in the LOB. 

For doing so, we compute $n_{e}(i,t_n)$ for the top ten quotes ($i\in [1,10[]$) at each time $t_n$ this quantity is changing. 
Fig. \ref{fig:emptyLevelsPerStock} shows the empirical distributions of this quantity for each level $i$ for the SIRI stock (large-tick) and for the BKNG stock (small-tick asset). We see that for a given asset the distribution of empty levels is similar across all depths of the LOB, moreover, we confirm that the smaller the relative tick-size, the sparser the LOB : the distributions are somewhat flat for small-tick assets and very concentrated around 1 for large-tick assets.  

A sparse LOB induces a lot of variablity in its shape. Thus another natural proxy for the sparsity of the LOB is to measure this variability. We choose to the empirically estimate the variance of four quantiles (10\%, 25\%, 75\%, and 90\%) of the shape of the LOB. 
Fig. \ref{fig:varPctiles} shows clearly that the larger the relative tick-size the smaller the variances. 

\subsection{{ Additional illustrations on the endogeneity of the LOB}}
\label{app:endogeneity}
\begin{figure}[h]
\centering
\begin{subfigure}[c]{0.49\textwidth}
    \includegraphics[width=\textwidth]{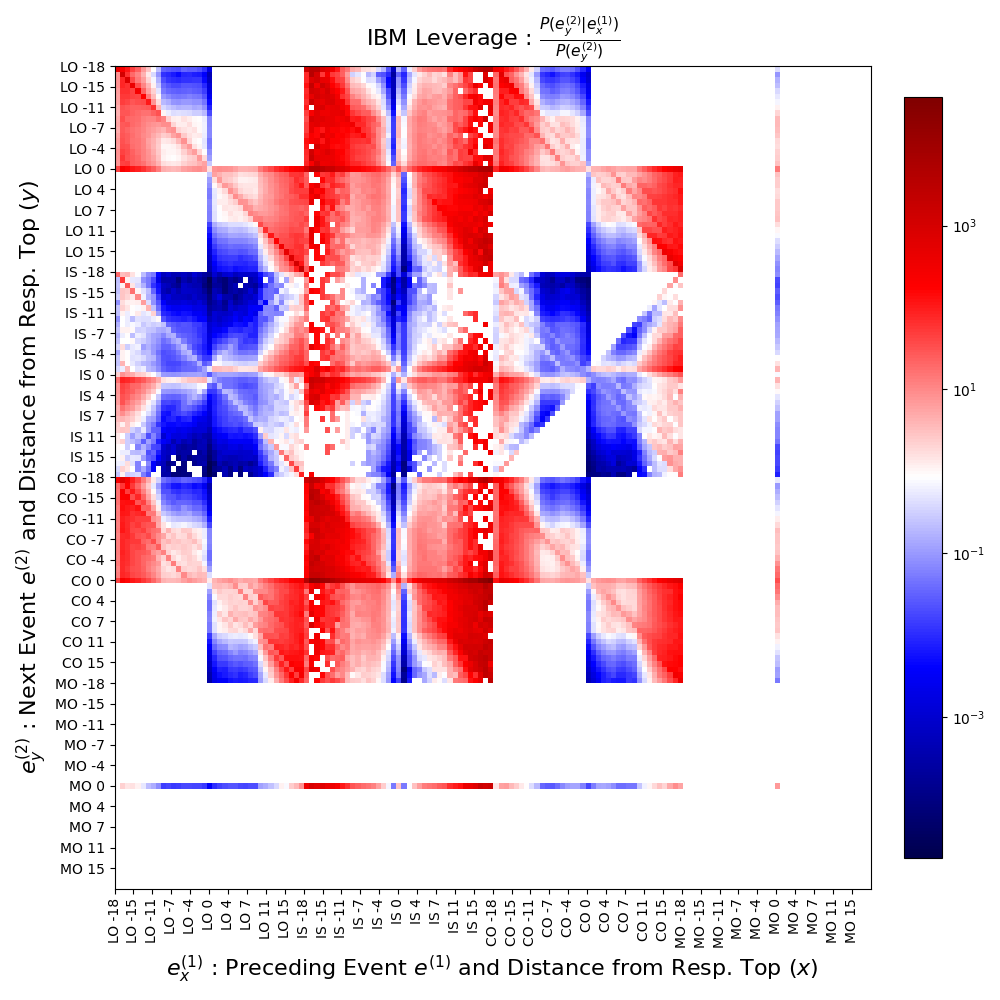}
    \caption{IBM (Medium-tick) }
    \label{fig:ibmlev}
\end{subfigure}
\begin{subfigure}[c]{0.49\textwidth}
    \includegraphics[width=\textwidth]{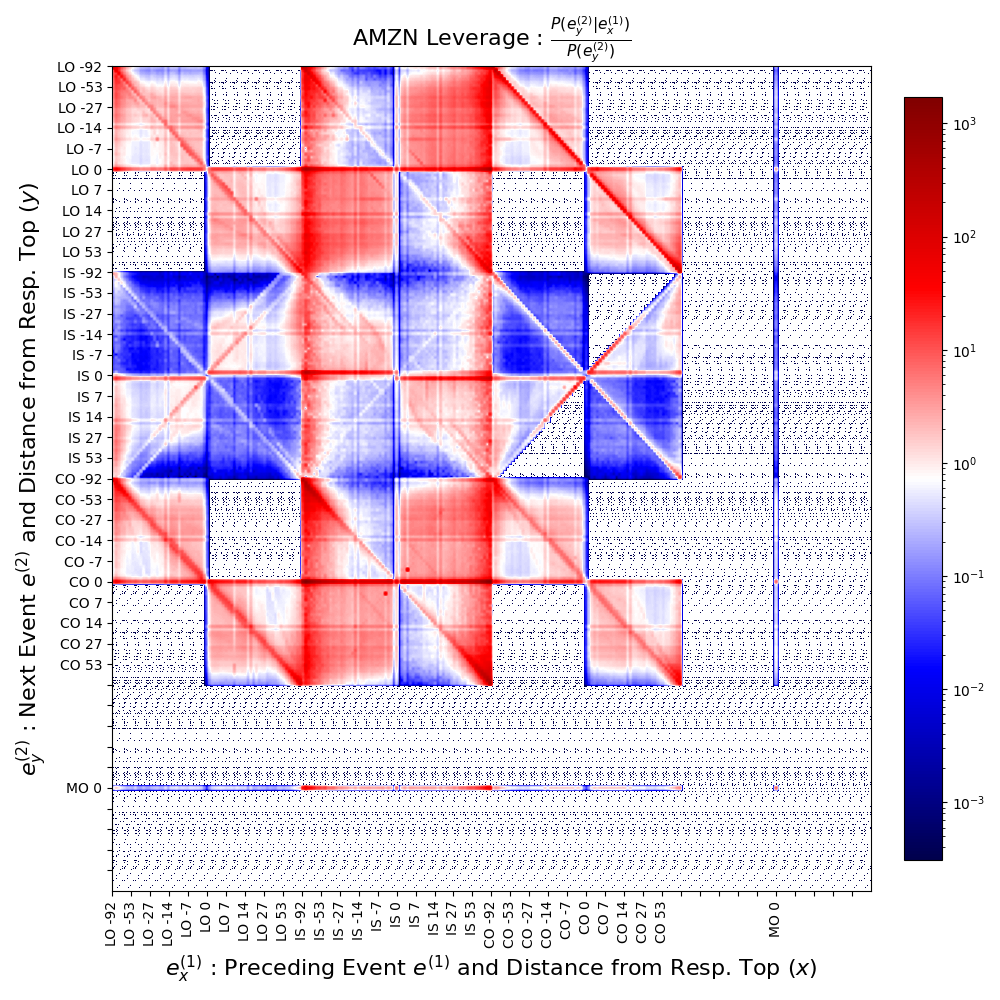}
    \caption{AMZN (Small-tick) }
    \label{fig:amznlev}
\end{subfigure}
 \caption{Leverage (from top): Darker the hue, more the leverage. { In continuation with Fig. \ref{fig:leverageTop_1}, we compare the leverage plots of a medium-tick asset, IBM, to a small-tick asset, AMZN. Note the remarkable similarity in the structure of these two plots as well as its similarity to Fig. \ref{fig:leverageTop_1}. }}
\label{fig:leverageTop}
\end{figure}

\highlighttwo{In Fig. \ref{fig:leverageTop}, we continue the depiction from Section \ref{leverage} of uniformity of leverage across all assets albeit with different scaling. { We can see here the observation that medium-tick $LO \rightarrow CO$ leverage plot has both diagonal and off-diagonal elements. This implies that the transition of the structure of $LO \rightarrow CO$ from large-tick to medium- and therafter small-tick is smooth. } }

\newpage
\section{Simulation Study Parameters (continued):} \label{params}

\begin{table}[h]
\centering
\begin{tabular}{|l|r|}
\hline
\textbf{Event}          & \textbf{$\mu^{(.)}$ ($s^{-1}$)} \\ \hline
$LO_{\text{Ask}_{D}}$             & 0.86           \\ \hline
$CO_{\text{Ask}_{D}}$             & 0.32           \\ \hline
$LO_{\text{Ask}_{T}}$              & 0.33           \\ \hline
$CO_{\text{Ask}_{T}}$              & 0.48           \\ \hline
$MO_{\text{Ask}}$                   & 0.02           \\ \hline
$LO_{\text{Ask}_{IS}}$         & 0.47           \\ \hline
$LO_{\text{Bid}_{IS}}$      & 0.47           \\ \hline
$MO_{\text{Bid}}$                  & 0.02           \\ \hline
$CO_{\text{Bid}_{T}}$              & 0.48           \\ \hline
$LO_{\text{Bid}_{T}}$              & 0.33           \\ \hline
$CO_{\text{Bid}_{D}}$             & 0.32           \\ \hline
$LO_{\text{Bid}_{D}}$             & 0.86           \\ \hline
\end{tabular}
\caption{Exogenous Intensity}
\label{tab:tabExo}
\end{table}

\highlighttwo{In Table \ref{tab:tabExo}, we provide the exogenous intensity part of the Hawkes Process parameters ($\mu^{(.)}$) and in Fig. \ref{fig:amznNorms}, we depict the Kernel Norm matrix according to standard research practice (\cite{bacry2016estimation, jain2023hawkes}). Since the number of parameters is quite large (3 $\times$ 72 for the 72 power law kernels), we do not report them in detail.} 
\begin{figure}[H]
\centering
\includegraphics[width=0.75\textwidth]{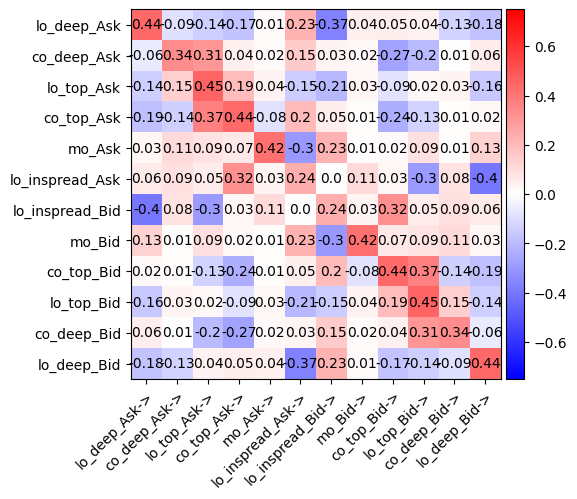}
\caption{Norms of Kernel: AMZN.OQ}
\label{fig:amznNorms}
\end{figure}

\section{Assumptions and Their Empirical Evidence or Counter-Evidence:} \label{assump}
In Section \ref{results} we underlined the universality of the model by illustrating, with numerical experiments, the fact that the model we propose is general enough so that controlling three parameters $\alpha, \beta$ and $\hat{\eta}_{(.)}$ enables us to simulate an LOB with Large or small-tick characteristics. However we made a  number of simplifying assumptions in the model construction to foster a straightforward analysis, an easier calibration, and an understandable simulation study. In the following we check if some of these assumptions are realistic as respect to empirical observations. 

\begin{assumption}
    Hawkes Parameters $(\alpha, \beta, \mu^{(.)}, \phi^{(.)}(t))$ and the LOB variables (i.e. LOB state variables $(m^{(.)}_{(.)}(t), Q^{(.)}_{(.)}(t))$, order sizes $(\kappa_{(.)})$, and order arrival distances from top $(\eta_{(.)})$ are independent of each other.
\end{assumption}

\begin{figure}[h]
\centering
\begin{subfigure}[c]{0.7\linewidth}
\includegraphics[width=\linewidth]{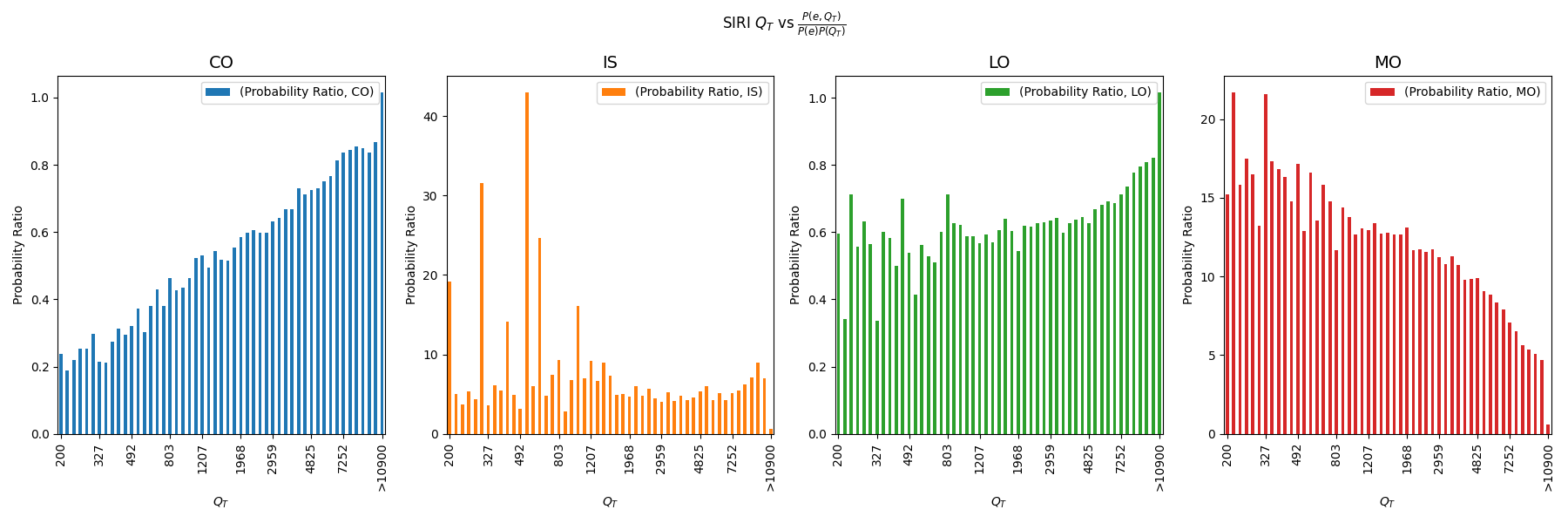}
\caption{SIRI $Q_T$ }
\label{fig:QT}
\end{subfigure}
\begin{subfigure}[c]{0.7\linewidth}
\includegraphics[width=\linewidth]{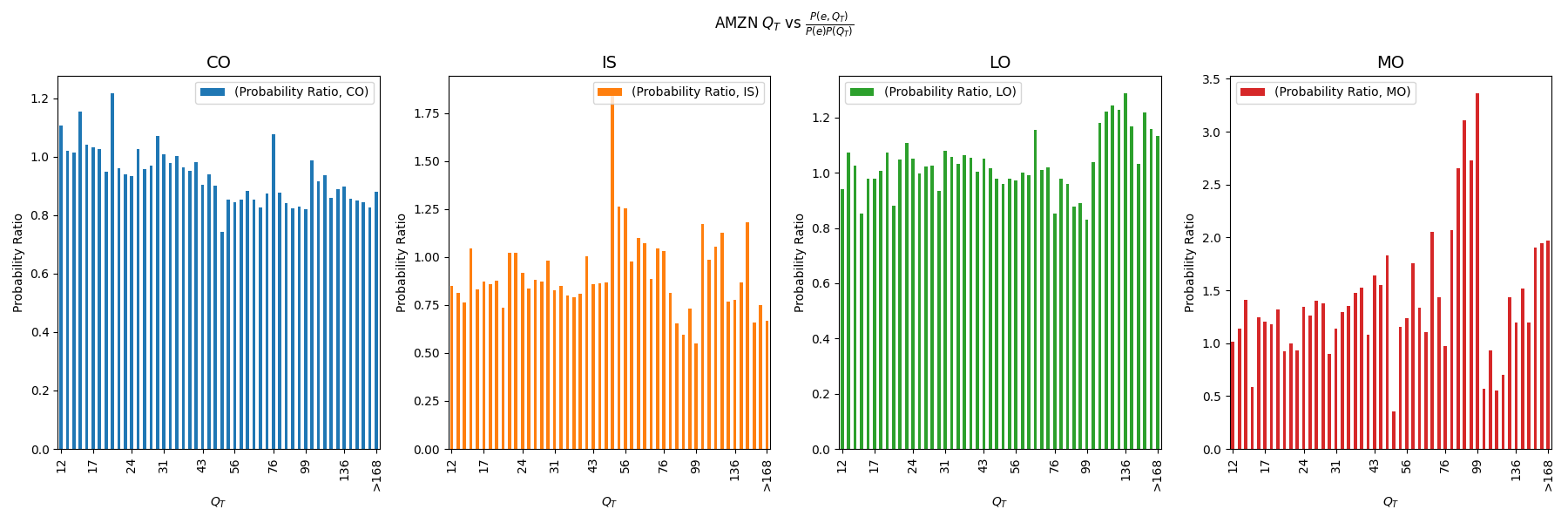}
\caption{AMZN $Q_T$ }
\label{fig:QT}
\end{subfigure}
 \caption{\highlighttwo{\textbf{Testing Assumption 1: Queue Reactiveness:} This independence test shows that for large-tick assets like SIRI, the intensity of events like COs, LOs and MOs, and the queue sizes at the top of the LOB are not independent variables. Unsurprisingly, the same is not true for the deeper volumes in a large-tick asset. For a small-tick asset like AMZN, the picture becomes less clear for $Q_T$ however for MOs, one can observe a clear dependence among the variables.  }}
\label{fig:qReact}
\end{figure}


Queue reactiveness in the LOB, i.e. the notion that event intensities depend directly on the current sizes of the queues in the LOB ($Q^{(.)}_{(.)}(t)$), has been relatively well established in literature (\cite{pakannen2022}, \cite{wu2019queue}, \cite{huang2015}). It is evident from these papers that the intensities of LOB events depend on the current queue sizes of the LOB. However, it is unclear whether the number of past events alone (i.e., without considering individual quantities) is a reasonable proxy for the cause of this dependence. We would like to note that the Hawkes Process methodology implicitly makes the intensity depend on the count of past events. 

We perform tests on this assumption in Fig. \ref{fig:qReact}
where we check the dependence of the LOB state variables to the events' likelihood by plotting the ratio of the joint probability of an event happening with the LOB state variable $v$ i.e. $P(e, v)$ with the product of the individual unconditional probabilities i.e. $P(e)P(v)$, against various categories of the width variable. This ratio informs us whether the two random variables are independent or not. While this test is not exhaustive or strict, it does inform us of the interdependence of two random variables in a simplified manner. If the ratio is near one, we have some albeit insufficient evidence to consider the two variables independent. However if the ratio is significantly different from one, we can conclude that the two variables are not independent. From this figure, we see that the queue reactiveness is very evident for top of the LOB queue size in large-tick assets however it is very less significant in the small-tick situation. 

Further we show the same analysis for $Q_D$ instead of $Q_T$ in the Fig. \ref{fig:QD}. Unsurprisingly, large-tick assets dynamics seem to be unreactive to the deeper quoted volumes but there is a definite structure, particularly in MOs, for small-tick assets. Additionally, we see that events of the type IS, CO and MO have a definitive trend with increasing $m_T$ (Fig. \ref{fig:mT}). There is a U-shape trend observed in IS and MO order types with increasing $m_D$ (see Fig. \ref{fig:mD}). Finally we observe a strong dependence of MO's probability with increasing $\eta_{IS}$ (Fig. \ref{fig:etaIS}). However we must note that the trend becomes significant only in the right tail of the quantities which are low probability states. We do not see any clear trend when we do the same analysis for $\kappa_{LO}, \kappa_{MO}$ (not illustrated). We observe similar characteristics in medium-tick assets and all other small-tick assets as well. 
\begin{figure}[h]
    \centering
\begin{subfigure}[c]{0.7\linewidth}
\includegraphics[width=\linewidth]{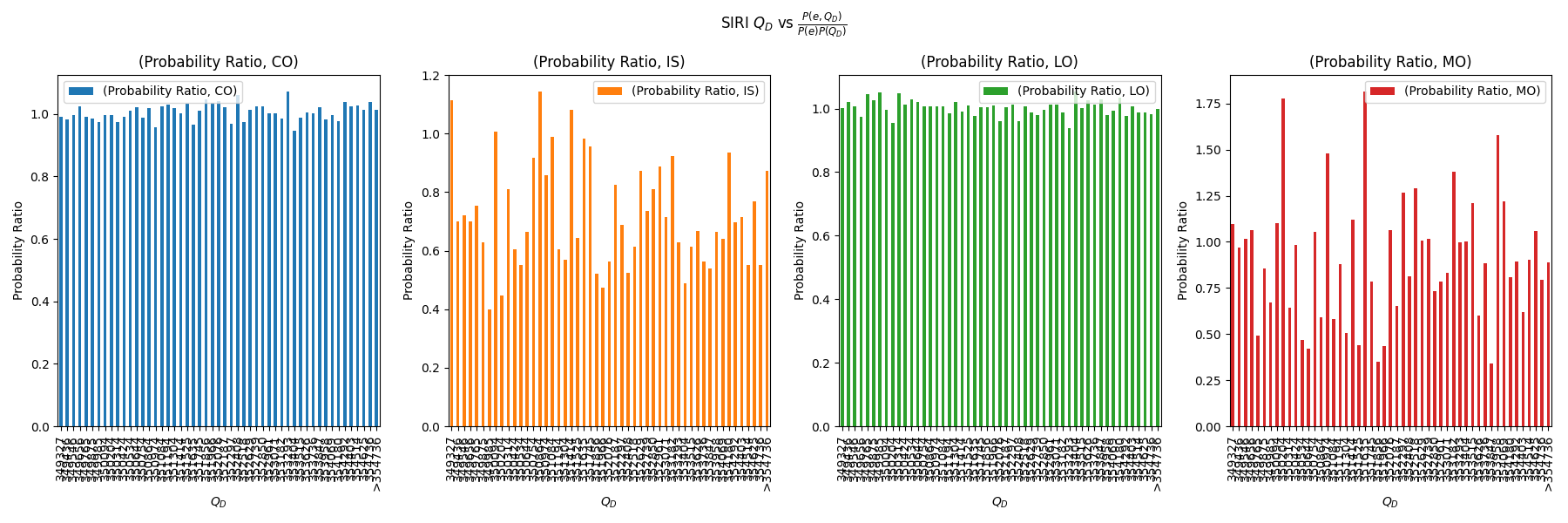}
\caption{SIRI $Q_D$ }
\label{fig:QD1}
\end{subfigure}
\begin{subfigure}[c]{0.7\linewidth}
\includegraphics[width=\linewidth]{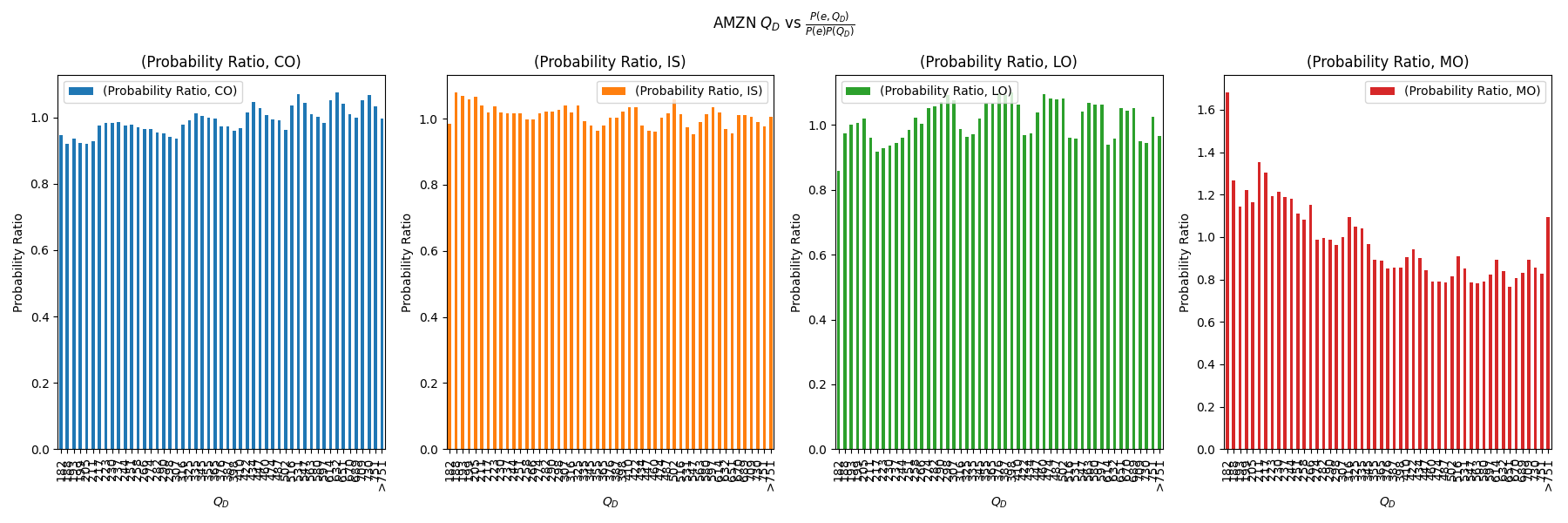}
\caption{AMZN $Q_D$ }
\label{fig:QD2}
\end{subfigure}
 \caption{\highlighttwo{\textbf{Testing Assumption 1: Queue Reactiveness:} 
 For $Q_D$ too we note that the MOs intensity is quite clearly not independent in the case of small-tick assets.  }}
 \label{fig:QD}
\end{figure}




\begin{figure}[h]
\centering
\begin{subfigure}[c]{.7\linewidth}
\includegraphics[width=\linewidth]{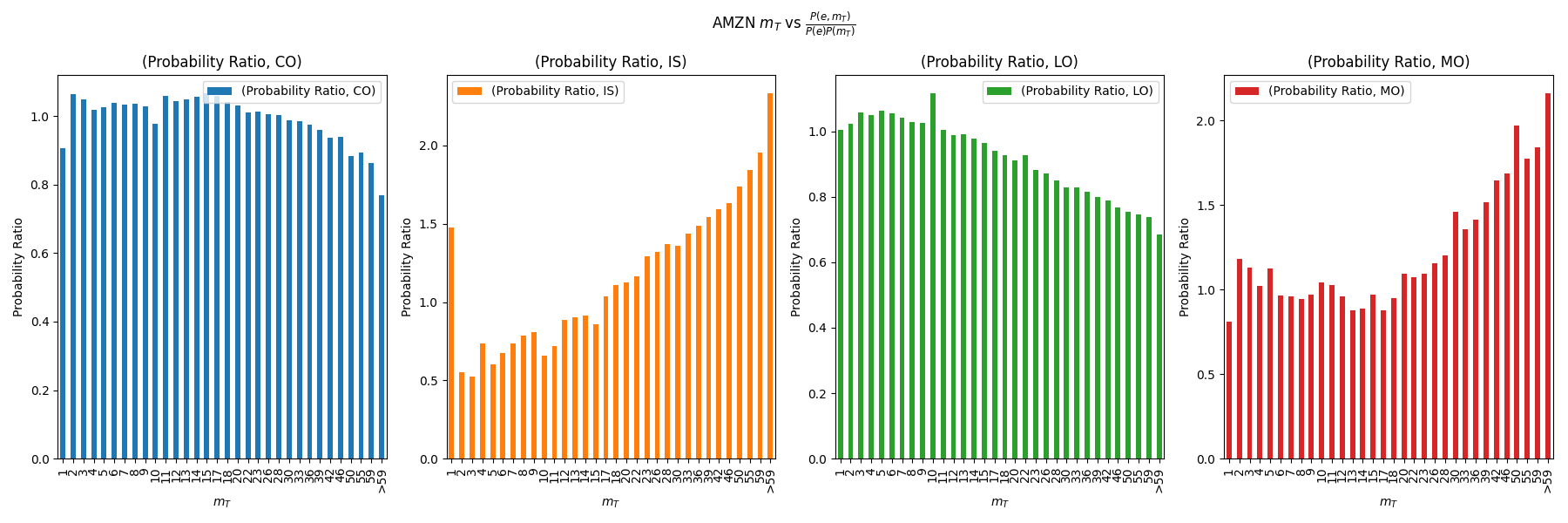}
\caption{$m_T$ }
\label{fig:mT}
\end{subfigure}
\begin{subfigure}[c]{.7\linewidth}
\includegraphics[width=\linewidth]{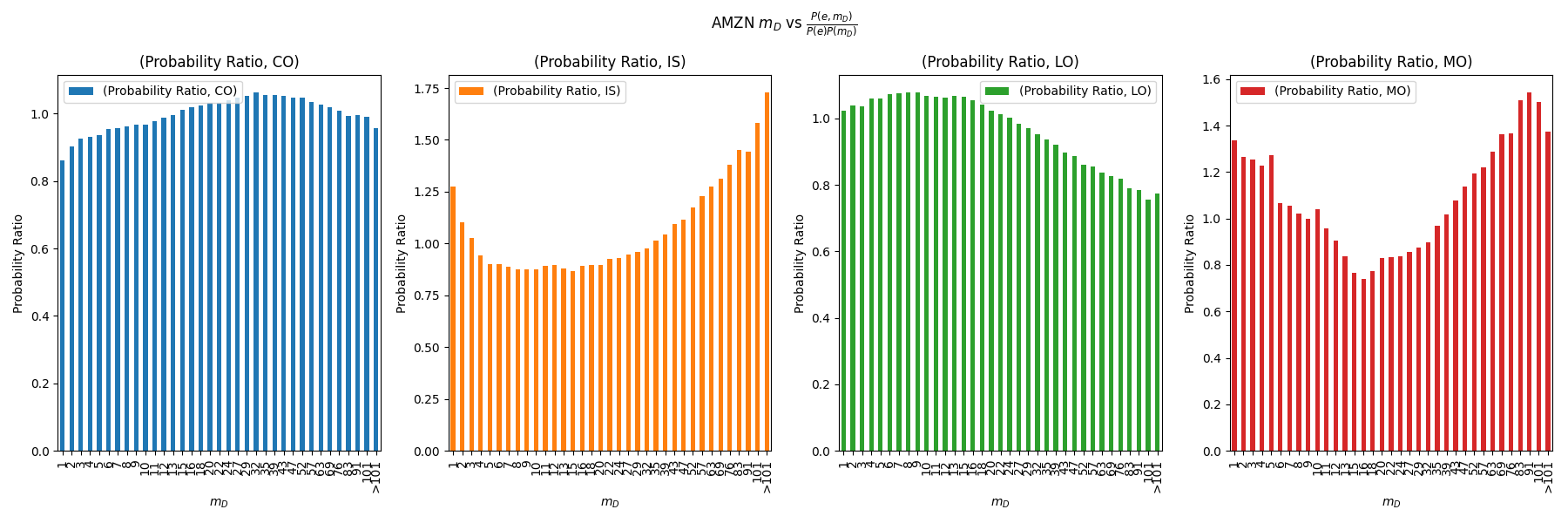}
\caption{$m_D$ }
\label{fig:mD}
\end{subfigure}
\begin{subfigure}[c]{.7\linewidth}
\includegraphics[width=\linewidth]{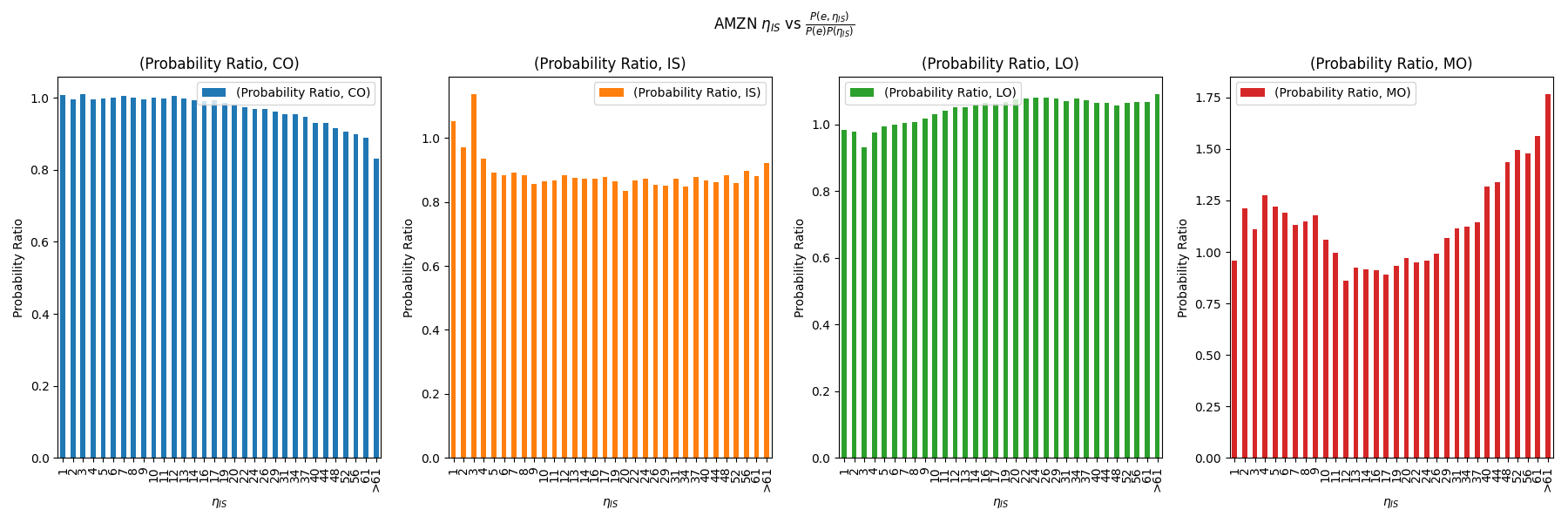}
\caption{$\eta_{IS}$ }
\label{fig:etaIS}
\end{subfigure}
 \caption{\highlighttwo{\textbf{AMZN: Testing Assumption 1:} We do the same analysis for a small-tick asset AMZN to check the dependence between the state variables and the intensity of various events. It is evident from figures (a) and (b), that all the order intensities are related to the current width of the top queue ($m_T(t)$) as well as the deeper queue ($m_D(t)$). However for the dynamic sparsity variable, we note that there seems to be a lot of evidence (except for when $\eta_{IS}$ is very large which is a low probability event) for independence between the intensity and the previous $\eta_{IS}$ which is the most recent price improvement (in ticks) made by an in-spread LO.}}
\label{fig:assum1}
\end{figure}
{Therefore we conclude that though for LOs and COs, we can assume the LOB state variables except $Q_T$ to be independent of their intensity functions, there exists a clear dependence of intensities of in-spread orders on the current $m_T, m_D$. MOs seem to also be very dependent on the state variables. Thus we identify Assumption 1 to be too strong for empirical data and reserve the direction of relaxing this assumption for future work.}

\begin{figure}[h]
\centering
\begin{subfigure}[c]{0.32\textwidth}
\includegraphics[width=\textwidth]{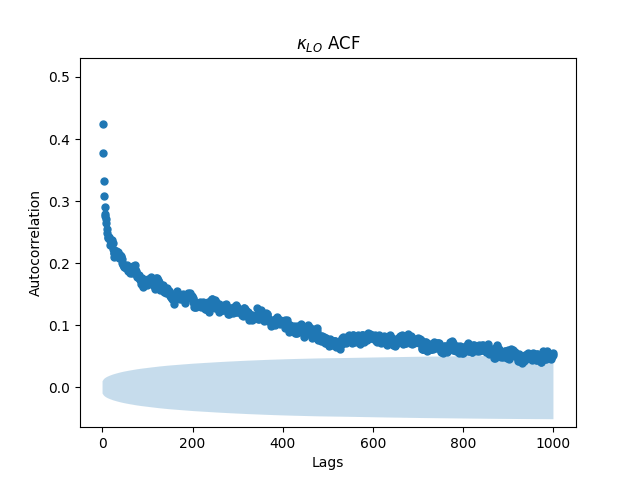}
\caption{$\kappa_{LO}$ }
\label{fig:q_LO}
\end{subfigure}
\begin{subfigure}[c]{0.32\textwidth}
\includegraphics[width=\textwidth]{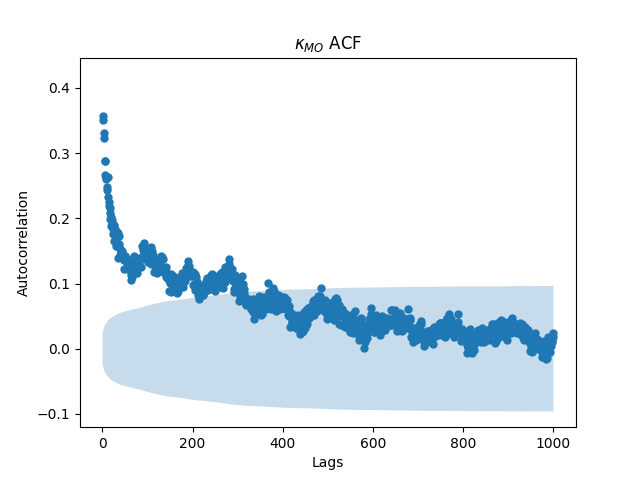}
\caption{$\kappa_{MO}$ }
\label{fig:q_MO}
\end{subfigure}
\begin{subfigure}[c]{0.32\textwidth}
\includegraphics[width=\textwidth]{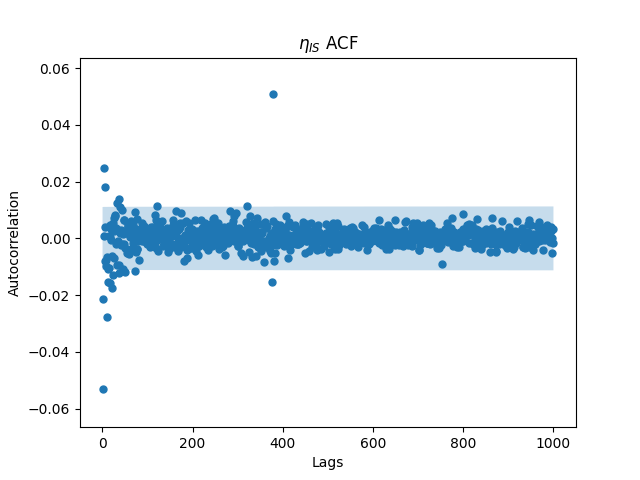}
\caption{$\eta_{IS}$ }
\label{fig:eta_IS}
\end{subfigure}
 \caption{\highlighttwo{\textbf{CHTR: Testing Assumption 2 (i), (ii) and (iii) - } Clearly the ACF of order sizes ($\kappa_{(.)}$) is non neglible for 10s of lags which means that the order sizes are probably not i.i.d. However we do see the ACF of price improvements by IS orders ($\eta_{IS}$) quickly dying down to 0, giving some support to the assumption 3.}}
\label{fig:assum2}
\end{figure}



\begin{assumption}
    (i) Order sizes, (ii) Order arrival distances and (iii) Unseen Queue sizes on queue depletion, are i.i.d. random variables.
\end{assumption}

To test the i.i.d. assumption, we measure the autocorrelations at various lags of the timeseries of the quantity. It is clear from Fig. \ref{fig:assum2} that Assumption 2 (i) is clearly violated by the empirical data. Assumption 2 (ii), however, seems to have some support from the data since the autocorrelations are absent. With regards to Assumption 2 (iii), we note that this is a modelling choice.











\end{document}